\documentclass[useAMS,usenatbib,usegraphicx]{mn2e}
\usepackage{epsfig}
\usepackage{amssymb}
\usepackage{times}
\usepackage{aas_macros} %needed to define journal macros generated by
\usepackage[usenames]{color}

\title[Variations in the roAp star HD~217522] {Short time-scale frequency and amplitude variations in the pulsations of an roAp star: HD~217522}
\author[R. Medupe et al.]
{R. Medupe$^{1}$\thanks{E-mail: rodney.medupe@nwu.ac.za}, D. W. Kurtz$^{2}$, V. G. Elkin$^2$, Z. Mguda$^{3,4}$, G. Mathys$^5$\\
$^{1}$Department of Physics, North West University, Dr. Albert Luthuli Drive, Mahikeng, 2735, South Africa\\
$^{2}$Jeremiah Horrocks Institute, University of Central Lancashire, Preston PR1~2HE, UK\\
$^{3}$Department of Astronomy, University of Cape Town, Rondebosch, 7701, South Africa\\
$^4$South African Astronomical Observatory, P.O. Box 9, Observatory, 7925, Cape Town, South Africa\\
$^{5}$European Southern Observatory, Casilla 19001, Santiago 19, 
Chile\\
}

\begin{document}

%\date{Draft \today ; Accepted . Received ; in original form }

%\pagerange{\pageref{firstpage}--\pageref{lastpage}} \pubyear{2008}

\maketitle

\label{firstpage}

\begin{abstract}
Photometric observations of HD~217522 in 1981 revealed only one pulsation frequency $\nu_1=1.21529$~mHz. Subsequent observations in 1989 showed the presence of an additional frequency $\nu_2=2.0174$~mHz. New observations in 2008 confirm the presence of the mode with $\nu_2=2.0174$~mHz. Examination of the 1989 data shows amplitude modulation over a time scale of the order of a day, much shorter than what has been observed in other roAp stars. High spectral and time resolution data obtained using the VLT in 2008 confirm the presence of $\nu_2$ and short term modulations in the radial velocity amplitudes of rare earth elements.  This suggests growth and decay times shorter than a day, more typical of solar-like oscillations. The driving mechanism of roAp stars and the Sun are different, and the growth and decay seen in the Sun are due to stochastic nature of the driving mechanism. The driving mechanism in roAp stars usually leads to mode stability on a longer time-scale than in the Sun. We interpret the reported change in $\nu_1$ between the 1982 and 1989 data as part of the general frequency variability observed in this star on many time scales. 
\end{abstract}

\begin{keywords}
asteroseismology -- stars: chemically peculiar -- stars: individual: HD~217522 -- stars: oscillations (including pulsations)
\end{keywords}

\section{Introduction}

HD~217522 is a member of the class of rapidly oscillating Ap (roAp) stars that are characterized by high overtone p-mode pulsations with periods ranging from 5.6 to 23~min and photometric peak-to-peak amplitudes under 13~mmag in Johnson $B$. Their radial velocity amplitudes in particular spectral lines range up to 5000~m~s$^{-1}$ \citep{elkin2005}. They have strong global surface magnetic fields up to 24.5~kG \citep{hubrig2005}. \citet{houk1978} classified this star as an Ap(Si)Cr star, although she also indicated that the feature she observed at $\lambda 4128$~\AA\ might also be Eu instead of Si. \citet{gelbmann1998} showed from his spectroscopic analysis that Eu in HD~217522 is overabundant by $10^3$ compared to the Sun, supporting the identification of the $\lambda 4128$~\AA\ feature as due to Eu \citep{kurtz1983}. The Str\"omgren colours \citep{martinez1993} indicate that HD~217522 is a cool Ap star. Martinez gives $V = 7.525$, $b-y=0.289$, $m_1=0.227$, $c_1=0.484$ and $\beta = 2.691$, yielding $\delta m_1 = -0.056$ and $\delta c_1 = -0.015$, both typical of the heavily line-blanketed cool Ap stars. A model atmosphere analysis of HD~217522 confirms that it is one of the coolest members of the roAp stars with $T_{\rm eff} = 6750 \pm 100$~K (\citealt{ryabchikova2004}). From our own spectra reported in this paper we estimate $T_{\rm  eff} = 6600 \pm 100$~K and $\log g = 4.2$.  

Interestingly, spectroscopic investigation of HD~217522 by \citet{hubrig2002} showed it to be very similar to Przybylski's star (HD~101065) to the point where they dubbed it one of the `Twins of Przybylski's star'. Przybylski's star is a 12-min period roAp star with one of the most complex spectra known, and a surface magnetic field of 2.3~kG \citep{cowley2000}. Its spectrum is rich with lines of lanthanides and suffers severe blending of iron-peak elements. The latest attempts to model the atmosphere of HD~101065 and to measure its abundances by \citet{shulyak2010} yielded $T_{\rm eff} = 6400$~K, but failed to yield a good value of surface gravity due to the difficulty in modelling the hydrogen lines. In light of progress made in modelling the atmospheres and rare earth element opacities of Ap stars in the last decade, \citet{shulyak2010} concluded that HD~101065 is not special among the roAp stars other than that it is the coolest.

Both HD~101065 and HD~217522 show no definite observable rotational spectral, magnetic or light variations that are typically seen in many Ap stars. This suggests that the two stars either have very long rotational periods (longer than two decades in the case of HD~217522) or that each has a rotation axis inclined nearly pole-on to the line-of-sight. Ap stars that show rotational light variations caused by permanent spots are known as $\alpha^2$~CVn stars. Recent long-term photometric monitoring (60 nights over a three year period) of HD~217522 by \citet{heerden2012} showed no evidence of rotational light variations, hence we have no information about the rotational period, other than the mild constraint of $v \sin i = 3$~km~s$^{-1}$ (section~\ref{sec:spatial}). 

The mean longitudinal magnetic field of HD~217522 was first measured by \citet{mathyshubrig1997} to be $-394 \pm 124$~G. Mathys (unpublished) derived a value of $-662\pm69$~G from an observation obtained 5.3 years later, which is not significantly different at the achieved accuracy. The mean quadratic field strength of $2017 \pm 403$~G determined by \citep{hubrig2002} is not inconsistent with the 1.5~kG upper limit of the mean field modulus reported by \citet{ryabchikova2008}. Like many other roAp stars, the H$\alpha$ lines of HD~217522 show core-wing anomaly \citep{cowley2001}. This anomaly happens in the line profiles of the hydrogen lines where their shapes are described by broad wings and narrow cores such that the wings and cores cannot be modelled by a single temperature. Their Ca~\textsc{ii} K lines show a wing-nib anomaly \citep{cowley2006} where the slope of the lines changes dramatically near the line cores to produce a `nib'. Both anomalies are explained by stratified abundances and unusual temperature structure (see \citealt{kochukhov2002}; \citealt{cowley2006}).

Detailed spectroscopic time-series analyses of 9 of the $\sim$60 known roAp stars show the presence of frequencies in the radial velocity amplitude spectra of the Pr~\textsc{iii} lines and H$\alpha$ line core that are not seen in the broadband photometry of these stars (\citealt{kurtz2006}; \citealt{kurtz2007a}). One of the possible explanations proposed by \citet{kurtz2006} is that the spectroscopically visible modes have nodes in the atmospheric layers that are sampled by the photometry. However, this seems improbable since this behaviour occurs in all of the 9 roAp stars observed. Furthermore, if the new frequencies are caused by pulsation modes, their separations should be half the large separation according to the asymptotic theory (see discussion below) and the resulting amplitude modulation will be periodic on the time scales of between 6 and 12~h. \citet{kurtz2007a} did not see evidence of any of this in their spectroscopic data on the roAp star HD~134214. The second possible explanation for frequencies seen in radial velocities but not in photometry is that the Pr~\textsc{iii} and Nd~\textsc{iii} spots that occur near magnetic poles act as spatial filters for higher $\ell$ modes that are visible in spectroscopy but are averaged out in the broadband photometry. The above two ideas suggest that the frequencies under discussion should be stable (although their amplitudes should not necessarily be so.)

Finally, it is also possible that the frequencies that are only seen in the radial velocities of Pr~\textsc{iii} and Nd~\textsc{iii} and H$\alpha$ lines are a result of losses and gains in the pulsation energy on time scales as short as a few pulsation cycles. In this case, it should be expected that long observing runs will yield frequencies that are not stable and reproducible, but that depend on the details of energy gains and losses at the times of observations. These ideas are important for understanding the short time-scale amplitude modulation in HD~217522 that we report in this paper.

The first photometric monitoring of HD~217522 over 74 nights in 1982 revealed the presence of a single pulsation frequency of $\nu_1 = 1.21509$~mHz \citep{kurtz1983}. Interestingly, when \citet{kreidl1991} observed this star in Johnson $B$ over a period of 13 nights in 1989 they discovered a second frequency at $\nu_2 = 2.0174~$mHz in addition to the one observed in 1982. This phenomenon was also noticed in HD~134214 where all previous photometric observations of this star revealed it to be singly periodic, and then  \citet{kurtz2007a} found a second period in the Johnson $B$ photometry of this star. They also found four other frequencies only visible in the amplitude spectra of radial velocities of rare element lines and the H$\alpha$ line. Another interesting star that shows non-periodic amplitude modulation is HD~60435 \citep{matthews1987}. HD~60435 also has a rich frequency spectrum with equally spaced frequencies that are consistent with models of asymptotic p-mode frequencies for A stars. \citet{matthews1987} found that the mode lifetimes for HD~60435 were shorter than the rotation period of this star of 7.6793~d \citep{kurtz1990}.

Not all roAp stars show mode instability; for example, the last extensive photometric study of HR~3831 by \citet{kurtz1997b} reported cyclic frequency changes (which might be indicative of a magnetic cycle), but no stochastic amplitude modulation on this star. Another example of a star with stable modes is $\alpha$~Cir. \citet{bruntt2009} did not find any short time scale amplitude modulation with 84~d of high precision WIRE satellite photometric data. The clearest case of this is for KIC~10195926, an roAp star discovered in the {\it Kepler} Mission data that shows high stability in both amplitude and frequency over the 4-yr {\it Kepler} data set (unpublished; see \citealt{kurtz2011} for an earlier analysis of this star). The pulsation amplitudes are stable in KIC~10195926 to a few $\umu$mag over 4~yr, showing also that there is no frequency variation. Thus a problem we discuss in this paper is that of the amplitude and frequency instability in HD~217522, even on a time scale of hours, when other roAp stars can be stable for years. 

There is another mystery. No other roAp star is like HD~217522. The other multiperiodic roAp stars have modes of alternating even and odd degree that are consecutive in frequency; i.e. the separations are half the large separation -- typically about $30-40$~$\umu$Hz. The best-studied example of this is HR~1217 \citep{white2011}. The two pulsation modes of HD~217522 (discussed above with frequencies of $\sim$1.2~mHz and 2.0~mHz) are separated by 800~$\umu$Hz. We expect, therefore, that there are many unseen or unexcited modes with frequencies between the two seen in photometry. 

Spectroscopic radial velocity studies of roAp stars have shown much higher signal-to-noise ratios for frequency analysis than broadband photometric studies. An extreme case showing this is the naked-eye star $\beta$~CrB for which no photometric pulsation variations were found, despite many attempts, yet spectroscopic radial velocity variations showing roAp pulsations with a period of 16.2~min were ultimately detected with amplitudes of only 30~m~s$^{-1}$ in individual lines with the best spectroscopic data available (\citealt{hatzes-mkrtichian2004}; \citealt{kurtz2007b}). 

On this basis, we predicted that high spectral resolution, high time resolution, high signal-to-noise spectra of HD~217522 would allow us to resolve and detect all of the apparently `missing modes' that lie between $\nu_1$ and $\nu_2$. Asymptotic theory for low degree, $\ell$, and high order $(n >> \ell)$ p~modes gives their eigenfrequencies as:
\begin{equation}
\nu_{n,\ell} \approx \Delta\nu \left(n+\ell/2+\epsilon \right)-
\left[\ell~(\ell+1) + \delta \right] A \frac{\Delta \nu^2}{\nu_{n,\ell}}
\end{equation}
(\citealt{tassoul1980}, \citealt{tassoul1990}) where $\epsilon$, A and $\delta$ are constants that depend on the equilibrium structure of the star. The quantity $\Delta \nu$ is the large frequency separation which is the inverse of twice the sound travel time between the surface and core of the star. From equation~(1) modes of even or odd $\ell$ values with consecutive $n$ are separated by the large separation. For HD~217522 we expect  $40 \le \Delta\nu \le 60$~$\umu$Hz, similar to other roAp stars and models. With an 800~$\umu$Hz separation between the two known modes, there could have been about dozens more mode frequencies between them, which would have given this star the richest frequency set for asteroseismic modelling of any roAp star. We obtained continuous spectroscopic observations for 10.1~h, which can resolve frequency separations down to 26~$\umu$Hz. We did {\it not} find the `missing modes'; they are not there at the precision we reached. Hence this remains a mystery for HD~217522. However, we did find that the amplitude and frequency of the pulsations are variable on a timescale of hours, giving further insight into the problem of understanding the photometric data. We show these results and discuss implications in this paper. 

The photometric data we investigate were obtained in 1982 and 1989 (\citealt{kurtz1983}; \citealt{kreidl1991}) and 2008 (this paper) -- a span of 26~yr. Our purpose is to illuminate the nature of the amplitude modulation first reported in HD~217522 by \citet{kreidl1991}. They suggested mode switching in this star. As we show in this paper, short mode lifetimes is a better model.

\section{Observations}

\subsection{Spectroscopic observations and reductions}

HD~217522 was observed for 10.1~h on 2008 September 11 with the UV-Visual Echelle Spectrograph (UVES) on the ESO Very Large Telescope (VLT). We used an image slicer (IS$\#3$) to optimally utilise the observing conditions at the instrument's maximum resolution with a 0.3~arcsec slit. Exposure times of 60~s were used, which together with readout and overhead times of $\sim$26~s provided a time resolution of 76~s. The camera uses two 2K $\times$ 4K CCDs with 15~$\umu$m pixels. We used the RED (600~nm) setting which covers the wavelength region $\lambda\lambda~4970 - 7010$~\AA, with a gap in the region $\lambda\lambda~5963 - 6032$~\AA\ caused by the space between the two CCD mosaic halves. The average spectral resolution is about $R = 110~000$. Raw CCD frames were processed using the UVES pipeline to extract and merge the echelle orders to 1D spectra that were normalised to the continuum. The average signal-to-noise ratio in the continuum, estimated from 1D spectra, is about 100. In total 440 spectra were obtained. 

\subsection{Photometric observations and reductions }

The reductions of the 1982 and 1989 data were described in \citet{kurtz1983} and \citet{kreidl1991}, respectively. The 2008 data were obtained on the nights 2008 September 04, 11 and 12 through a Johnson $B$ filter on the South African Astronomical Observatory (SAAO) 0.5-m telescope. A 30-arcsec aperture and 10-s integration times were used. A journal of the 2008 observations is given in Table~\ref{tabjournal}. The data were sky-subtracted and corrected for dead-time losses and extinction. In addition, heliocentric time corrections, accurate to $10^{-5}$~d were applied to each data point. Low frequencies due to sky transparency variations were removed. 

The observations were planned to be contemporaneous with the VLT spectroscopic observations to allow us to compare directly the photometric and spectroscopic radial velocity amplitudes.

\begin{table}  
\caption{A journal of the 2008 observations of HD~217522. Column 1 gives the date of observations, column 2 the corresponding Julian date, the column 3 gives the number of data points and column 4 gives the duration of each observations in hours. The observer was ZM.}\label{tabjournal} 
\begin{tabular}{cccc} 
UT date & Julian Date & N & $\Delta t$ hr \\ 
 2008 September 04 & 2454714 & 1715   & 5.41  \\ 
 2008 September 11 & 2454721 & 2512   & 7.39  \\ 
 2008 September 12 & 2454722 & 1906  & 6.02  \\ 
              &       &       & $\Sigma=18.82$ 
\end{tabular} 
\end{table} 

\section{Photometric Analysis}

\begin{figure*}
\centering
\epsfxsize 5cm\epsfbox{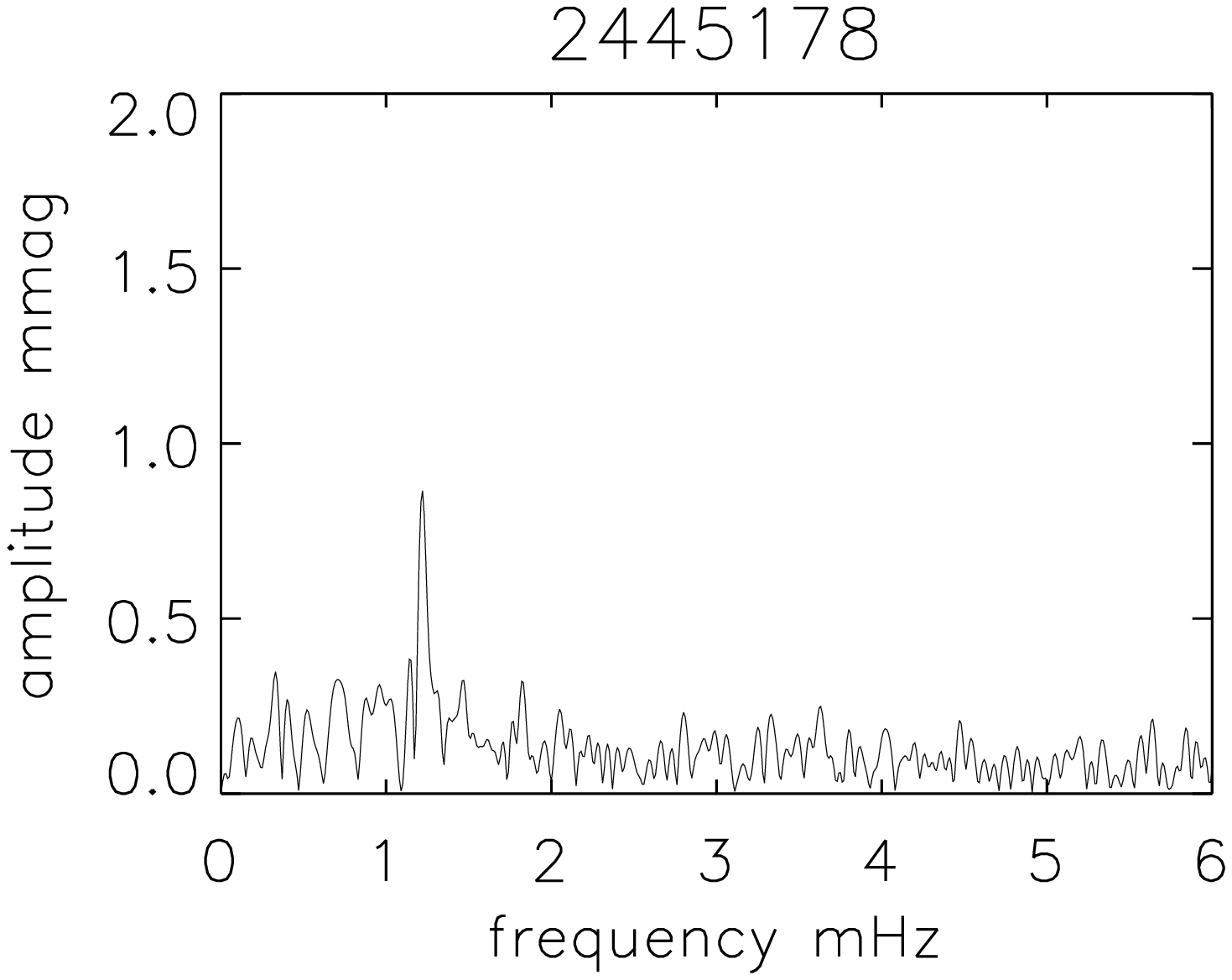}
\epsfxsize 5cm\epsfbox{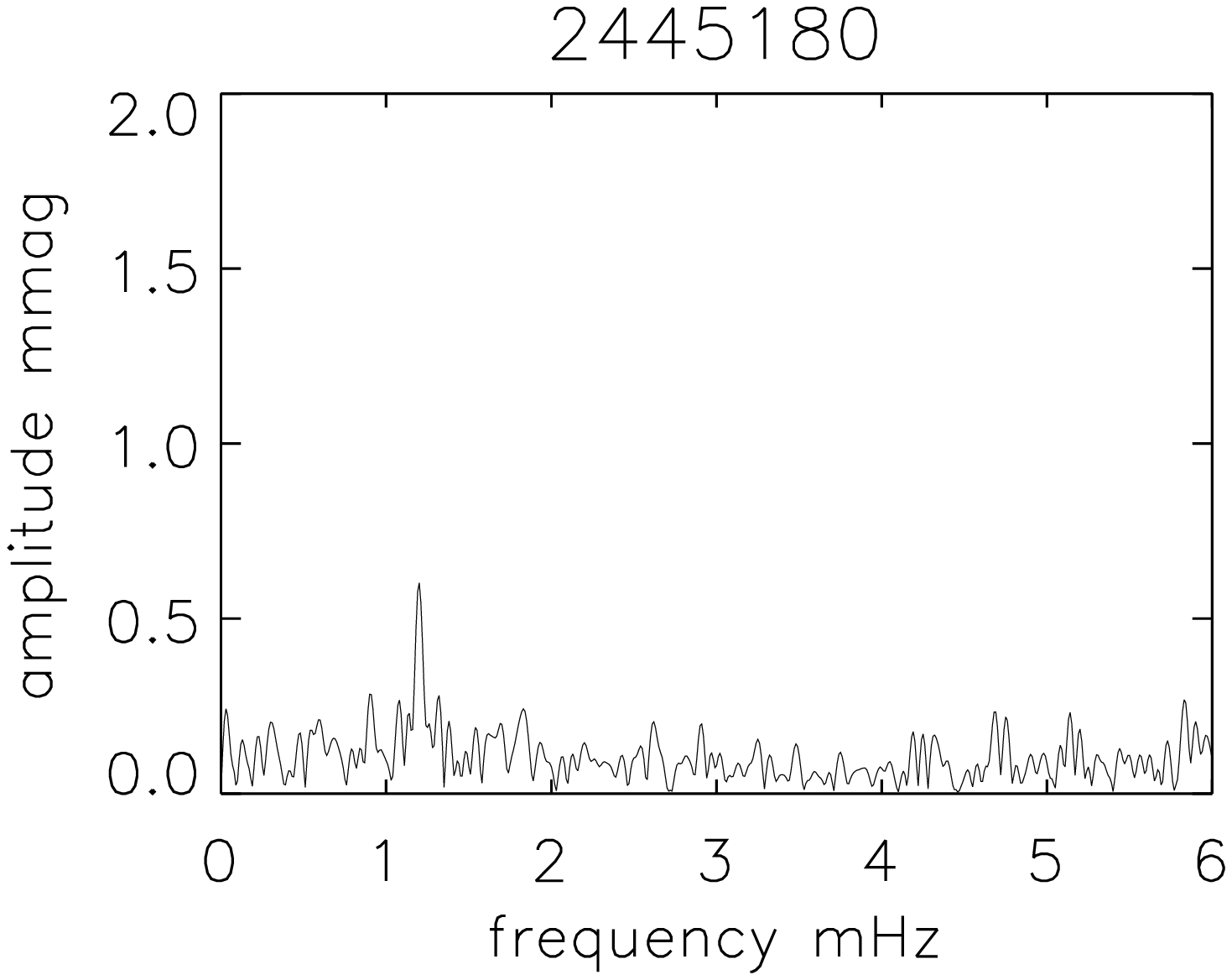}
\epsfxsize 5cm\epsfbox{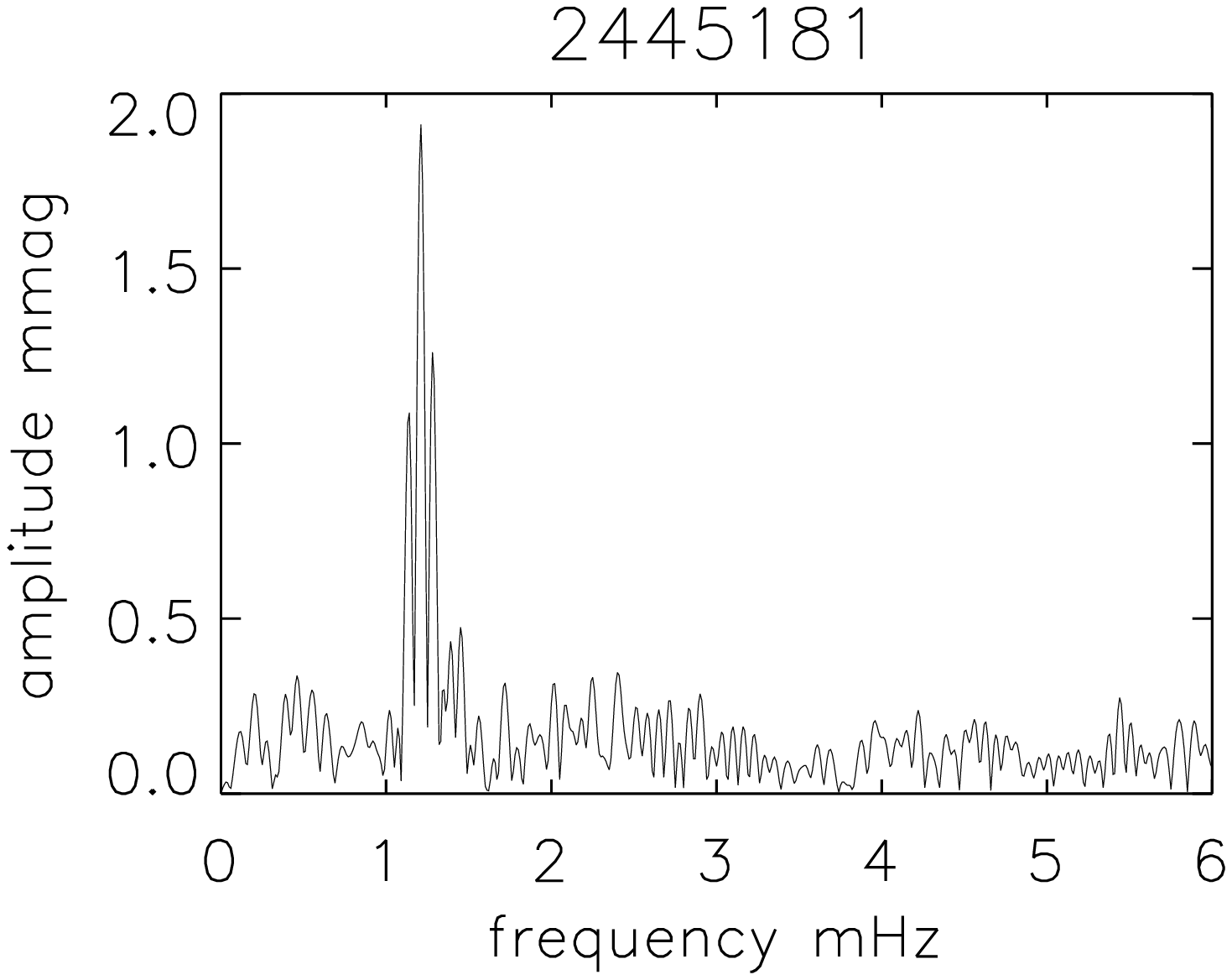}
\epsfxsize 5cm\epsfbox{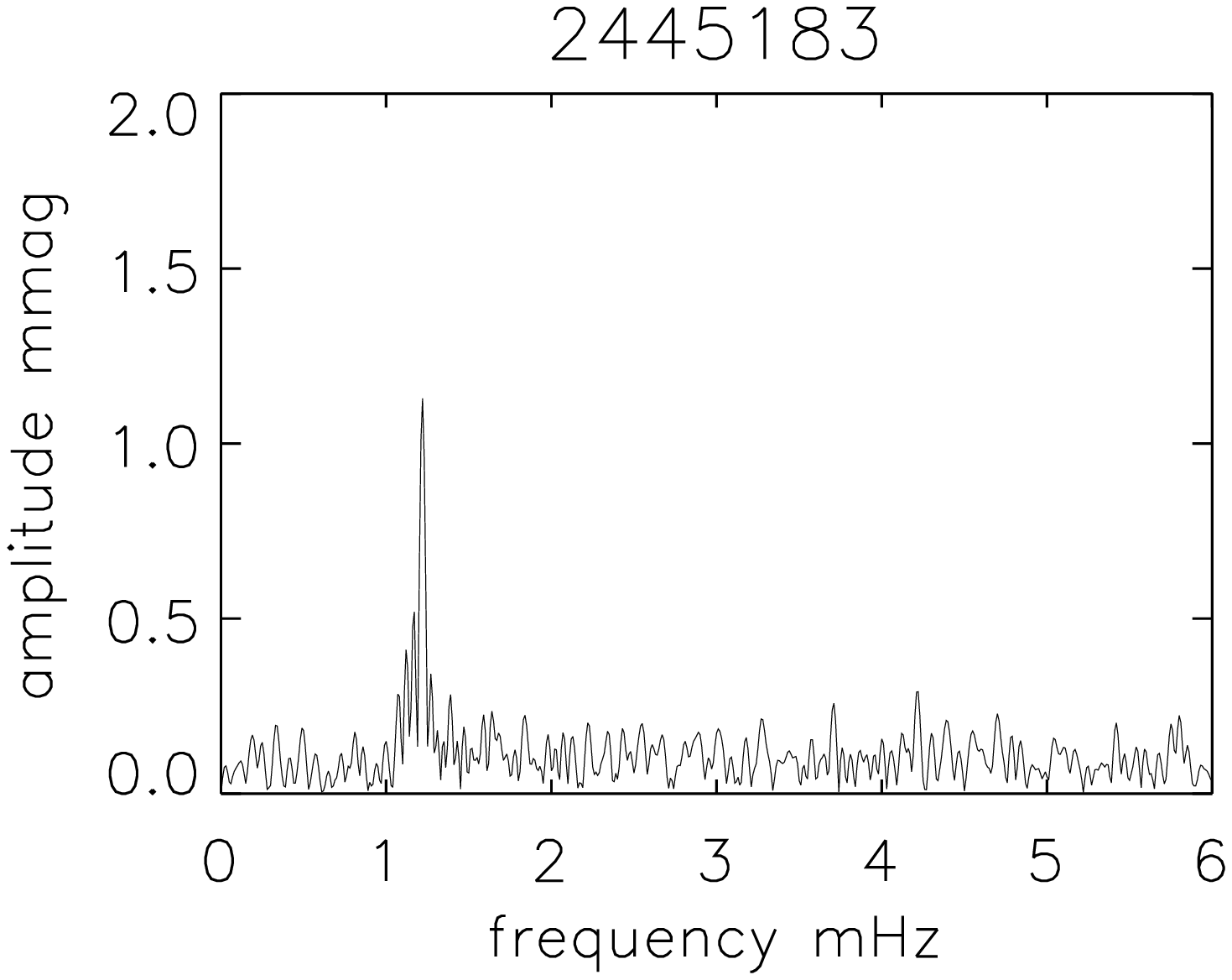}
\epsfxsize 5cm\epsfbox{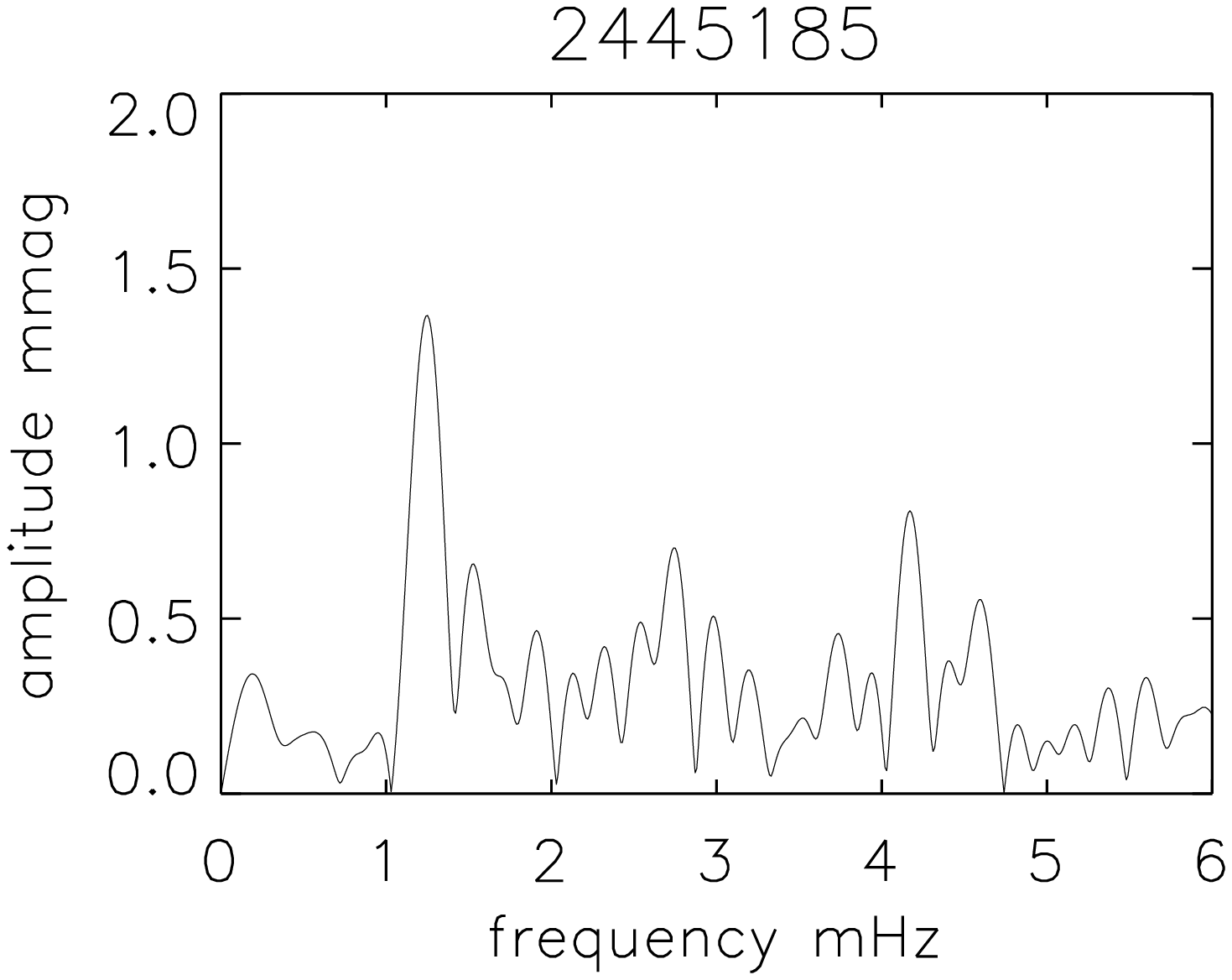}
\epsfxsize 5cm\epsfbox{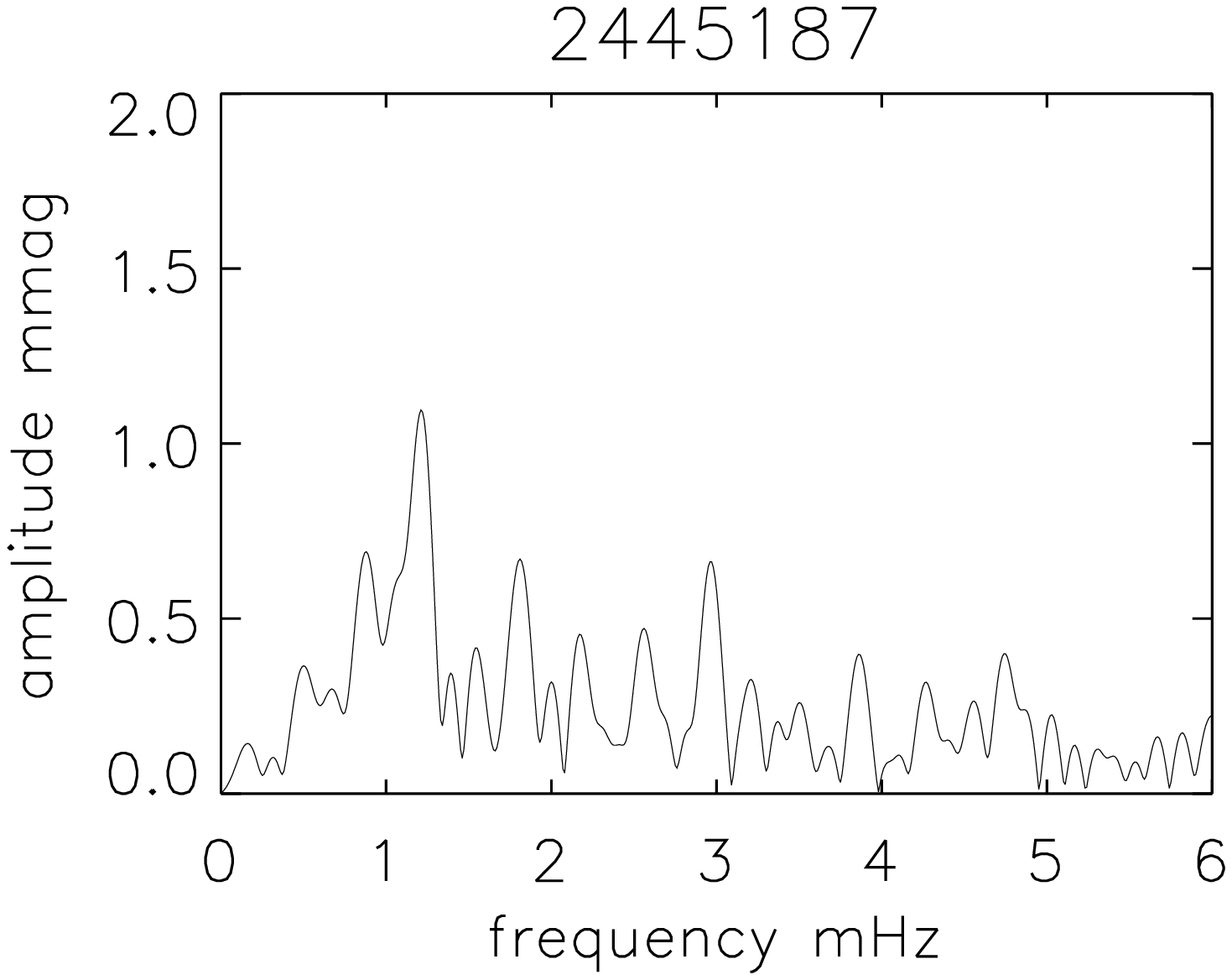}
\epsfxsize 5cm\epsfbox{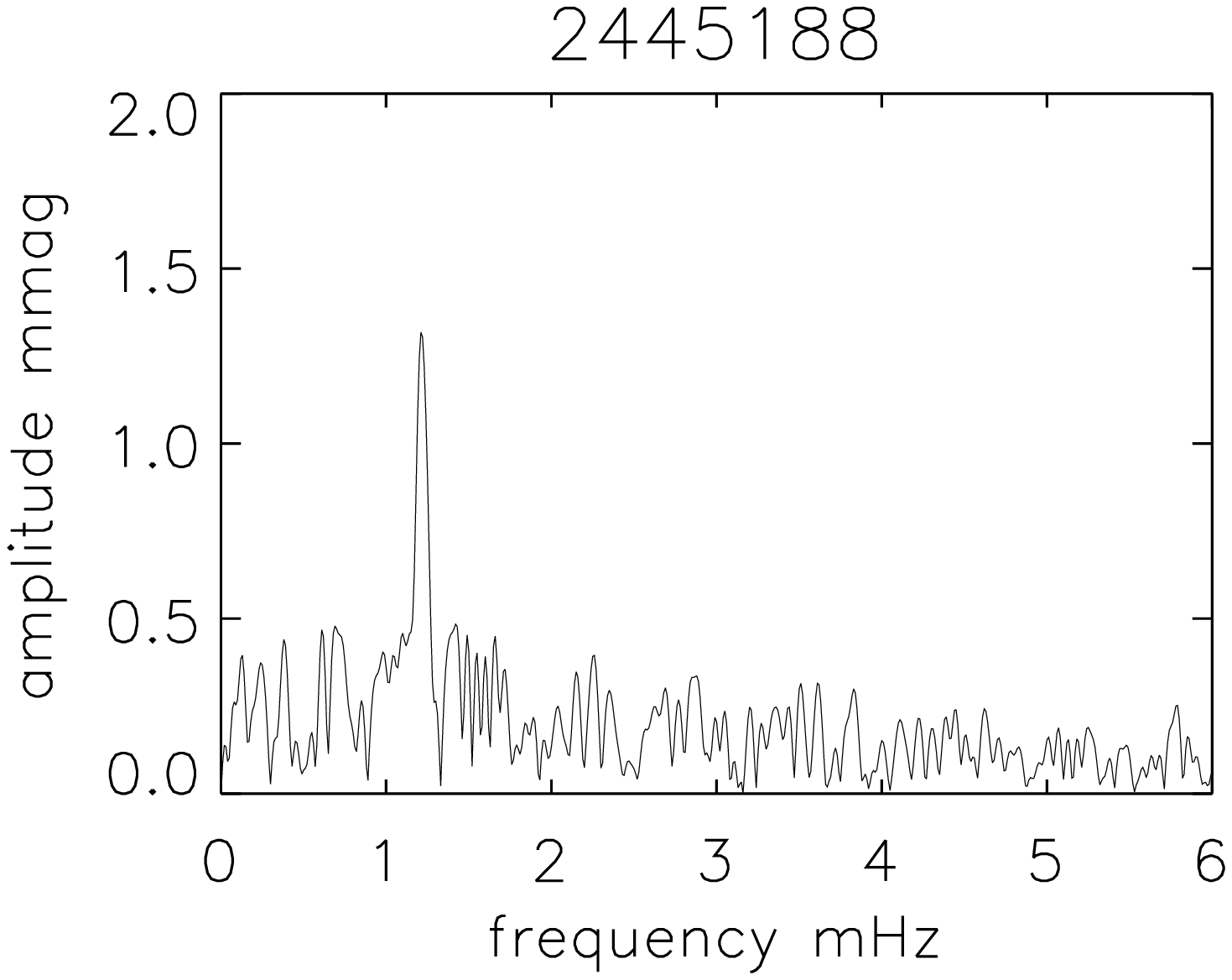}
\epsfxsize 5cm\epsfbox{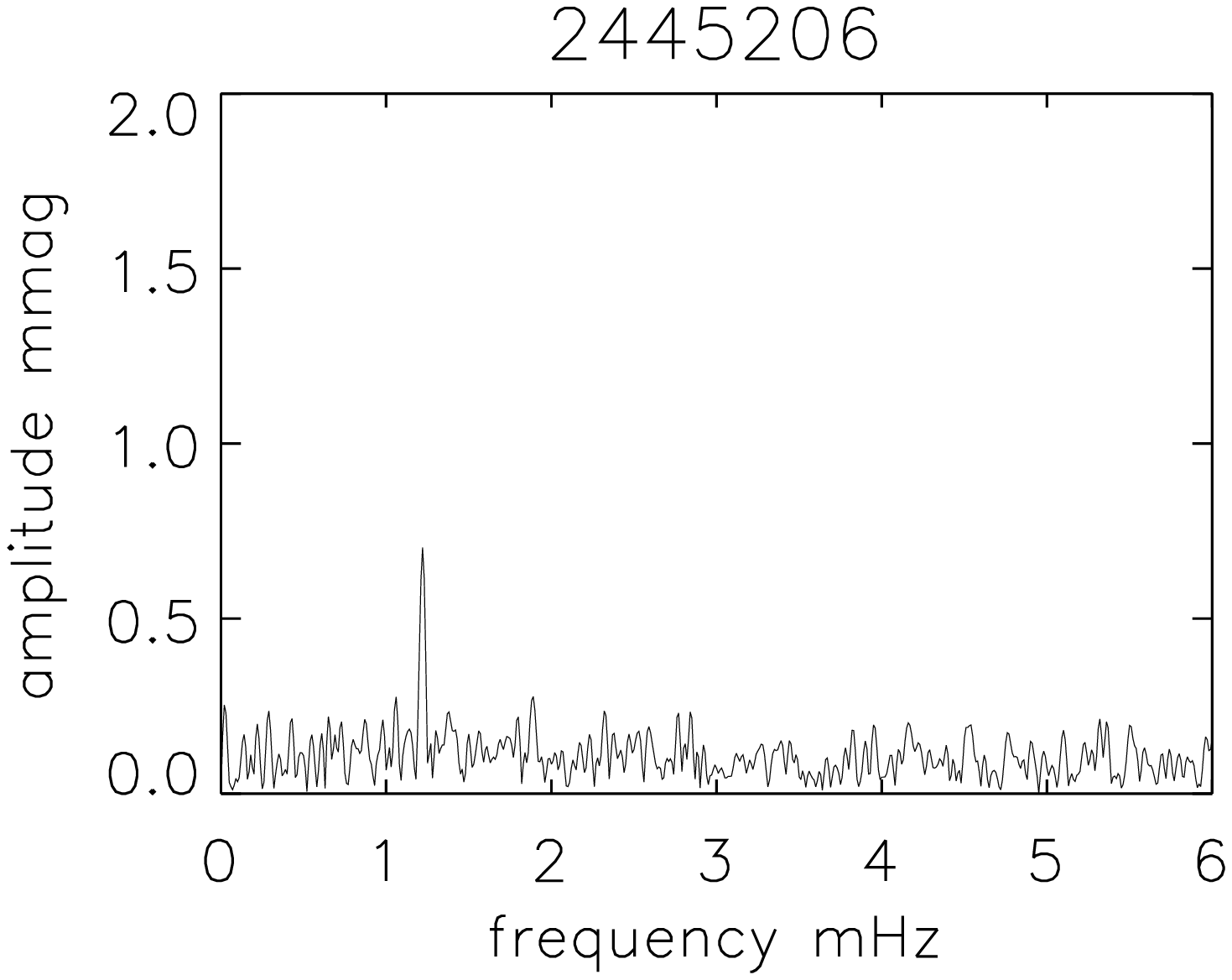}
\epsfxsize 5cm\epsfbox{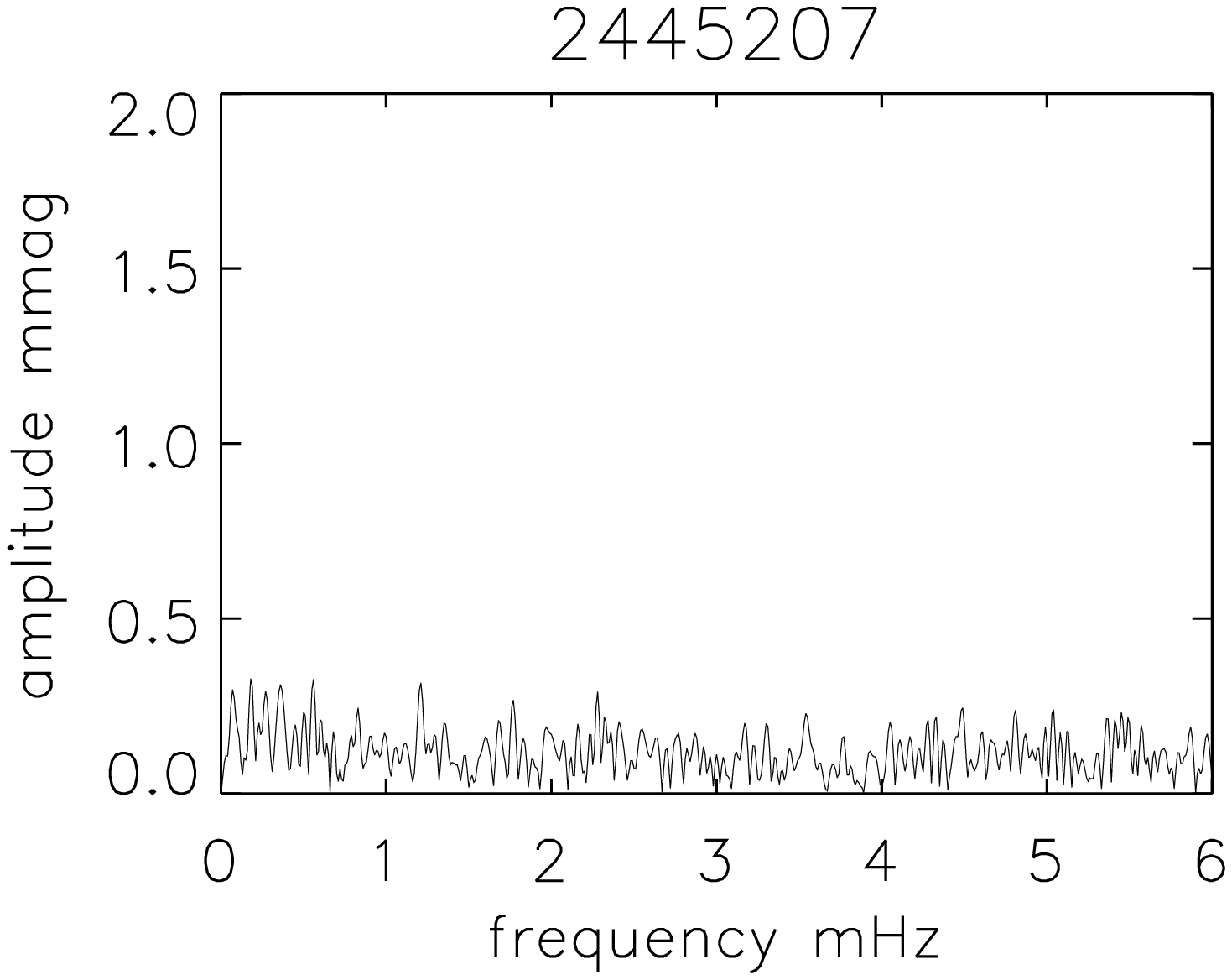}
\epsfxsize 5cm\epsfbox{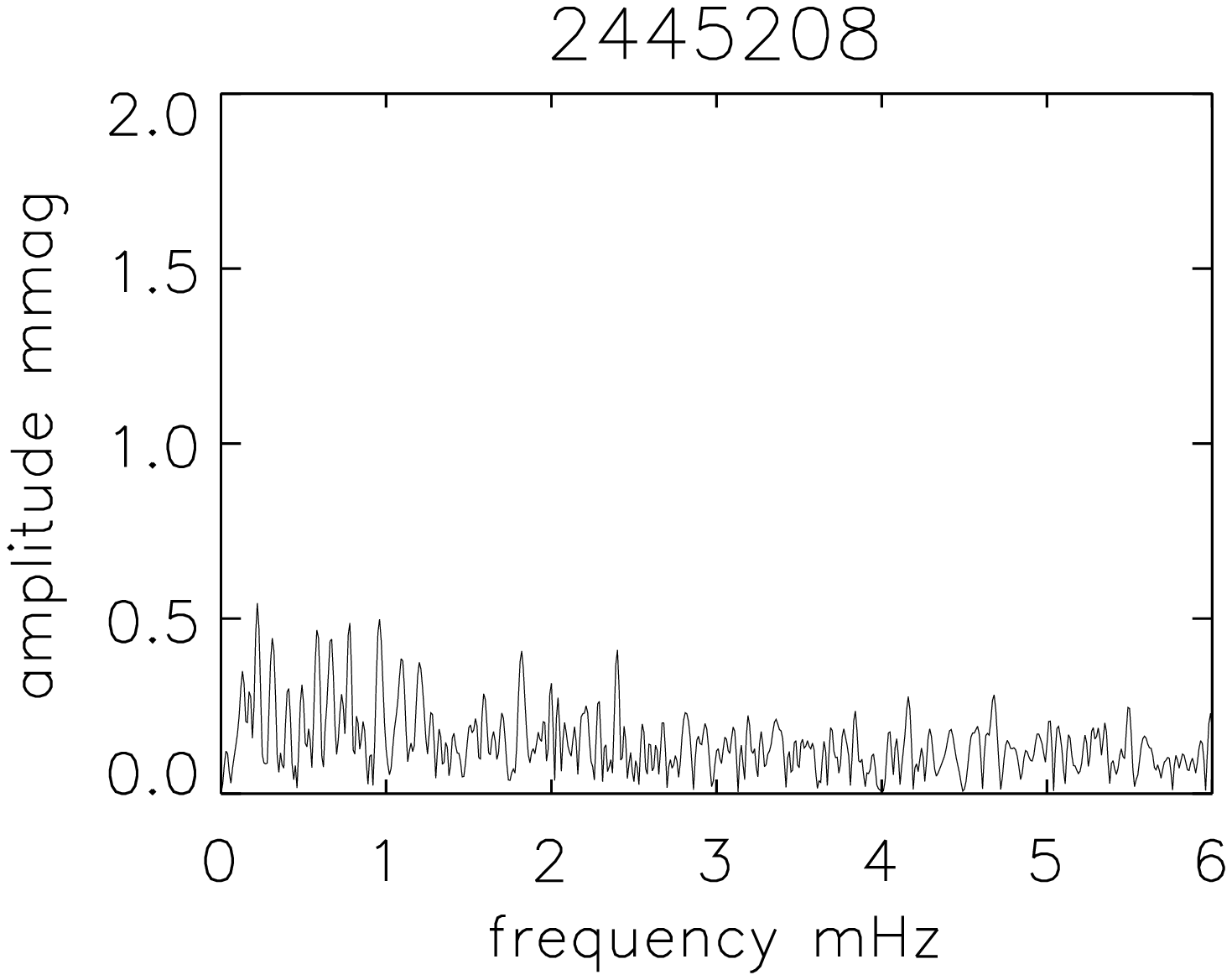}
\epsfxsize 5cm\epsfbox{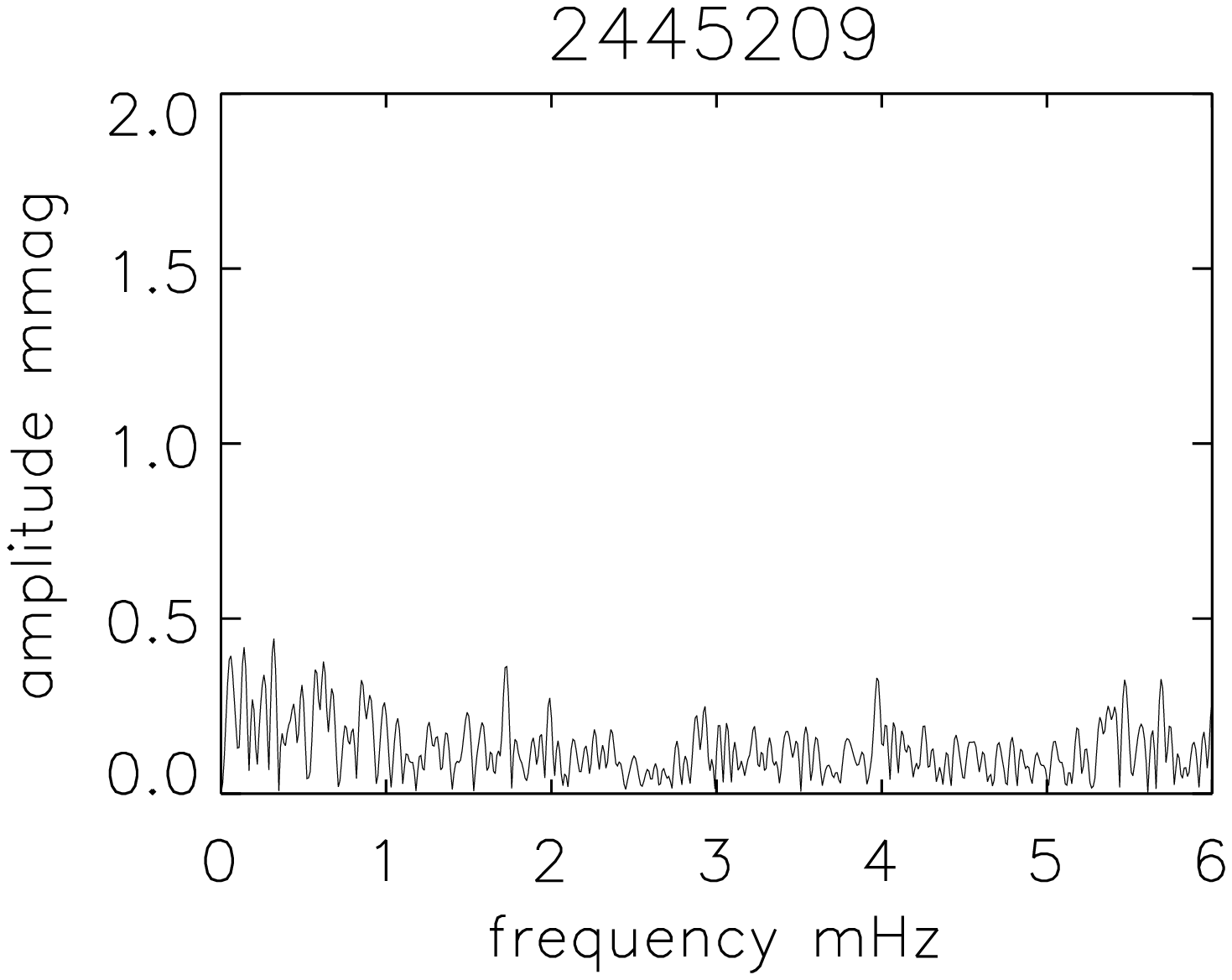}
\epsfxsize 5cm\epsfbox{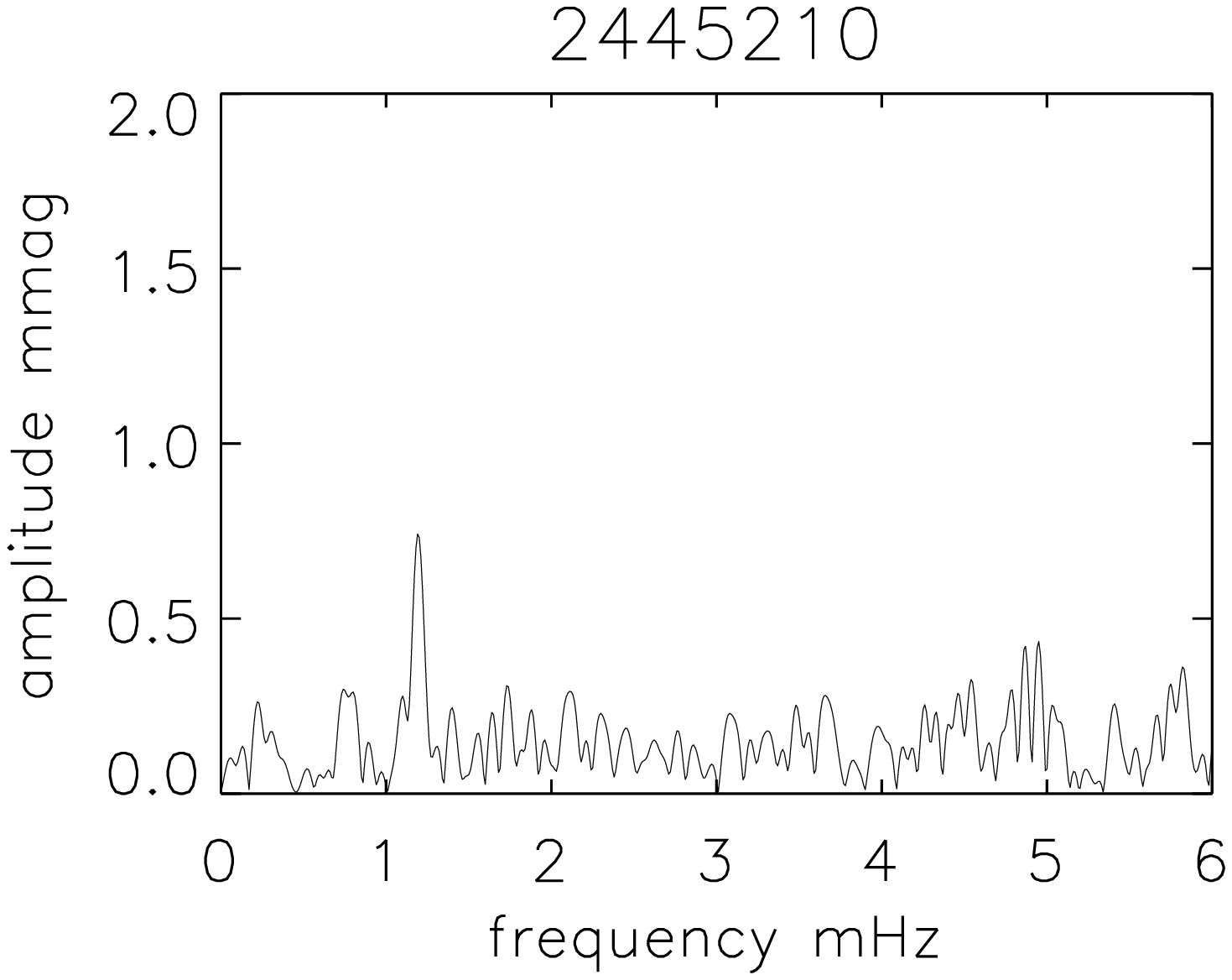}
\epsfxsize 5cm\epsfbox{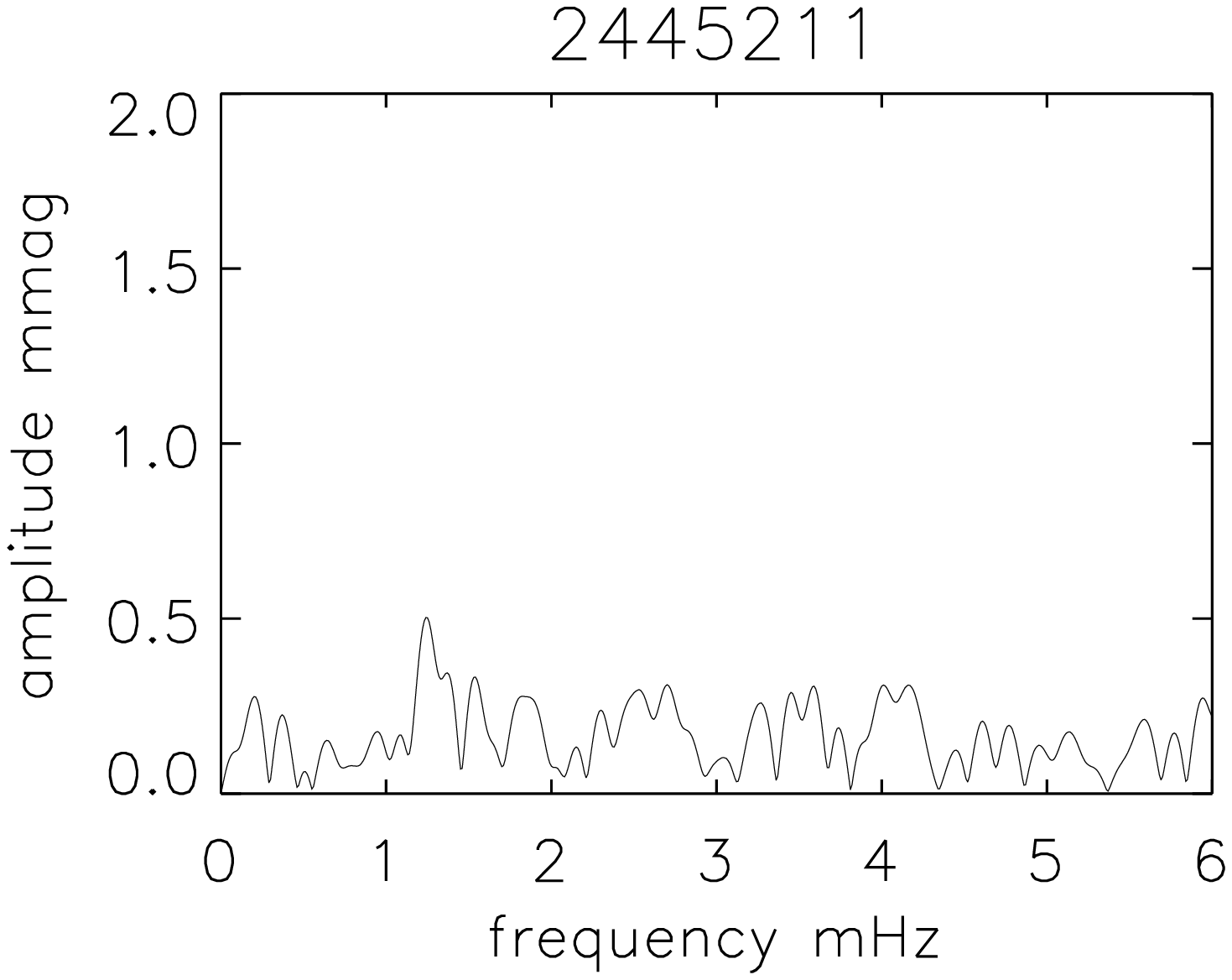}
\epsfxsize 5cm\epsfbox{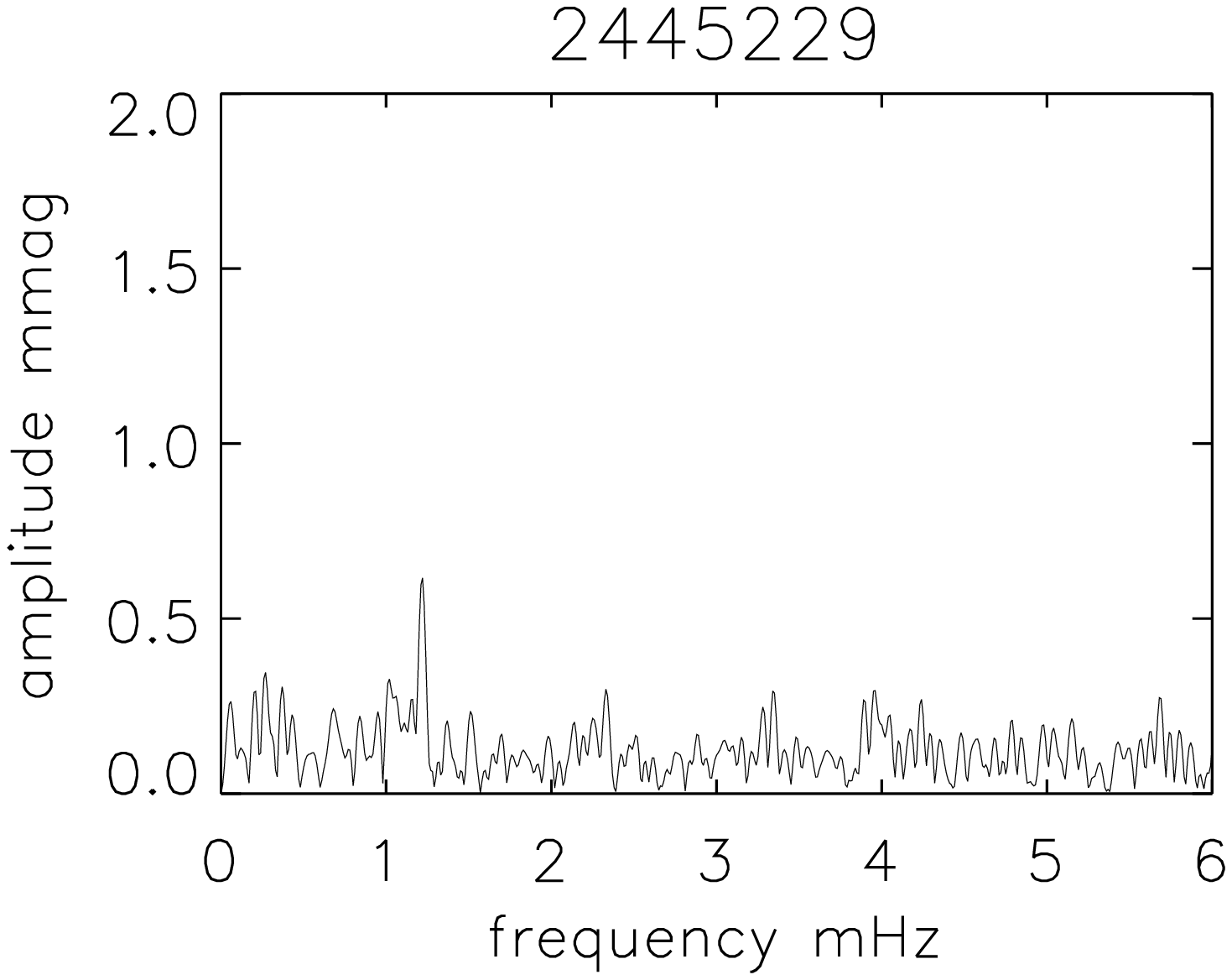}
\epsfxsize 5cm\epsfbox{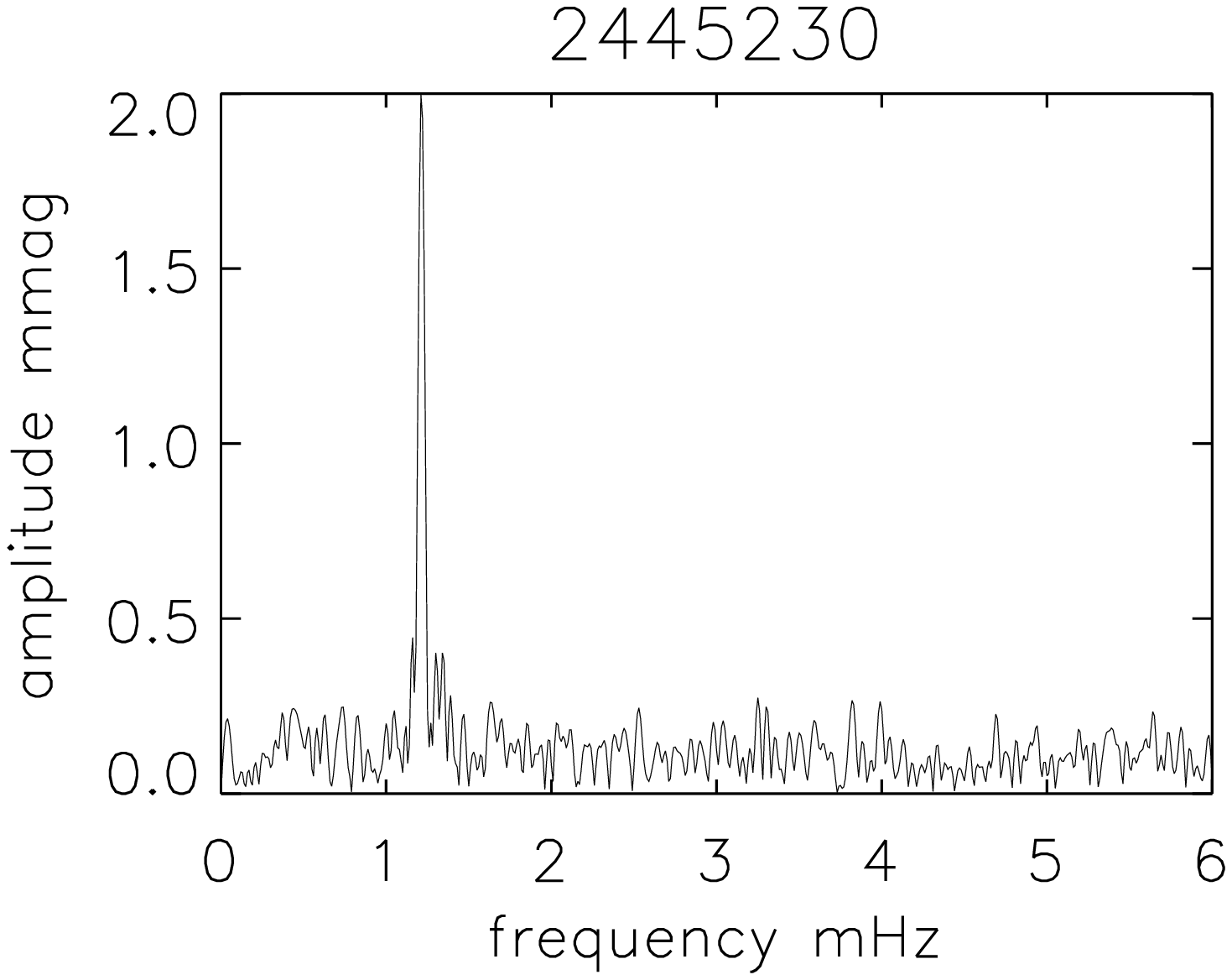}
\epsfxsize 5cm\epsfbox{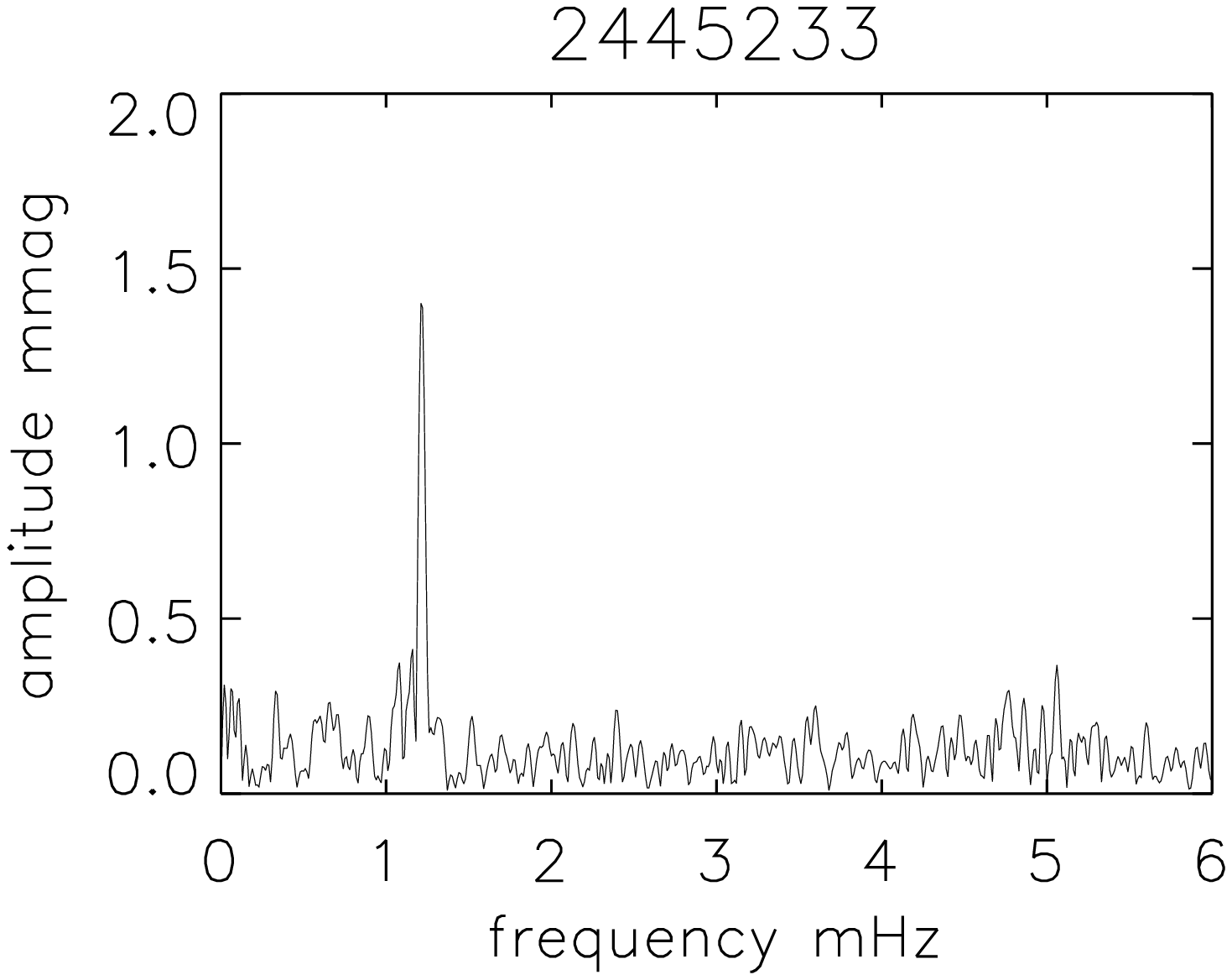}
\epsfxsize 5cm\epsfbox{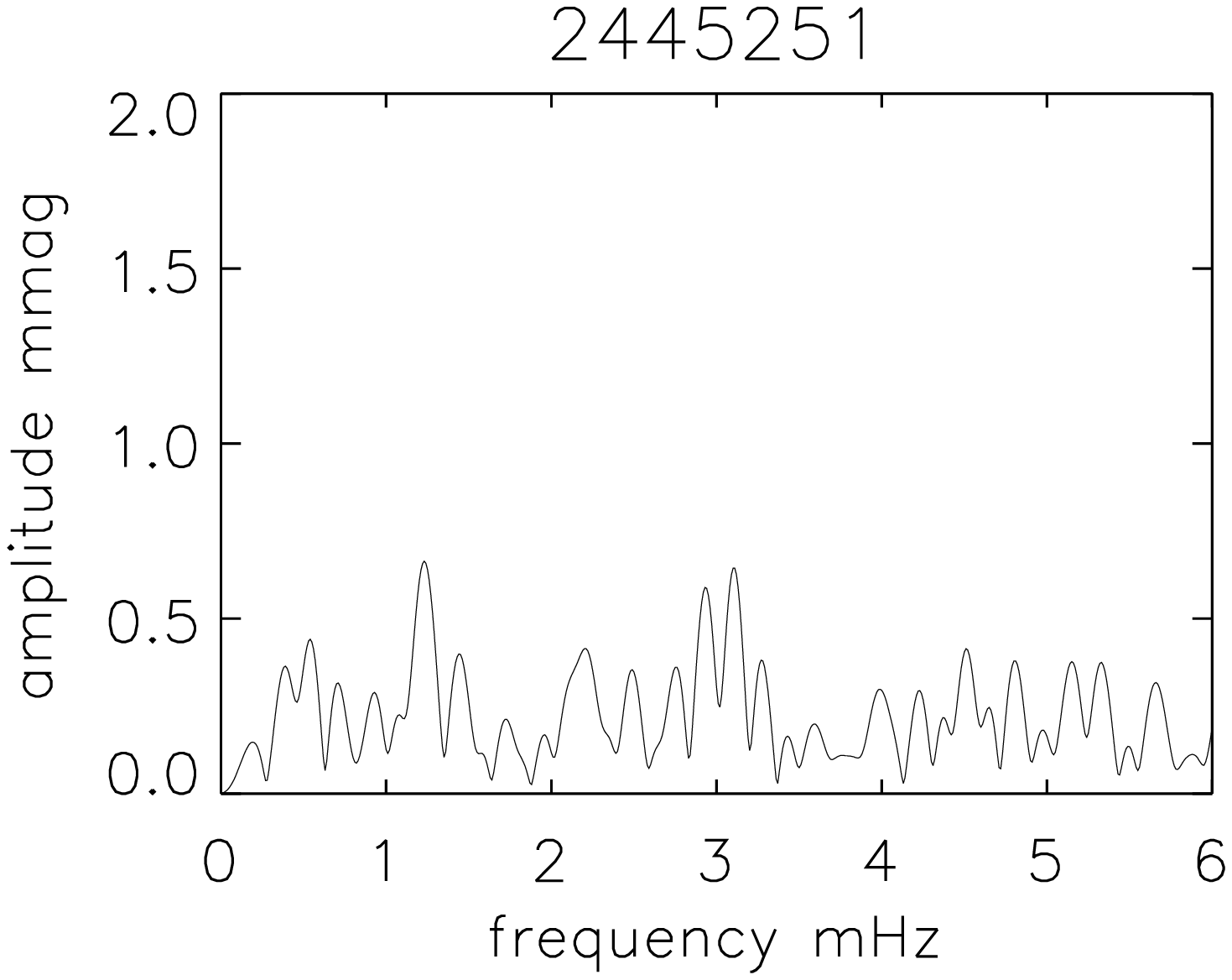}
\epsfxsize 5cm\epsfbox{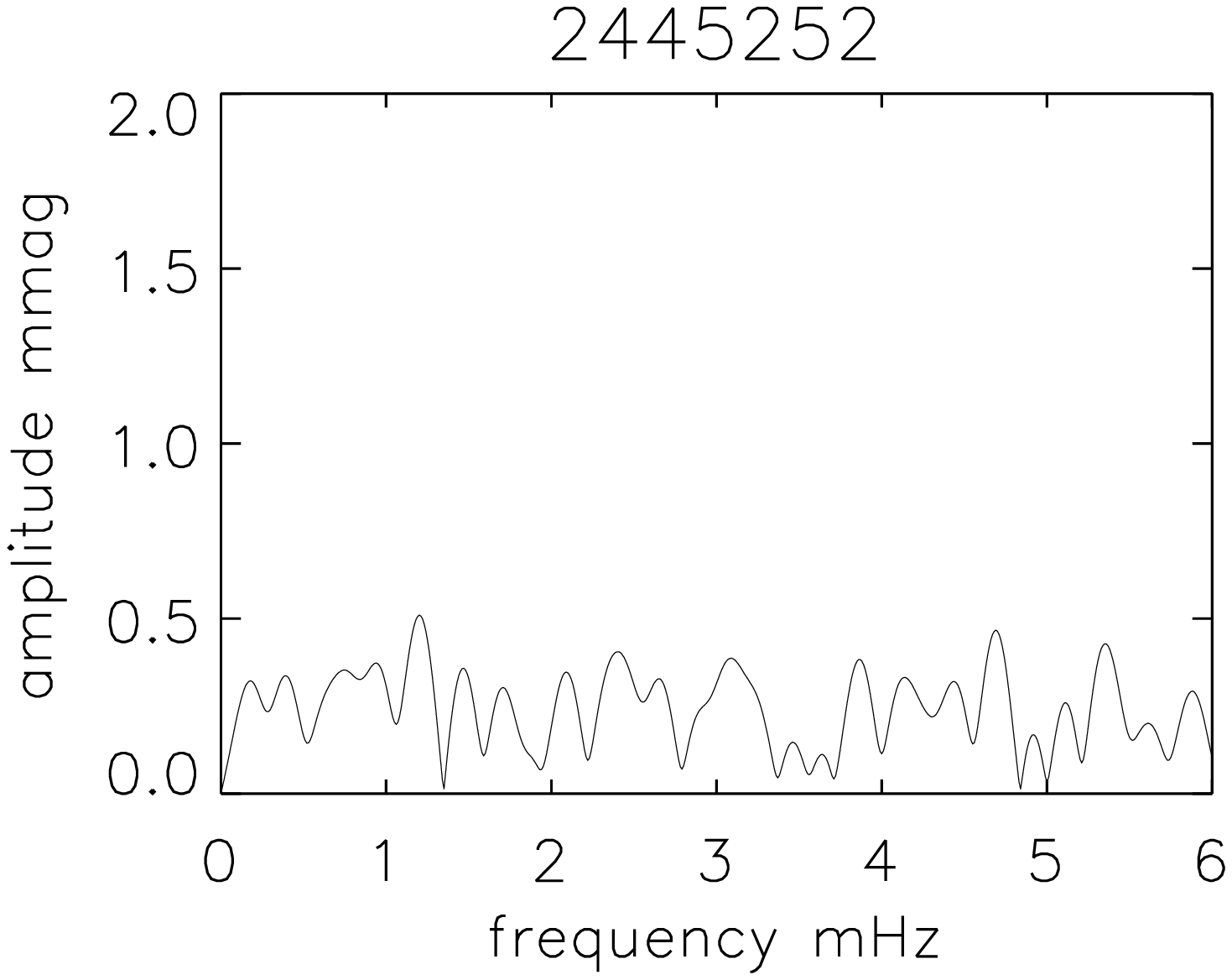}
\caption{\label{ft1982}
Amplitude spectra of the 1982 Johnson $B$ data. In these data, presented in \citet{kurtz1983}, one can clearly see only one frequency whose amplitude varies on a daily basis, and is absent on some nights, such as on the night of JD~2445211.}
\end{figure*}

\begin{figure*}
\centering
\epsfxsize 5cm\epsfbox{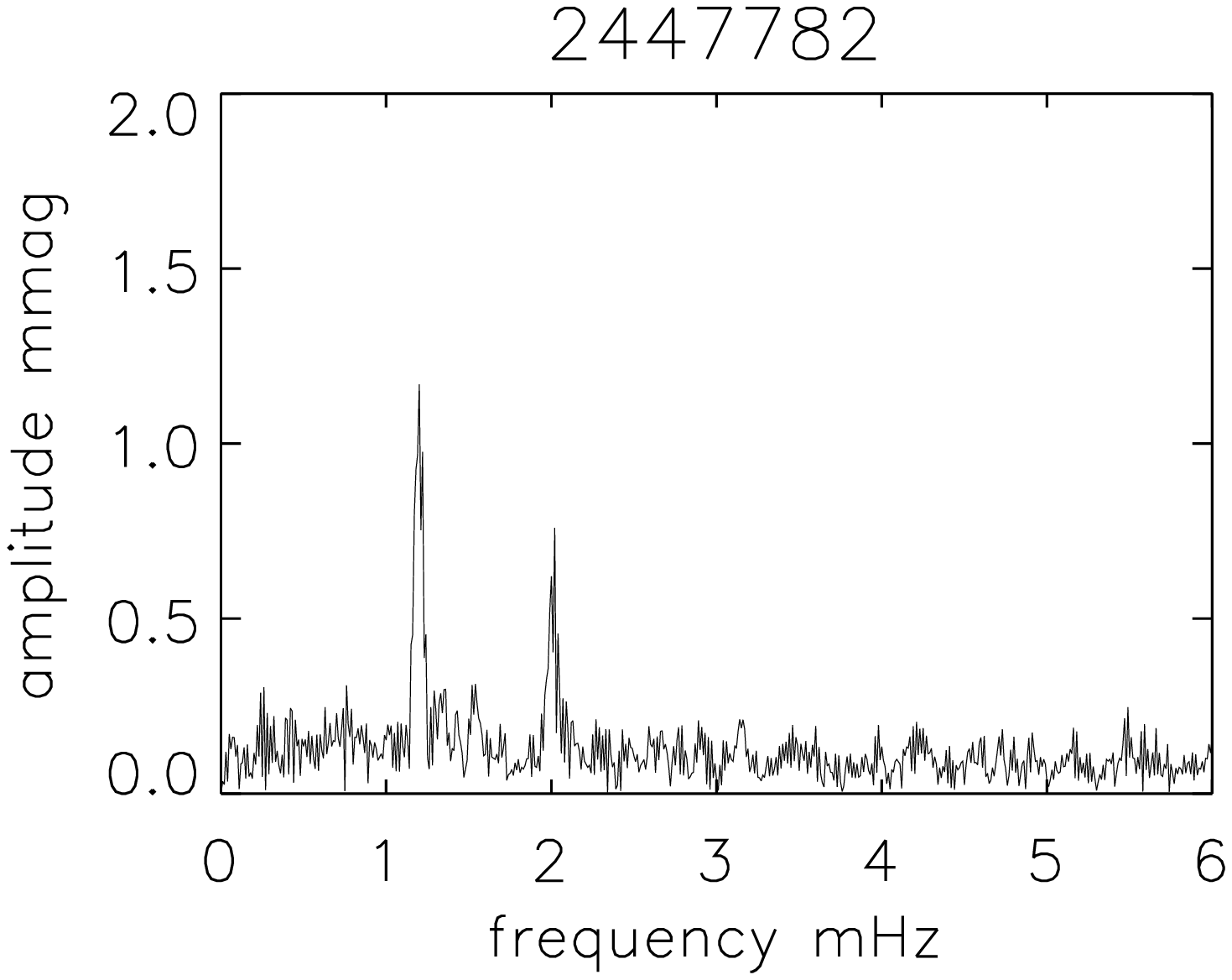}
\epsfxsize 5cm\epsfbox{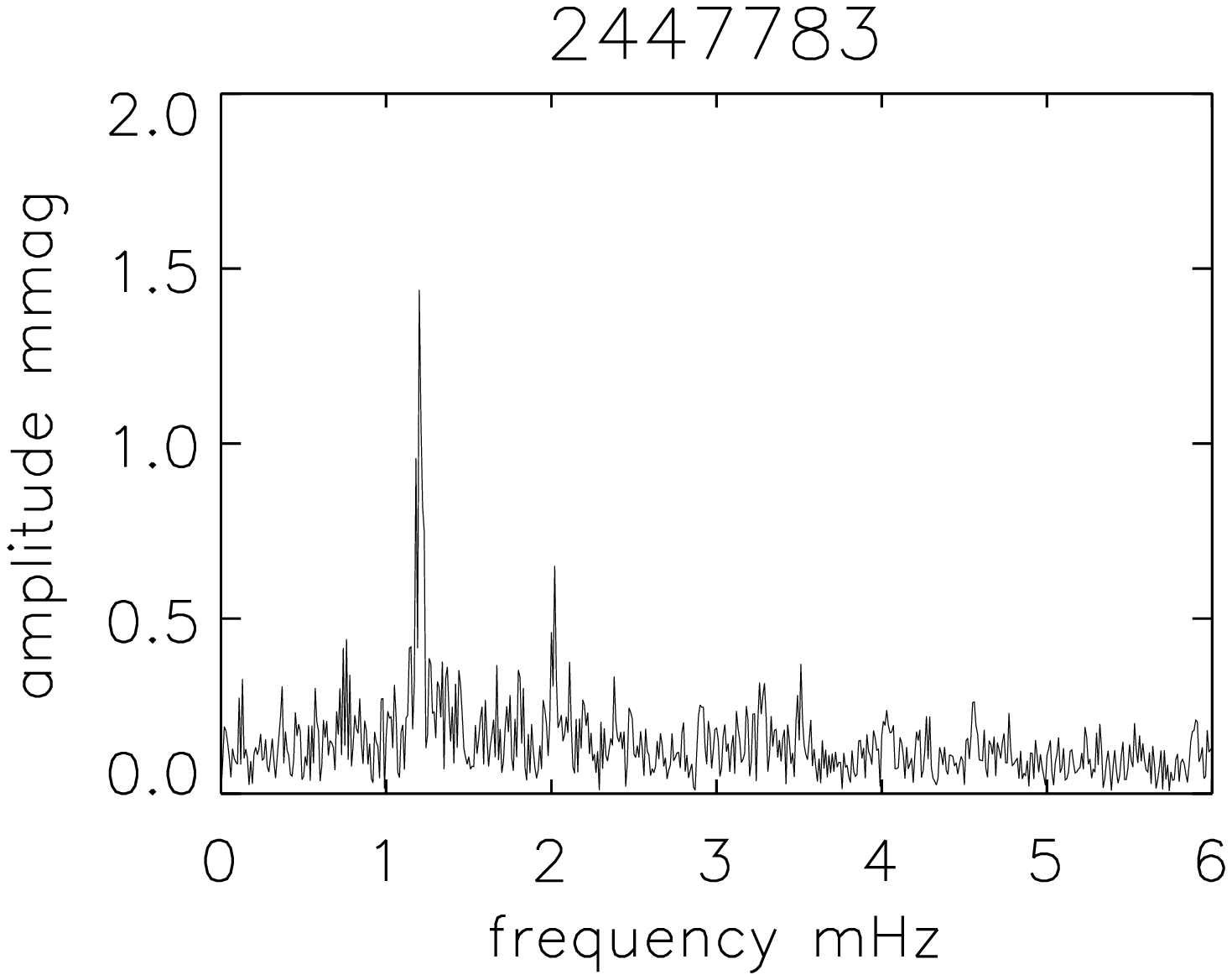}
\epsfxsize 5cm\epsfbox{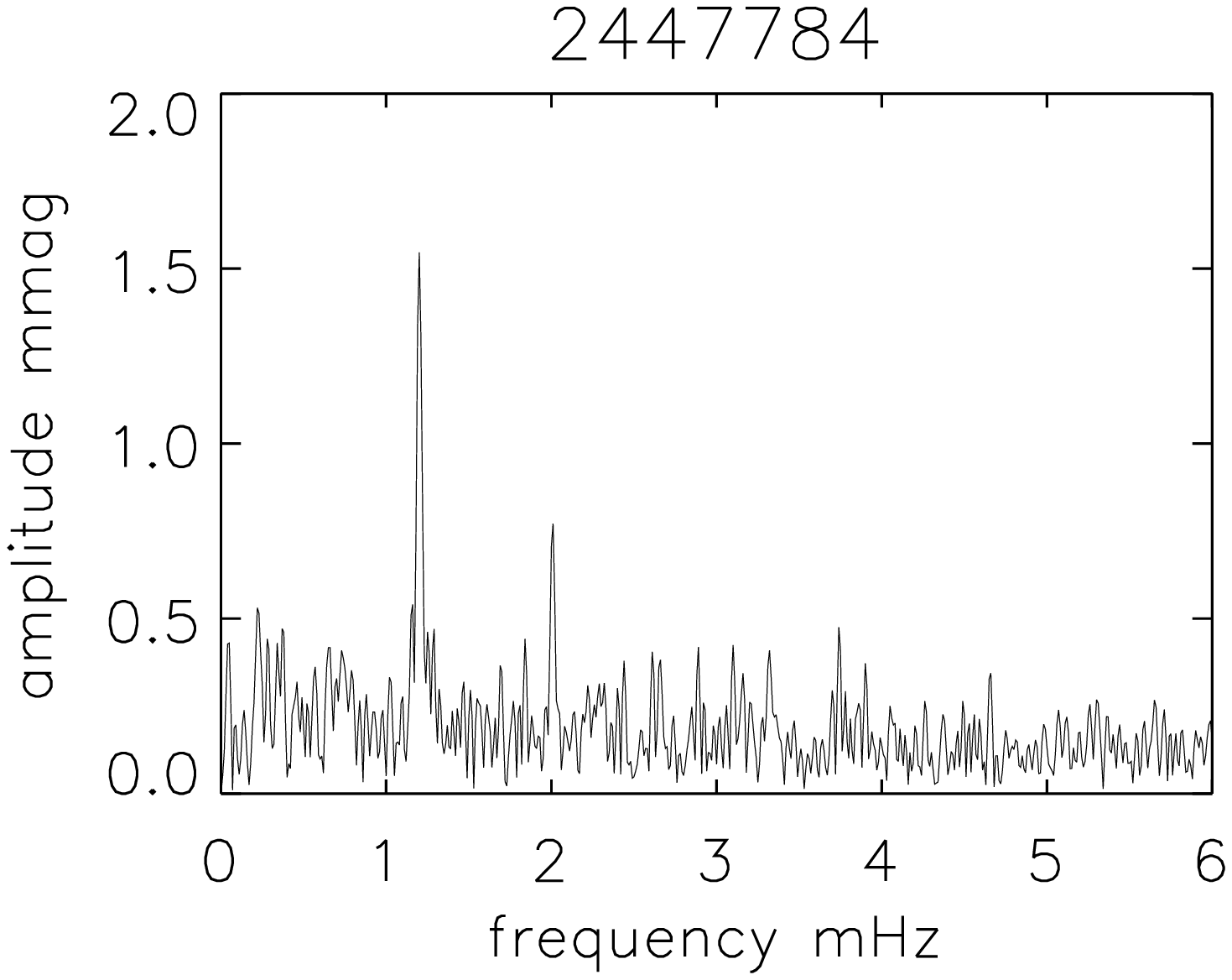}
\epsfxsize 5cm\epsfbox{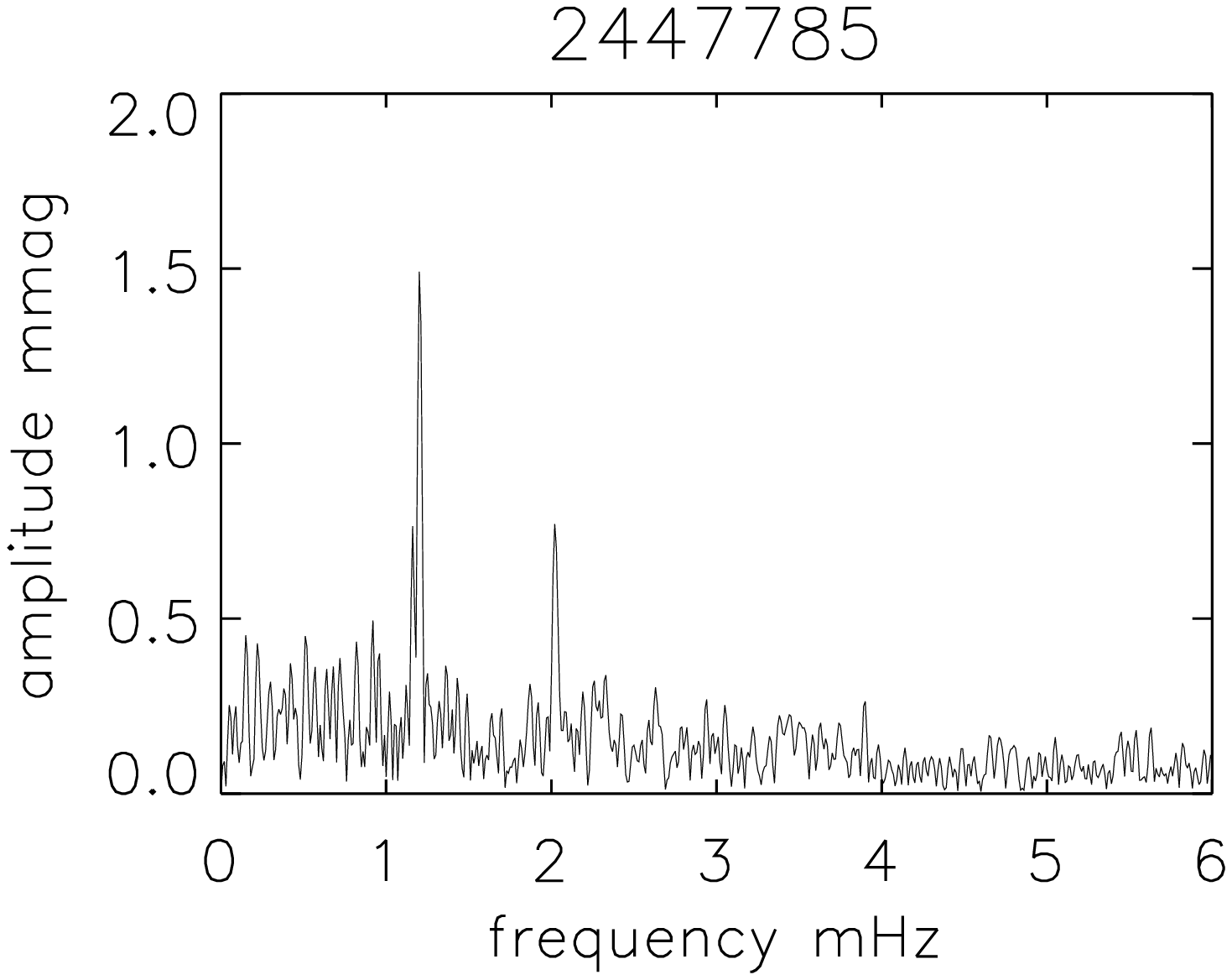}
\epsfxsize 5cm\epsfbox{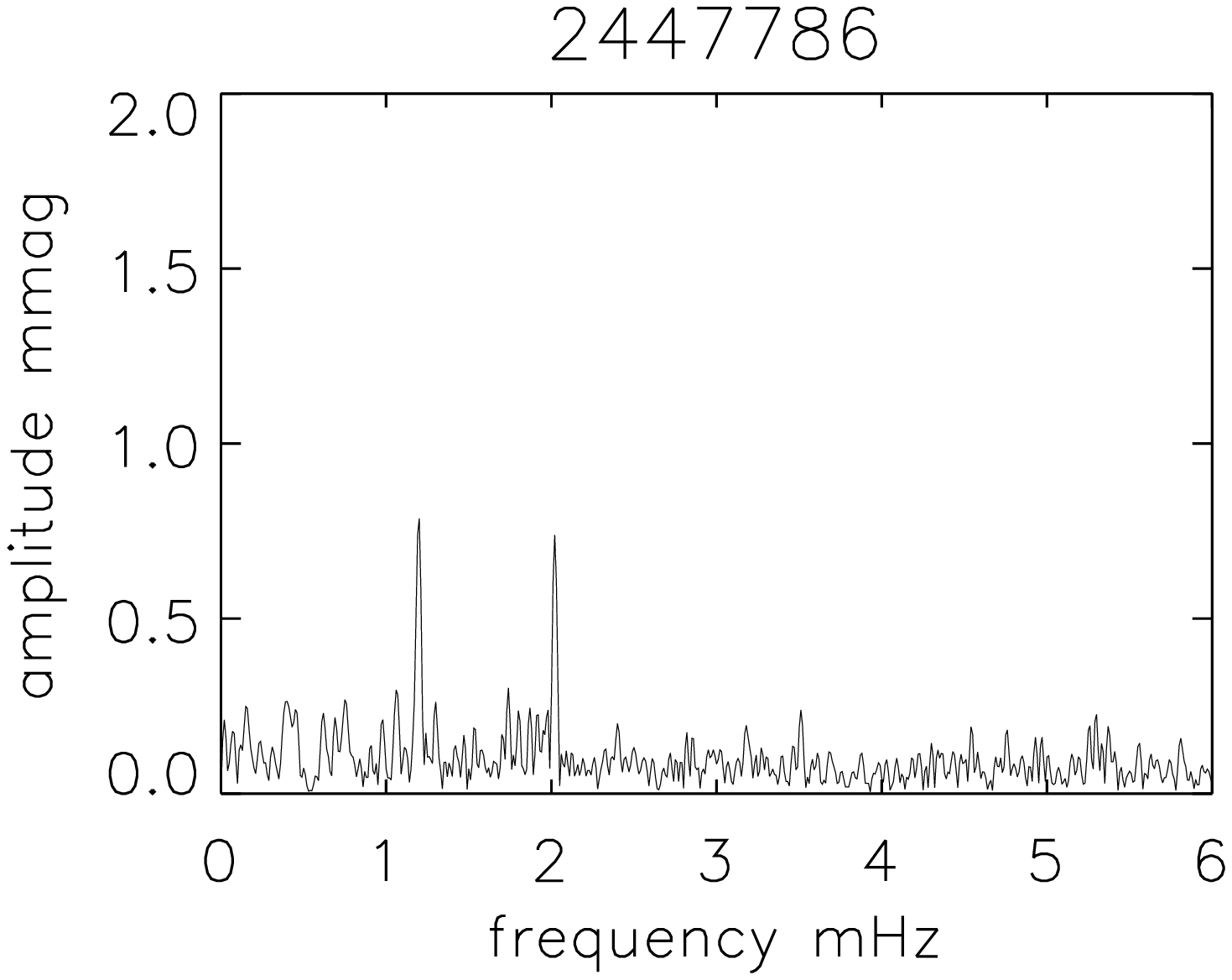}
\epsfxsize 5cm\epsfbox{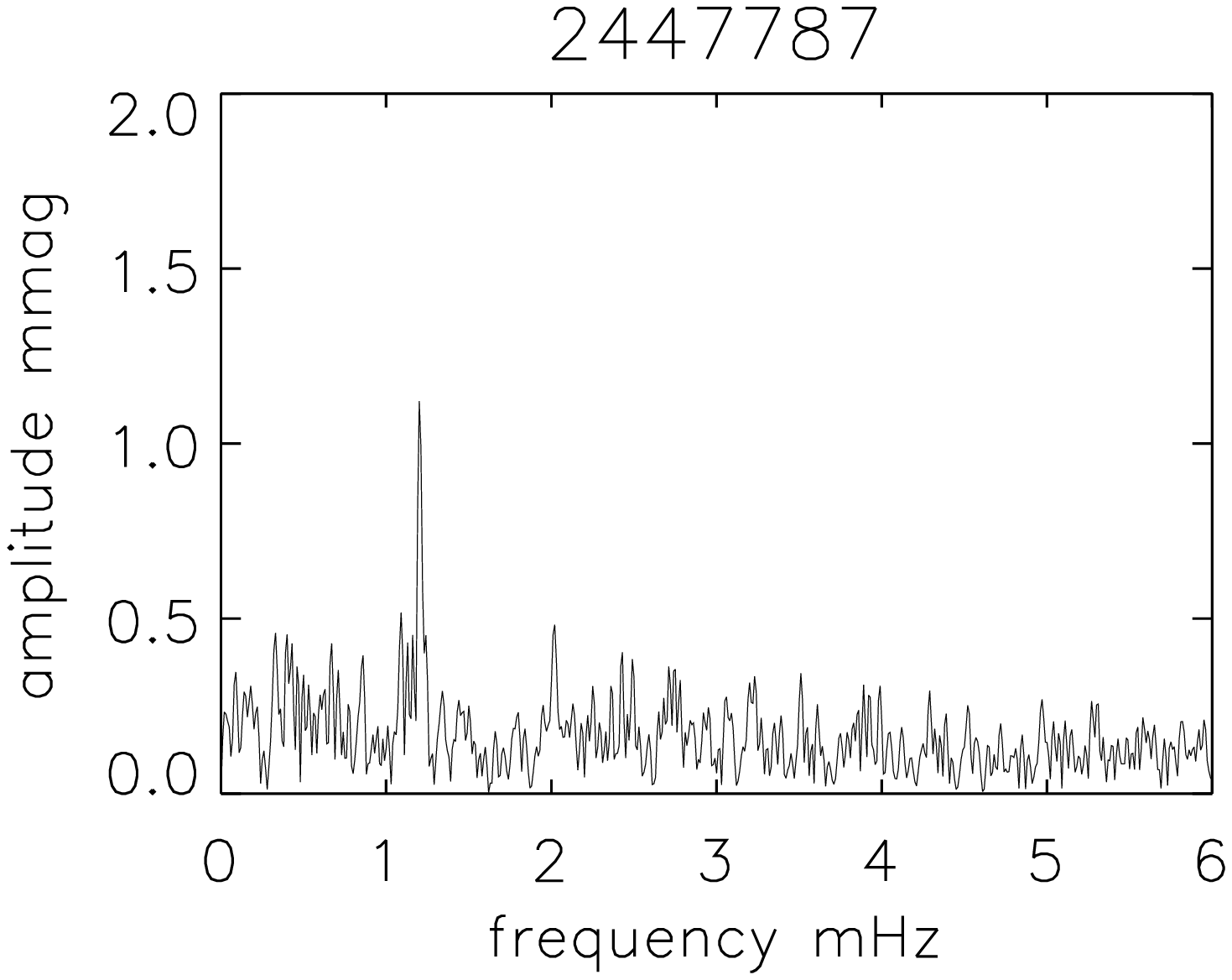}
\epsfxsize 5cm\epsfbox{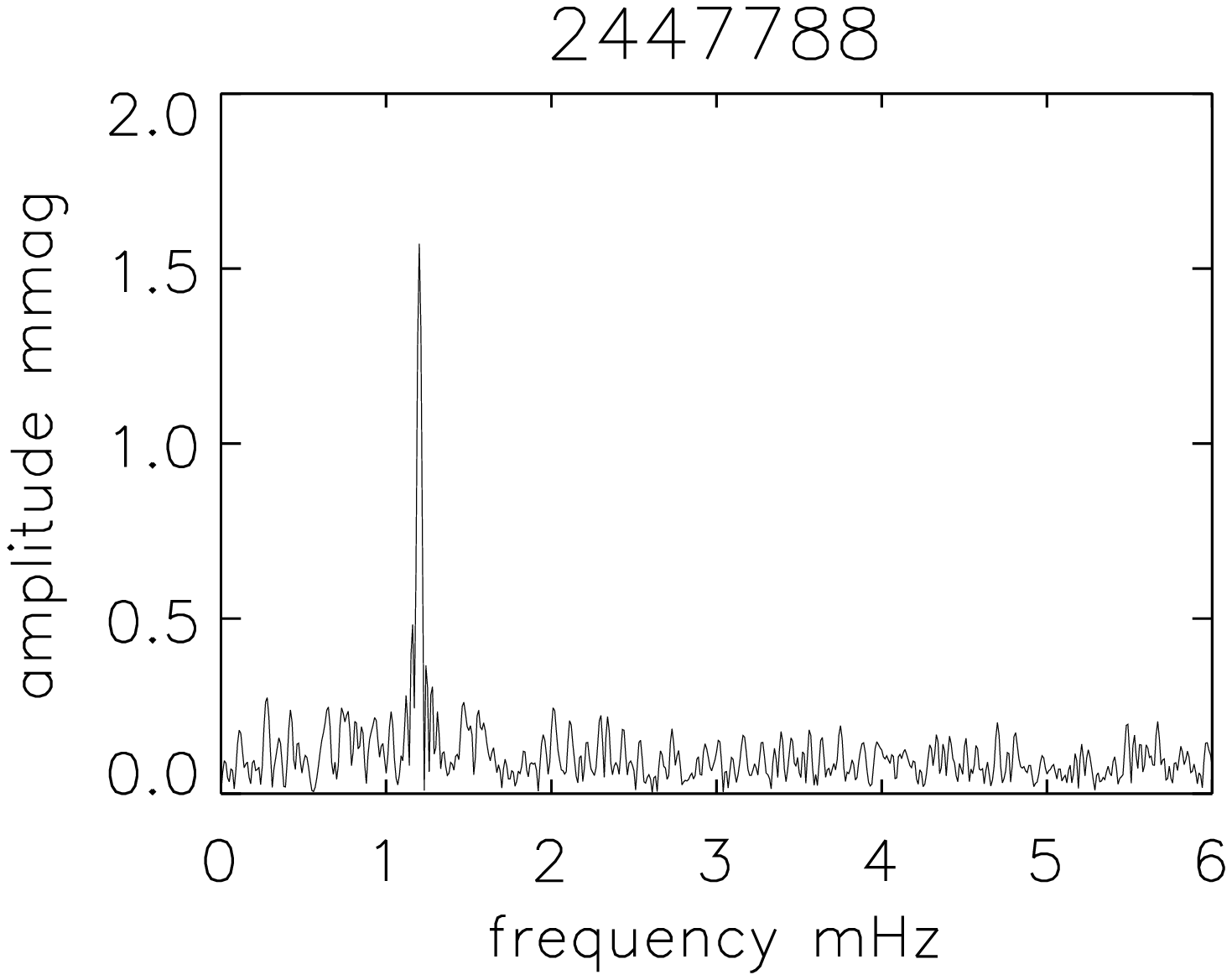}
\epsfxsize 5cm\epsfbox{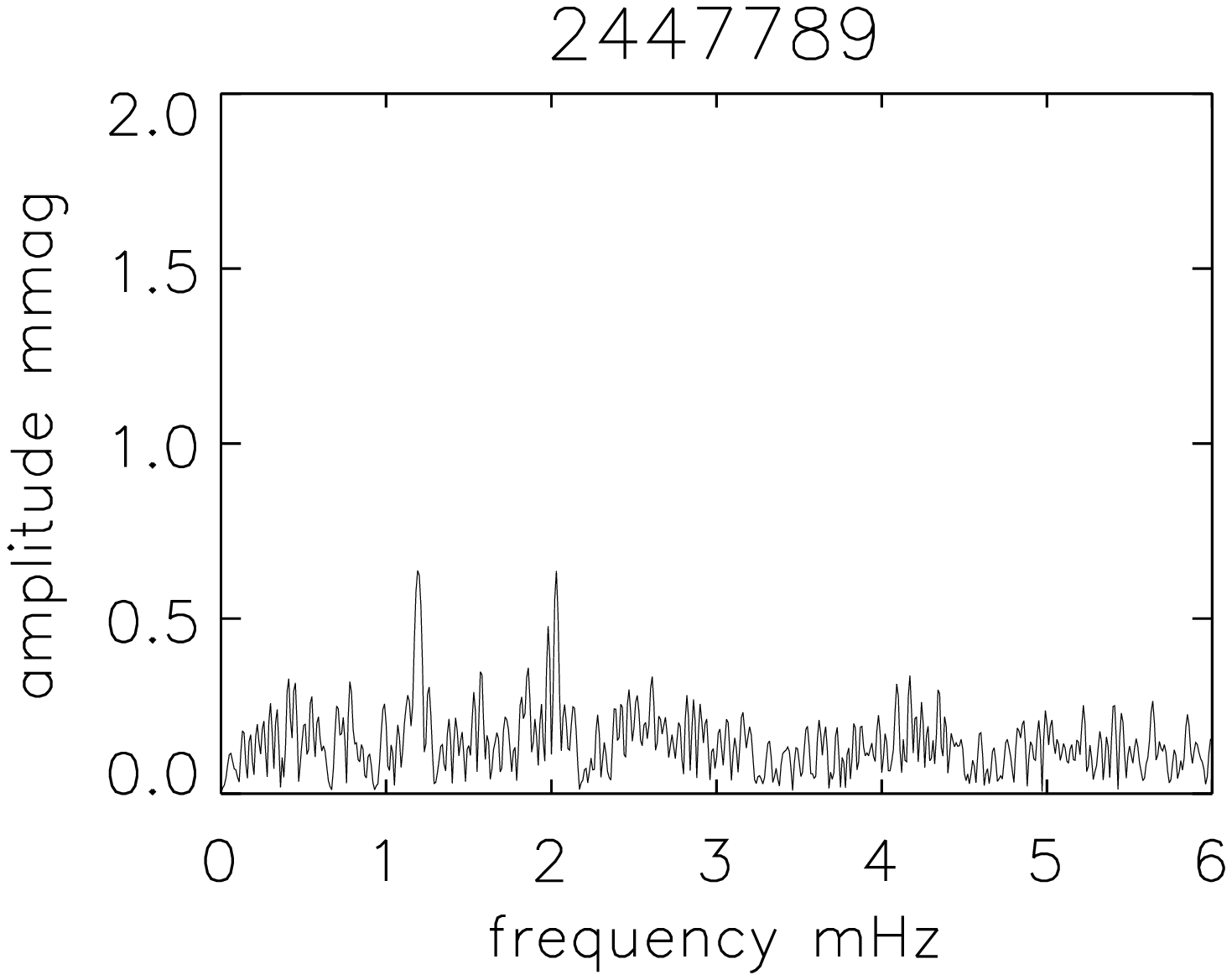}
\epsfxsize 5cm\epsfbox{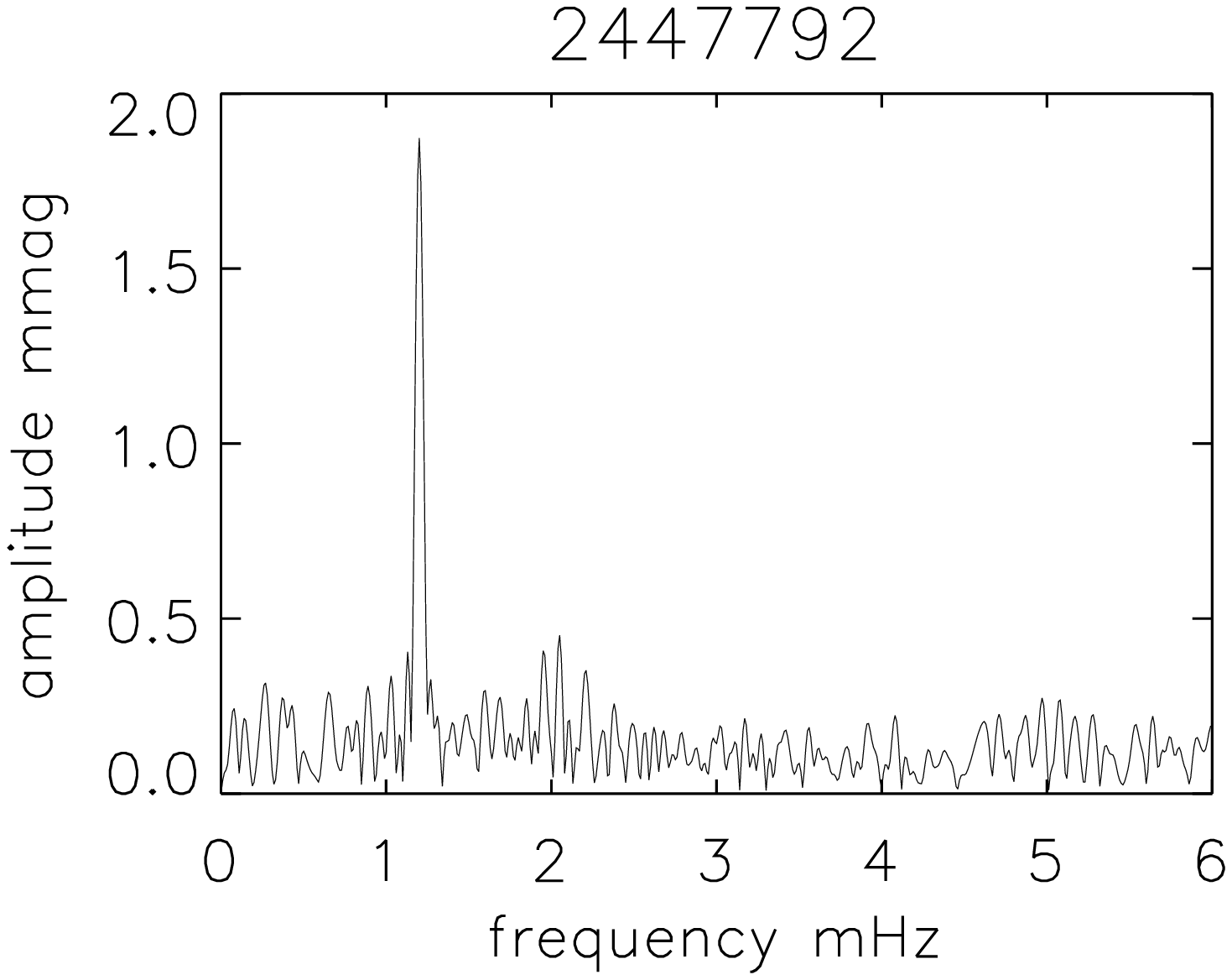}
\epsfxsize 5cm\epsfbox{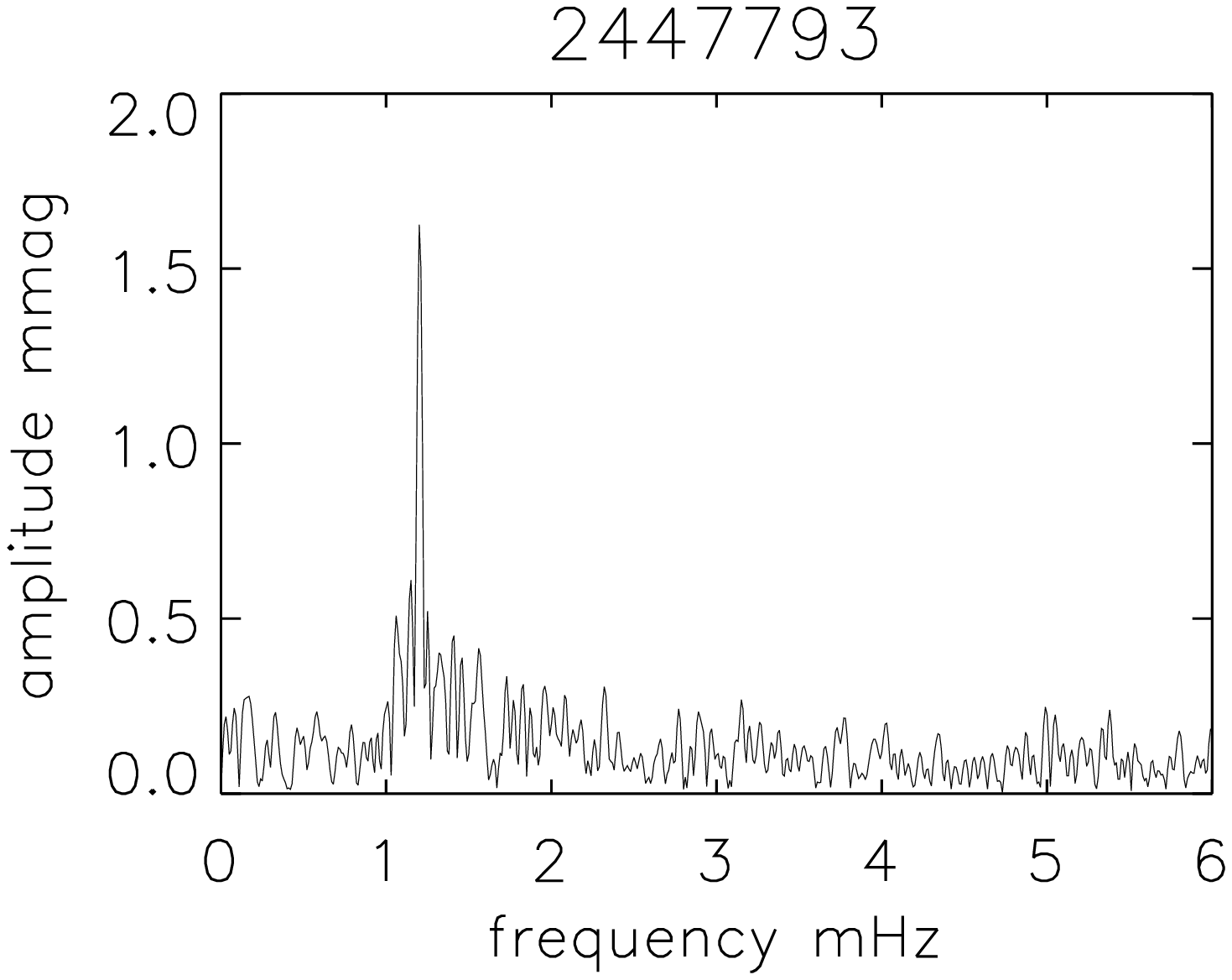}
\epsfxsize 5cm\epsfbox{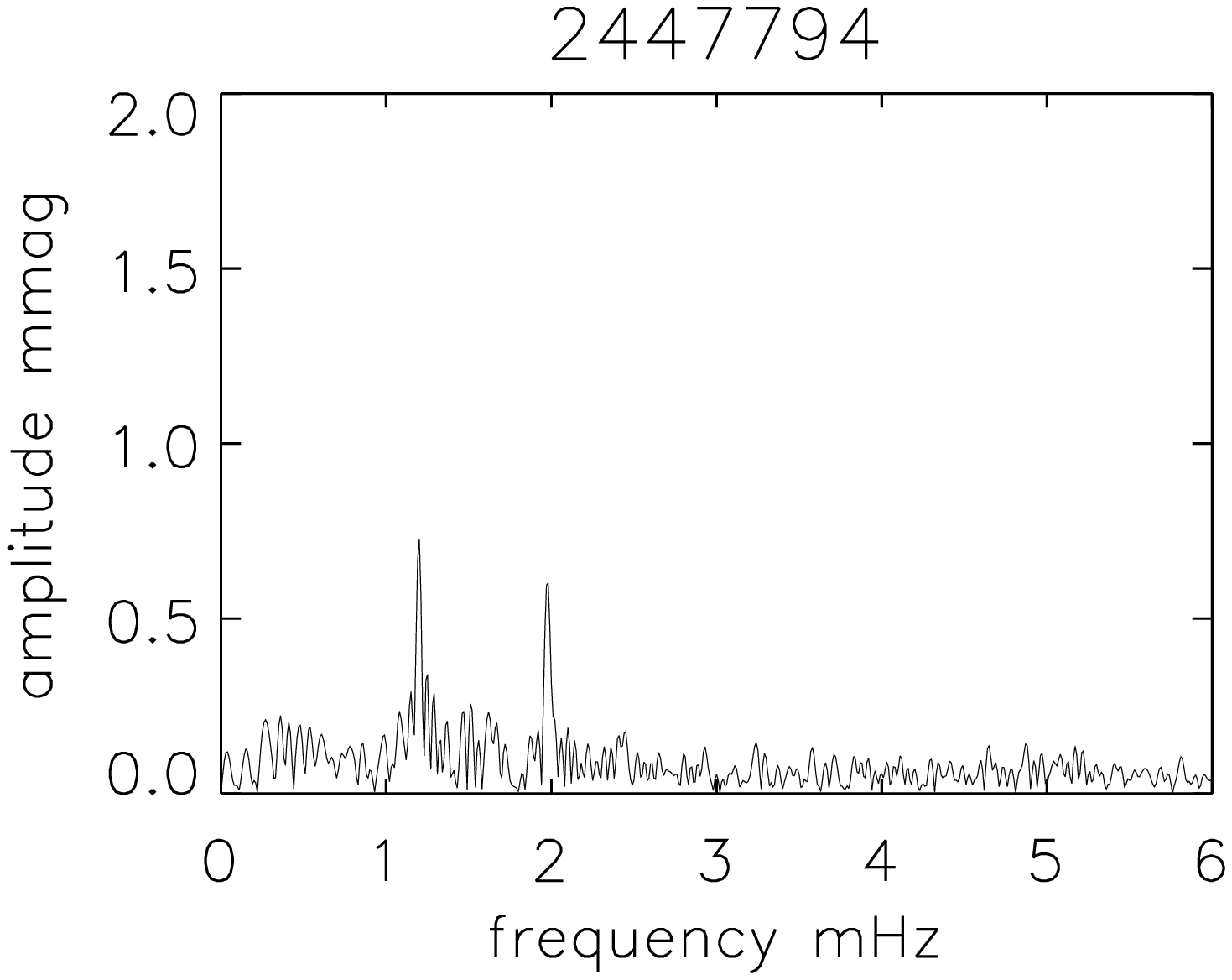}
\caption{\label{ft1989}
Amplitude spectra of the 1989 Johnson $B$ data. This data, obtained by \citet{kreidl1991} show the presence of the second frequency near 2~mHz. The relative amplitudes of the frequencies vary from night-to-night. On the night of JD 2447789 the amplitudes of both frequencies are equal, and the second frequency cannot be seen on the nights JD~2447788 and JD~2447793.}
\end{figure*}

\begin{figure*}
\centering
\epsfxsize 5cm\epsfbox{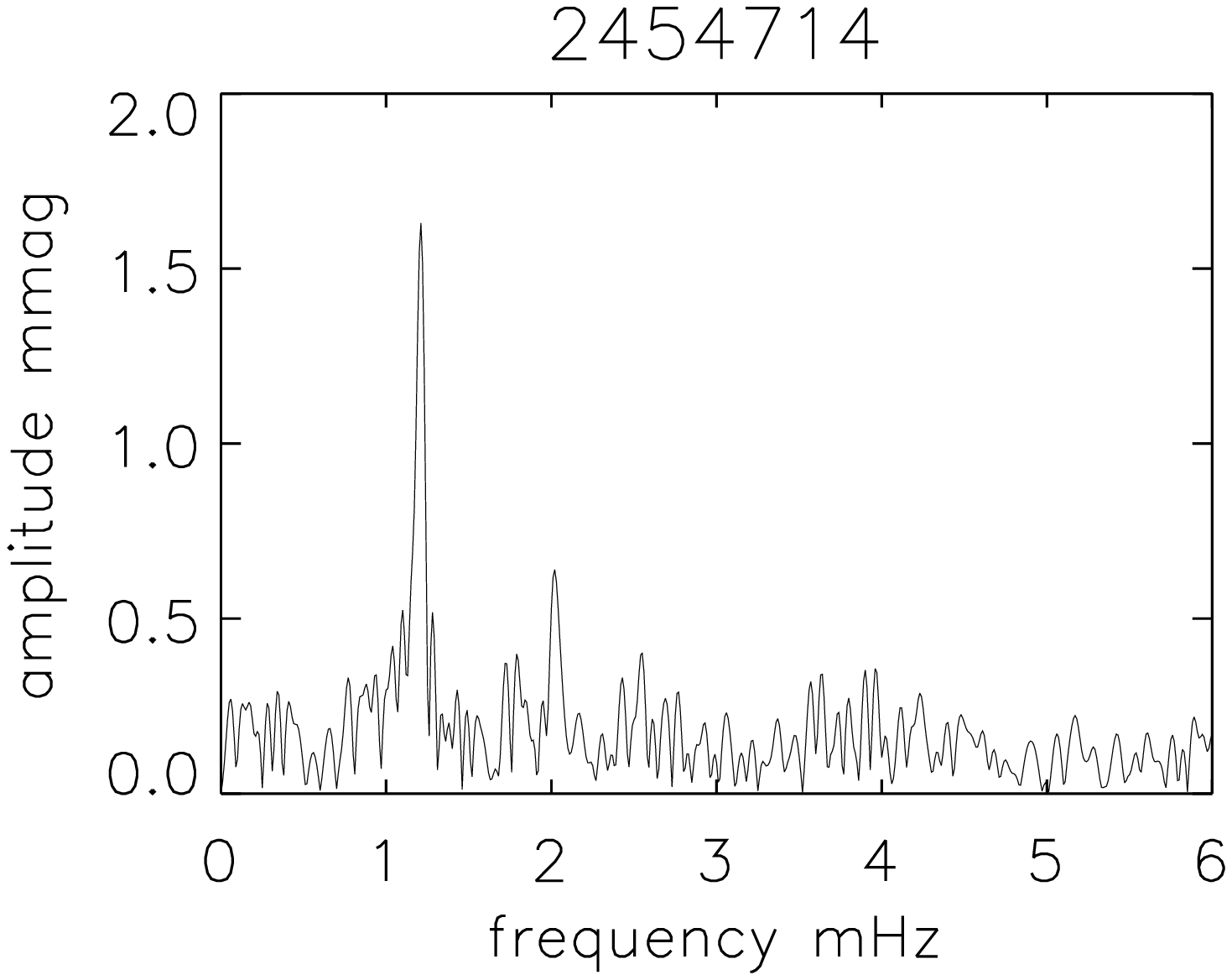}
\epsfxsize 5cm\epsfbox{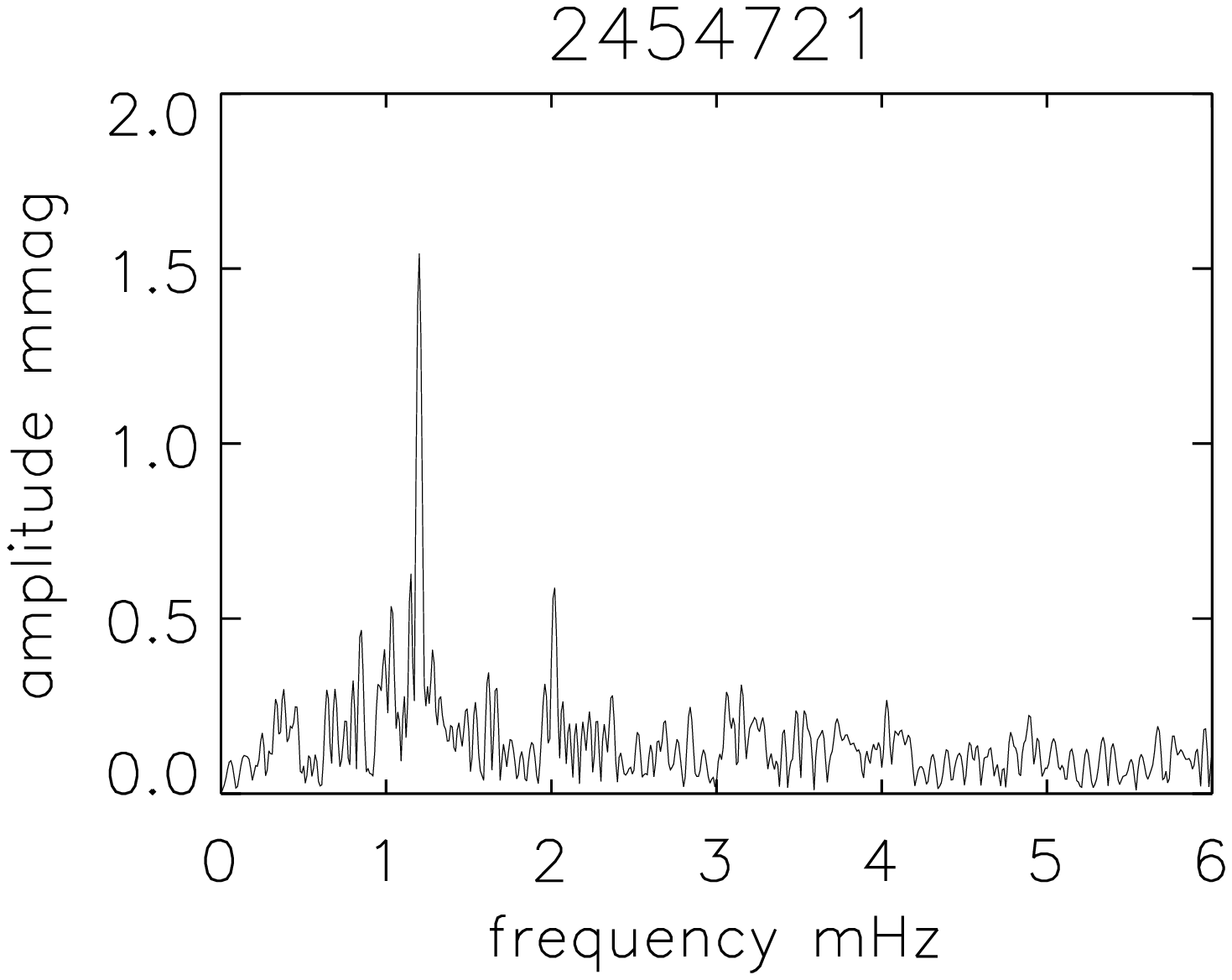}
\epsfxsize 5cm\epsfbox{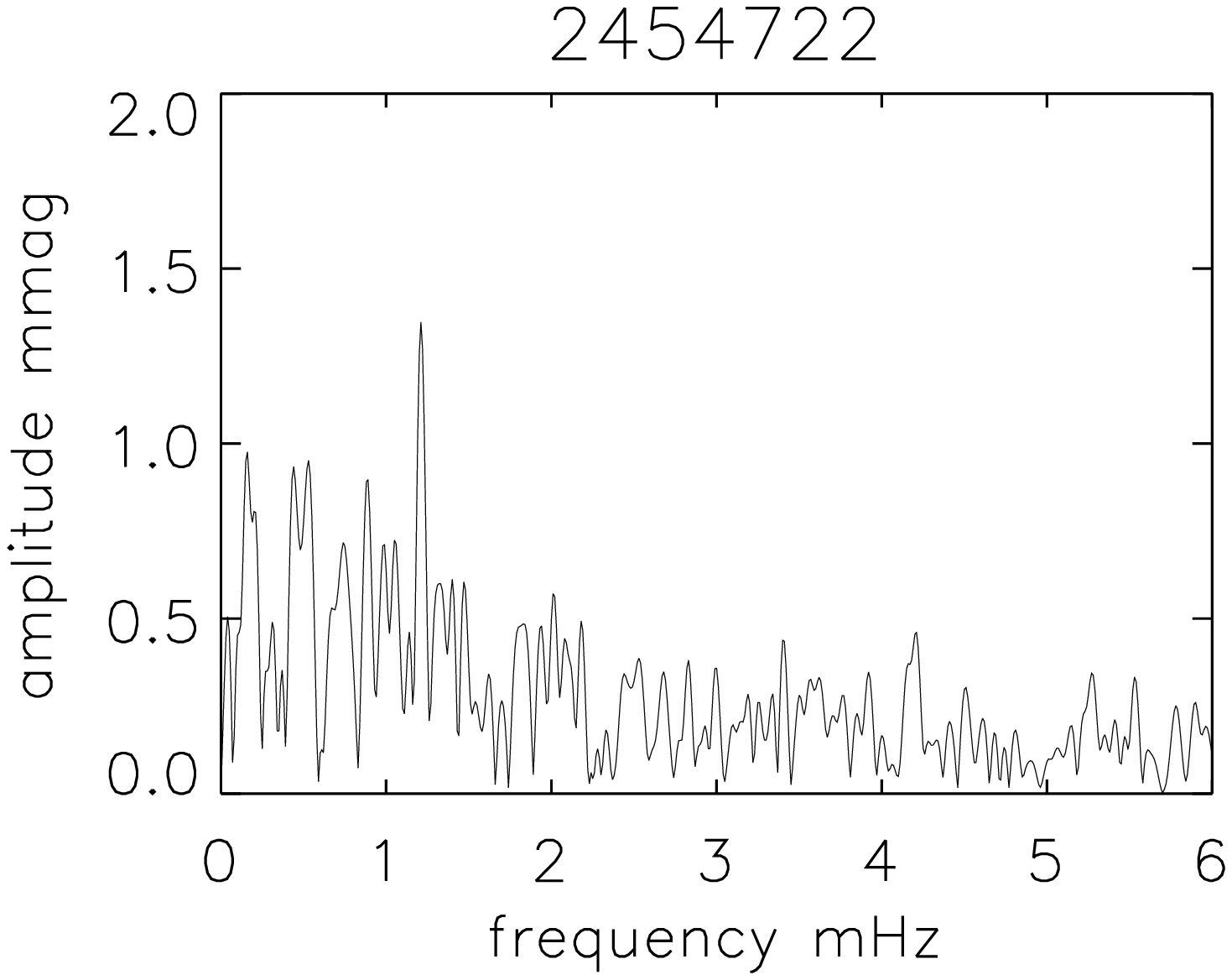}
\caption{\label{ft2008}
Amplitude spectra of the 2008 Johnson $B$ data (this paper) again showing the presence of both frequencies and amplitude modulation.}
\end{figure*}

In Figs~\ref{ft1982}, \ref{ft1989} and \ref{ft2008} we present the Fourier spectra of the data obtained in 1982, 1989 and 2008, respectively. Clearly, frequency $\nu_2$ is not present in the 1982 data, but is obvious in the 1989 data (as reported in \citealt{kreidl1991}) and in the new 2008 data. In Fig.~\ref{ft1989}, the amplitude spectra show clear amplitude modulation night-to-night. In particular, note that on the night 2447788 the frequency $\nu_2$ is not visible at all, and the noise level is low. In none of the amplitude spectra is there any evidence for the `missing modes' between $\nu_1$ and $\nu_2$. The photometric observations therefore suggest strong amplitude modulation on a time-scale of 1~d, or less.

\section{Spectroscopic analysis}

\subsection{The spatial variability of the two pulsation modes}
\label{sec:spatial}

We studied the radial velocity pulsations for HD~217522 by measuring centre-of-gravity shifts of individual lines in the spectra. Because of the stratification of ions in the observable atmosphere of roAp stars, each spectral line samples a different line formation region, giving information on both the stratification and the pulsation mode geometry and behaviour. Line bisector measurements on many roAp stars show significant amplitude and phase variations as a function of atmospheric depth (e.g., \citealt{kurtz2005}; \citealt{ryabchikova2007}).

\begin{figure*}
\centering
\epsfxsize 5.cm\epsfbox{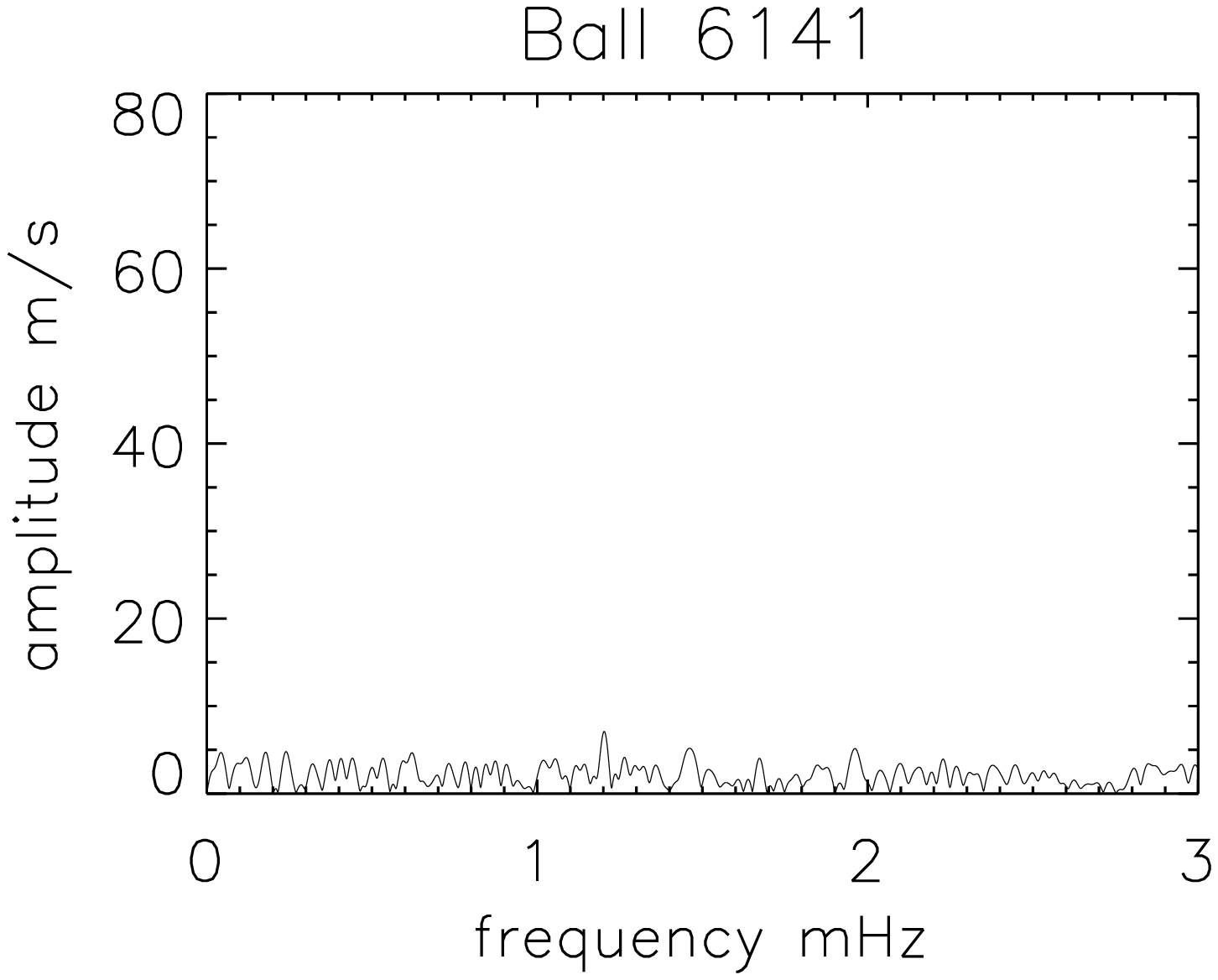}
\epsfxsize 5.cm\epsfbox{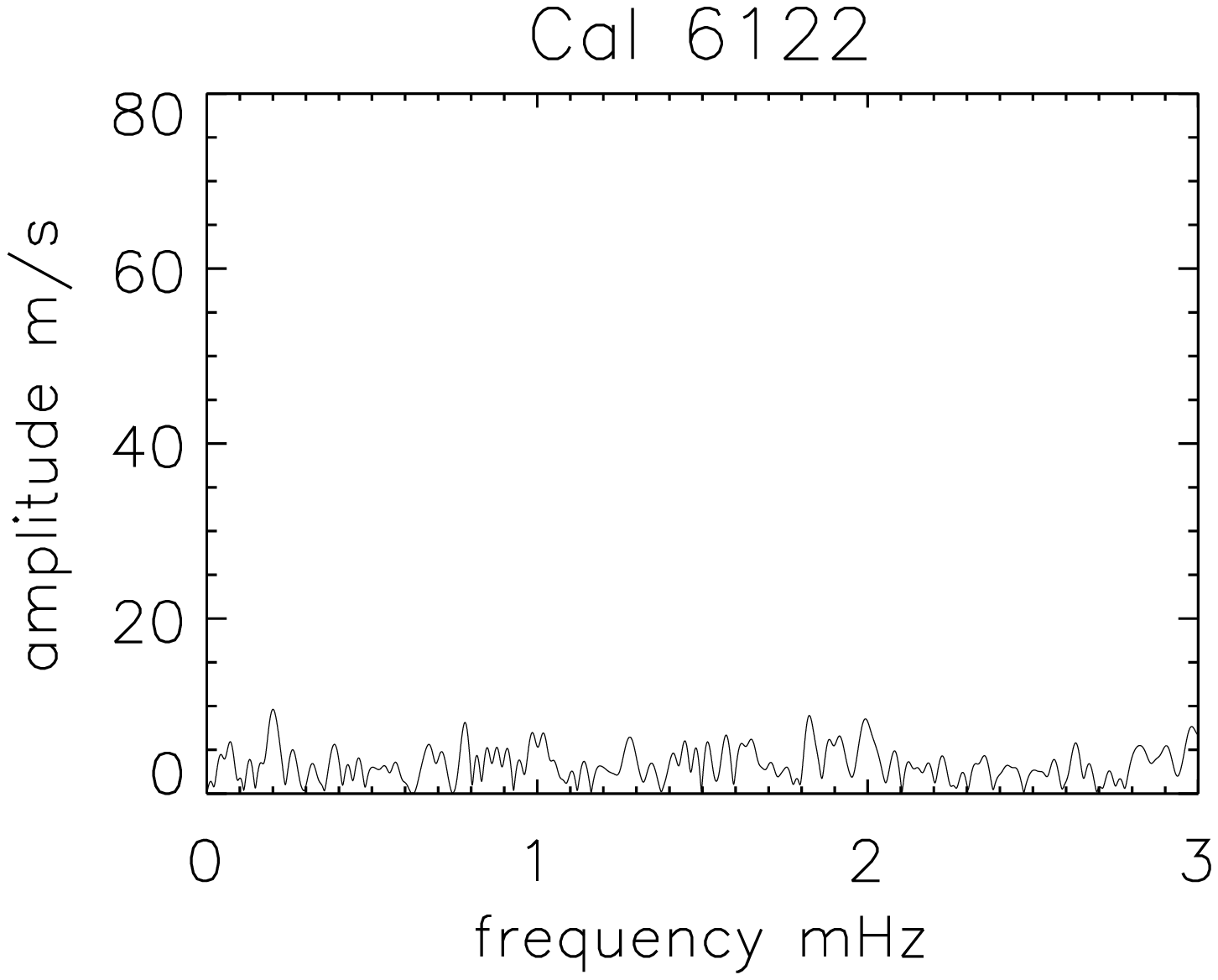}
\epsfxsize 5.cm\epsfbox{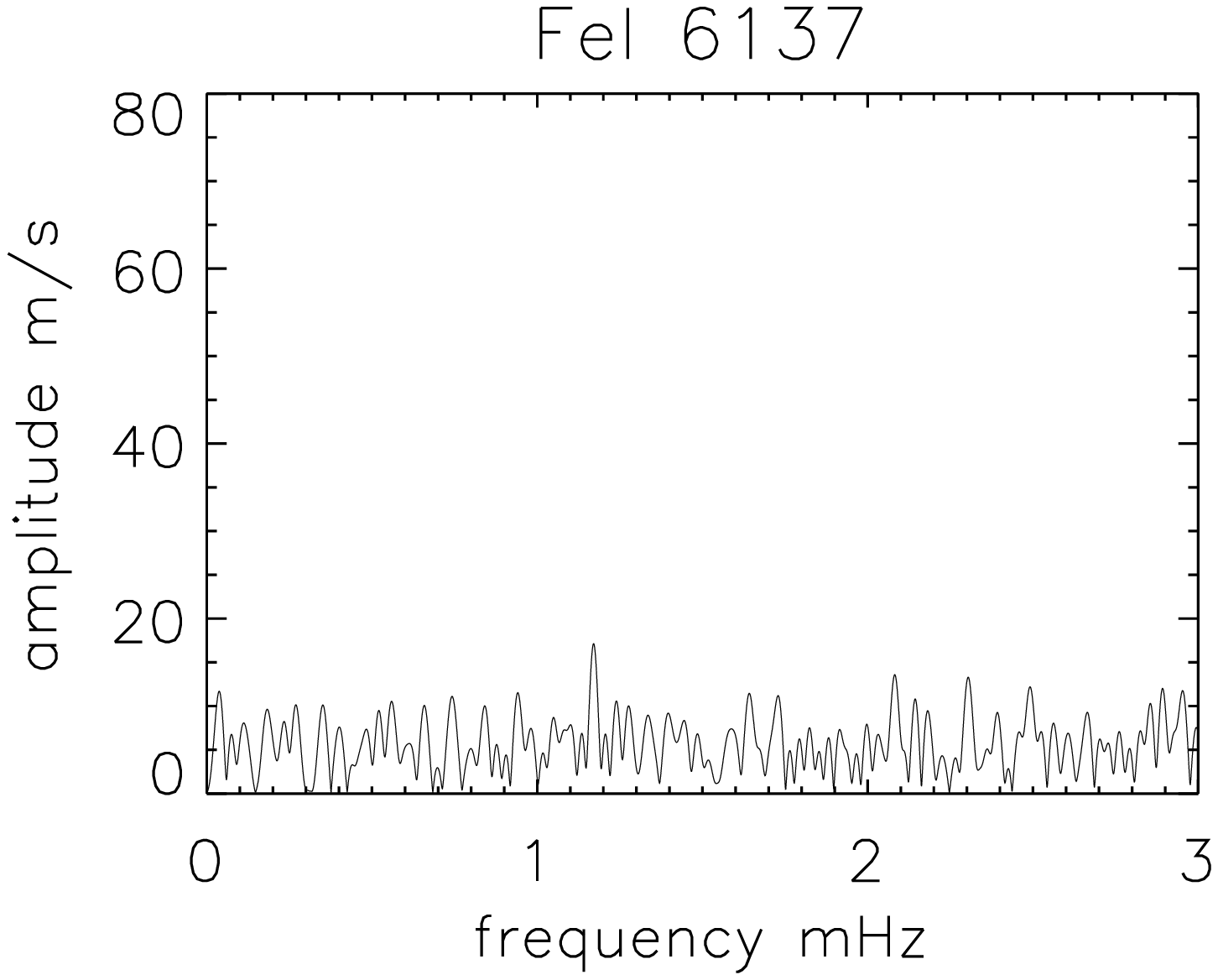}
\epsfxsize 5.cm\epsfbox{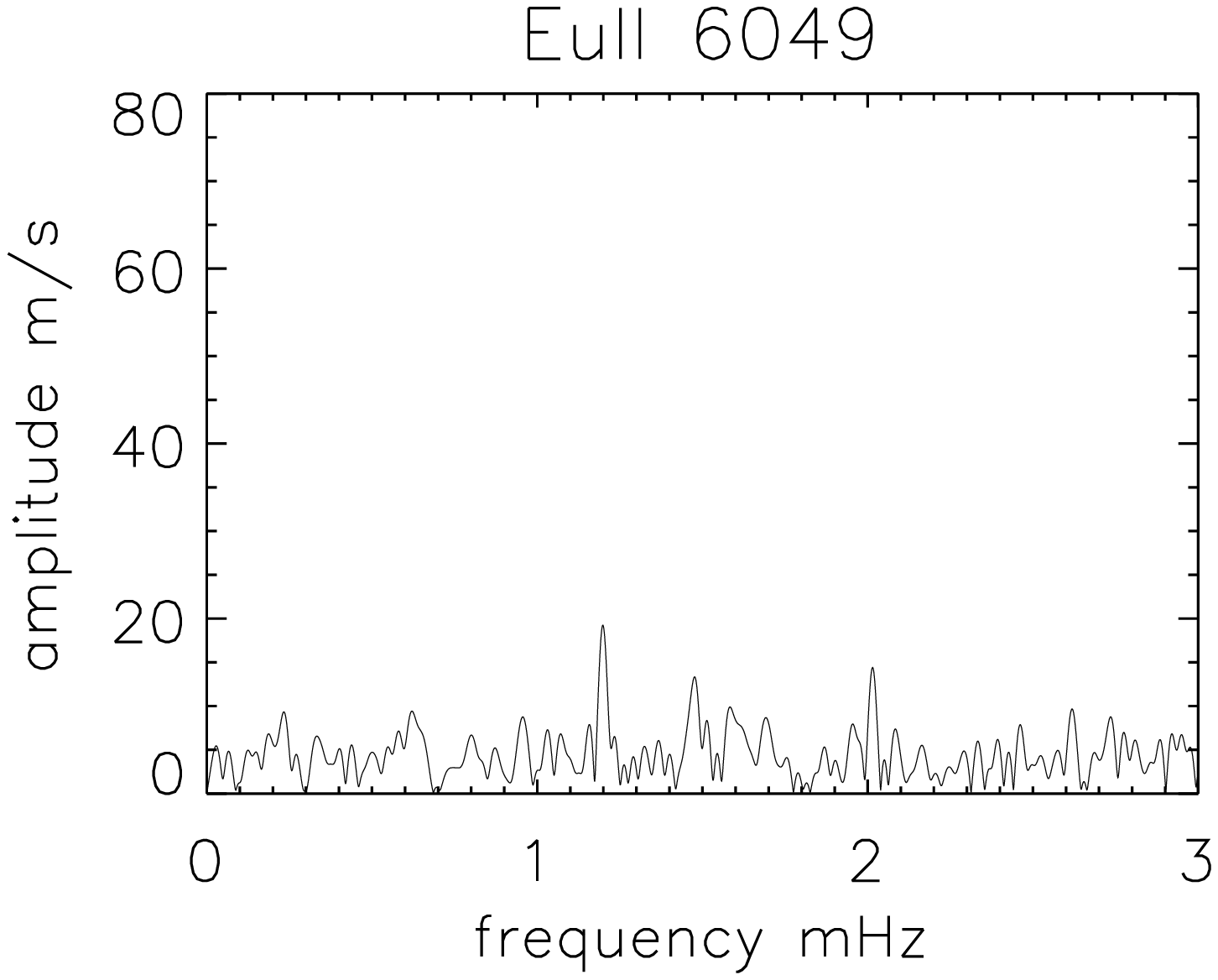}
\epsfxsize 5.cm\epsfbox{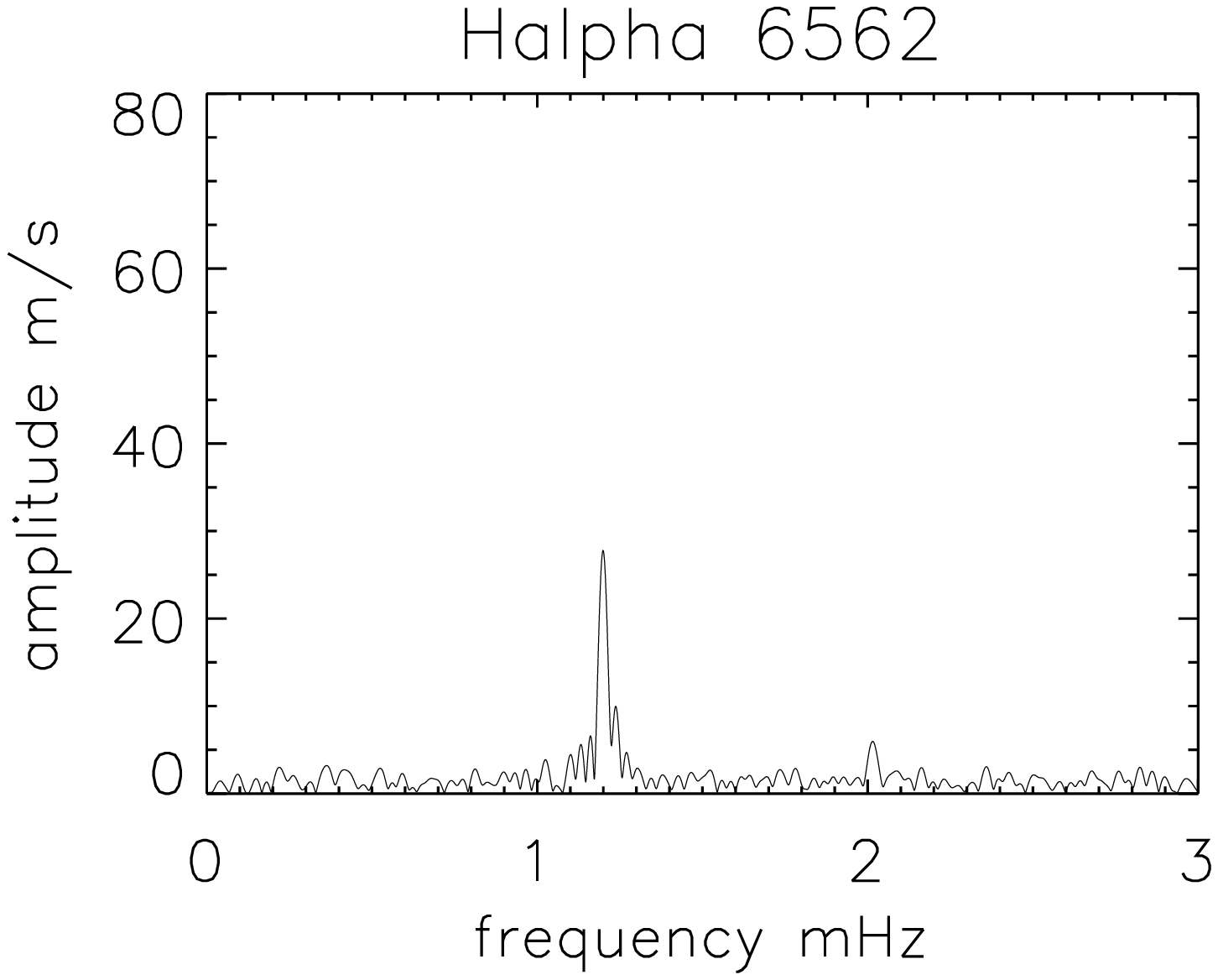}
\epsfxsize 5.cm\epsfbox{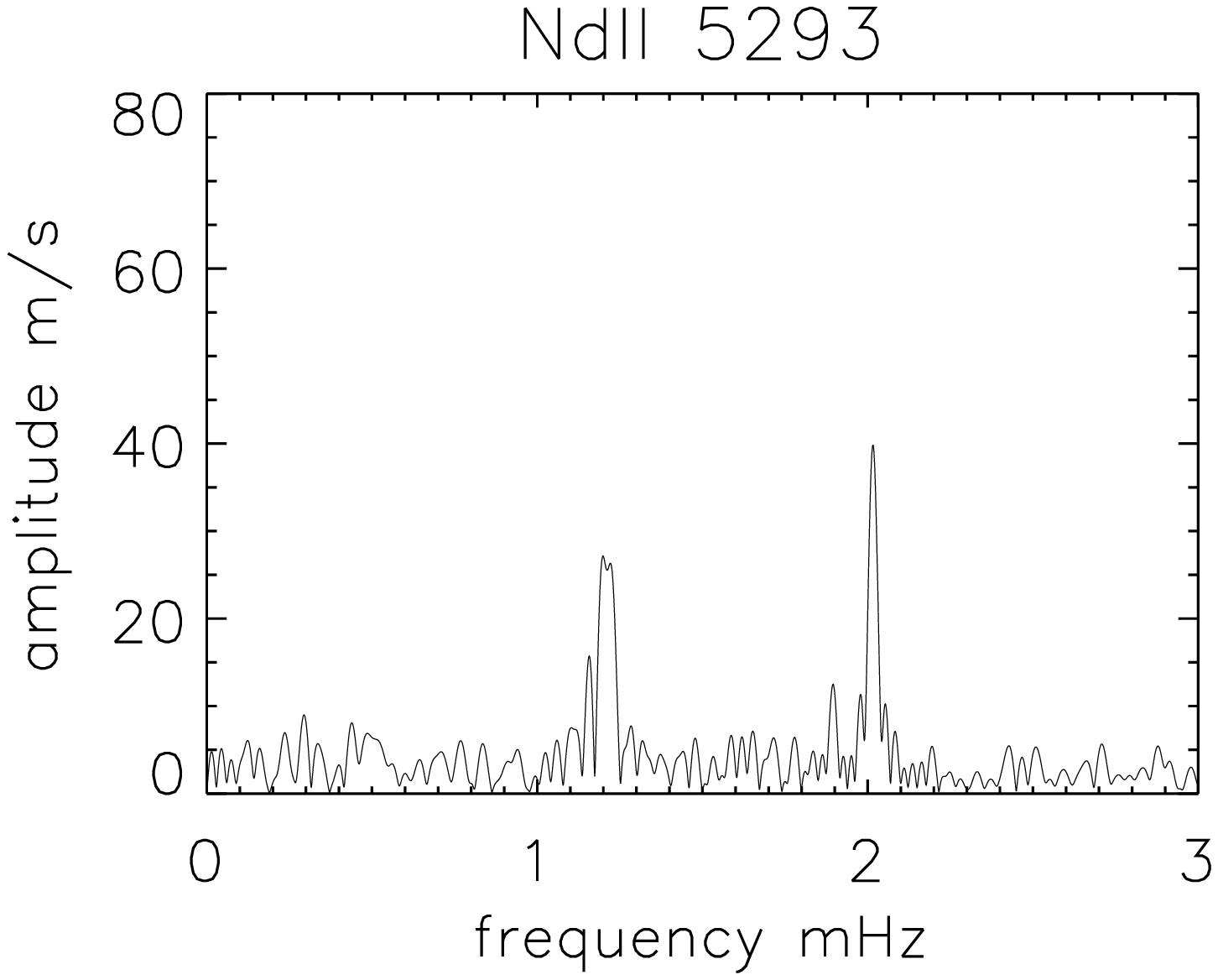}
\epsfxsize 5.cm\epsfbox{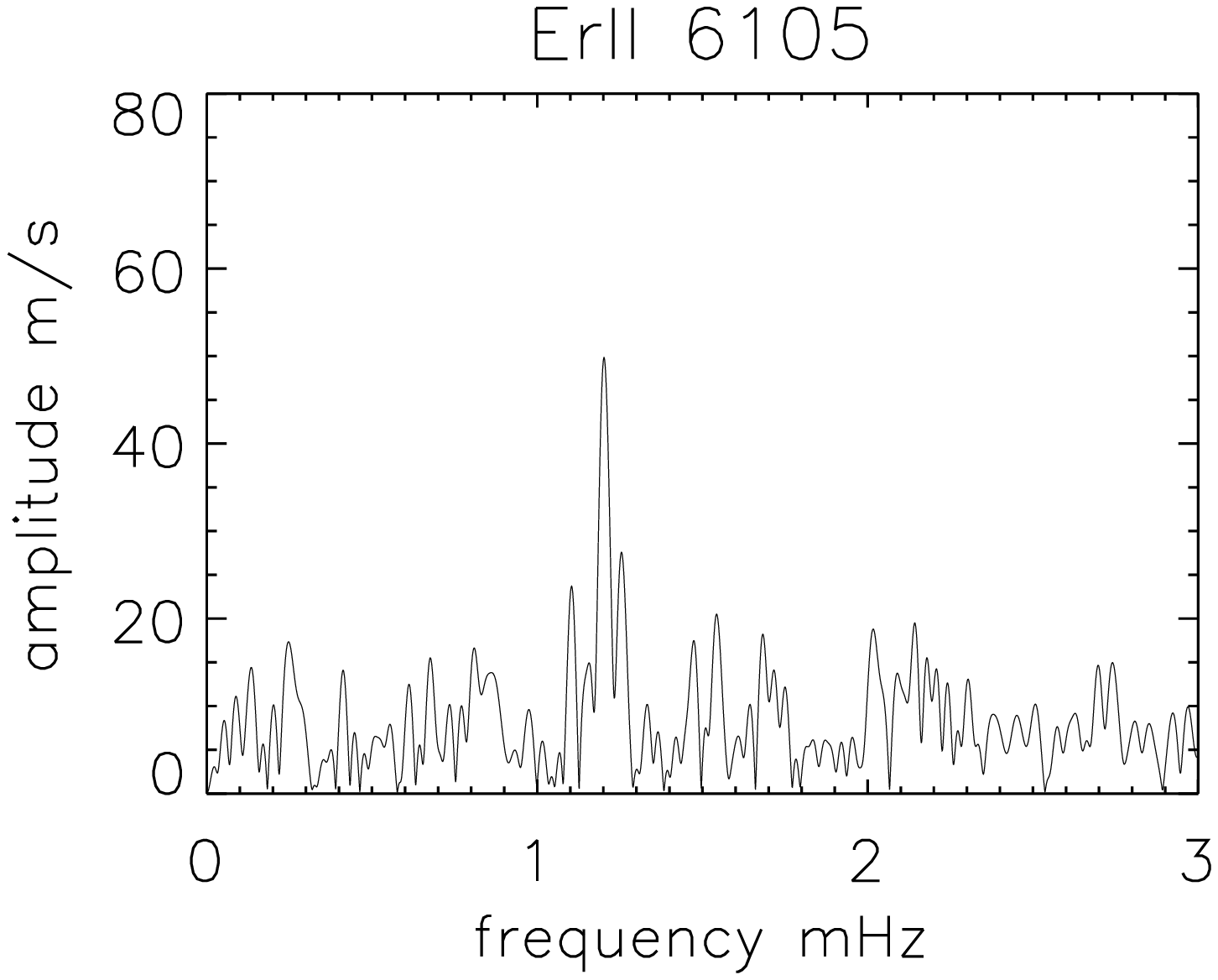}
\epsfxsize 5.cm\epsfbox{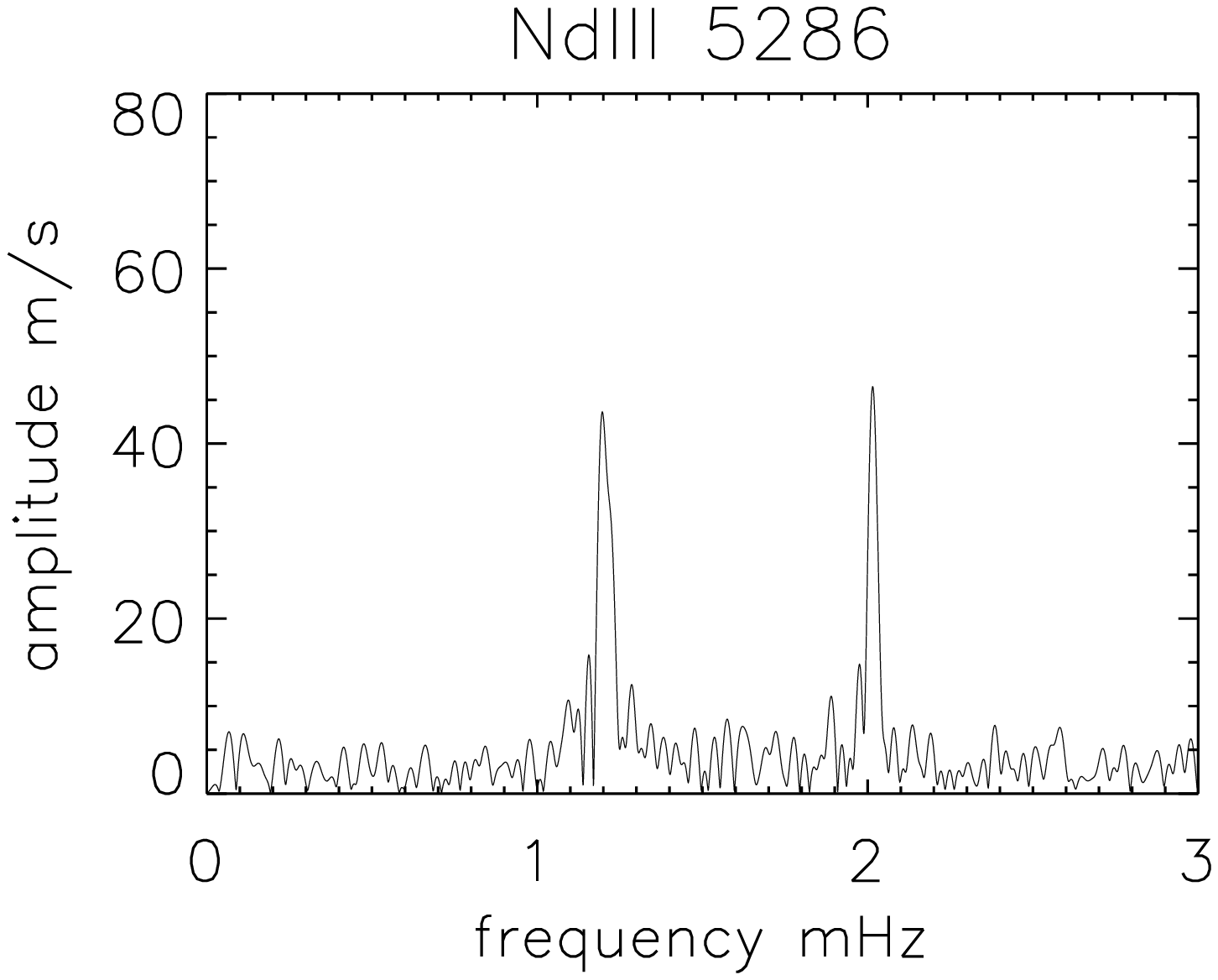}
\epsfxsize 5.cm\epsfbox{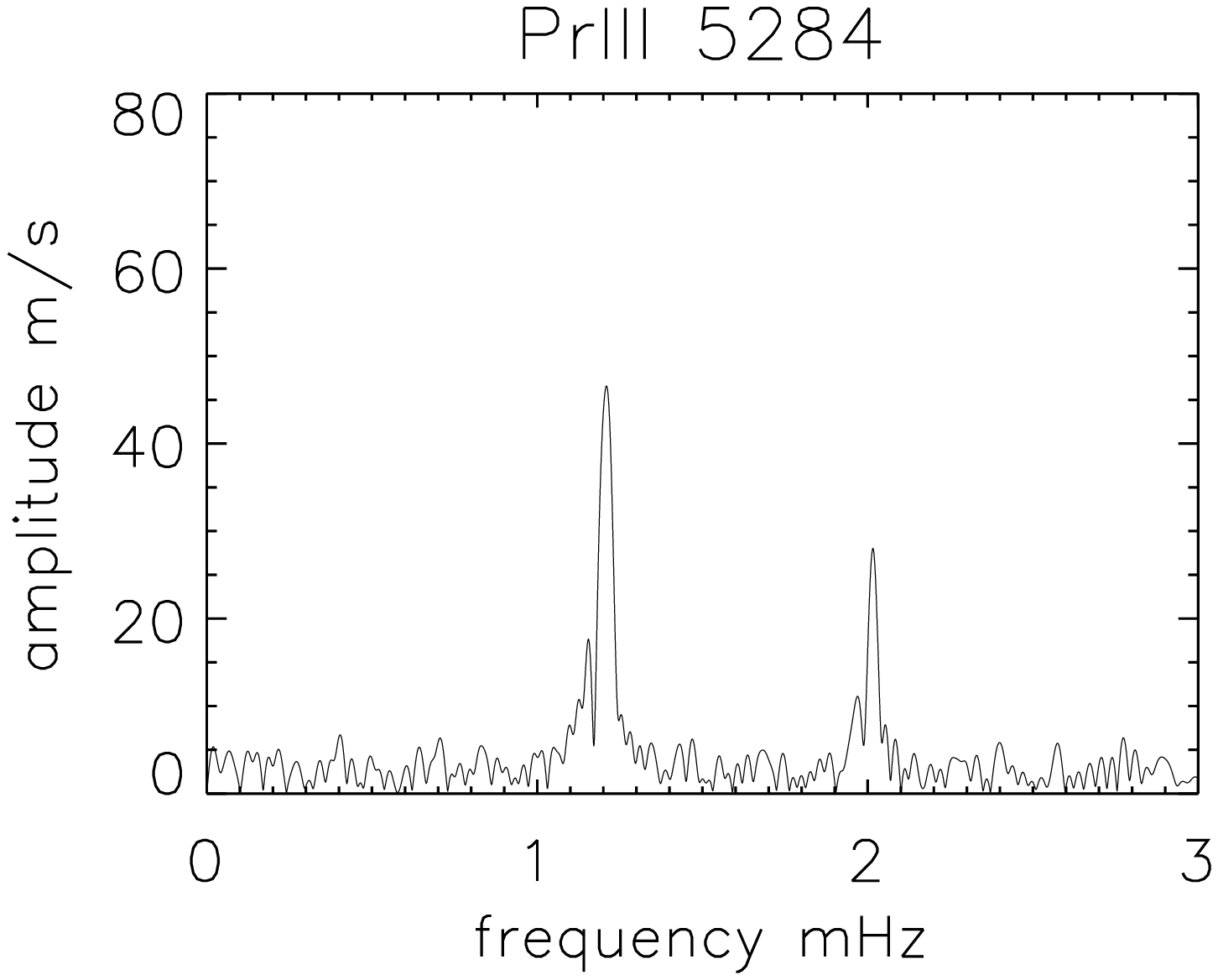}
\epsfxsize 5.cm\epsfbox{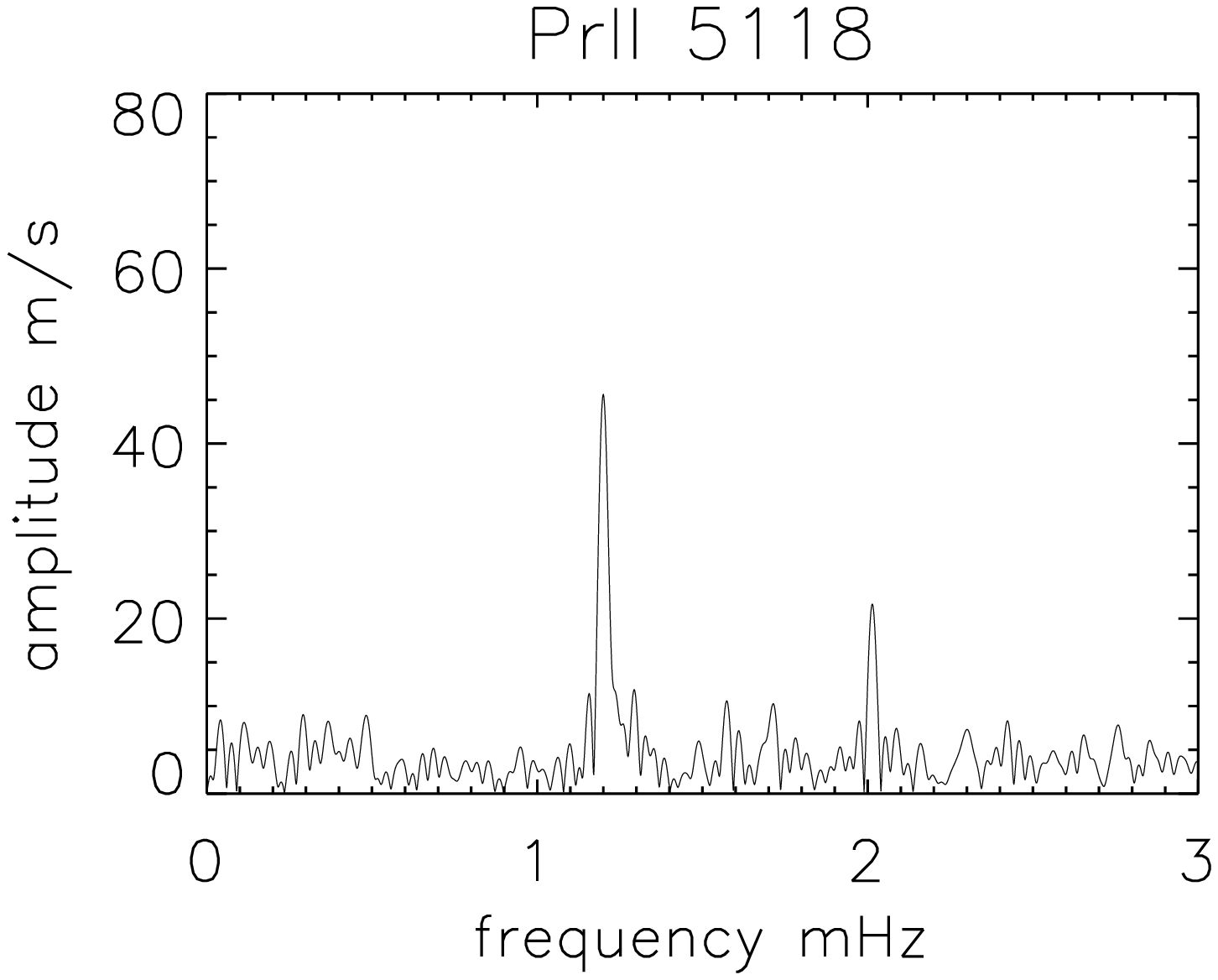}
\epsfxsize 5.cm\epsfbox{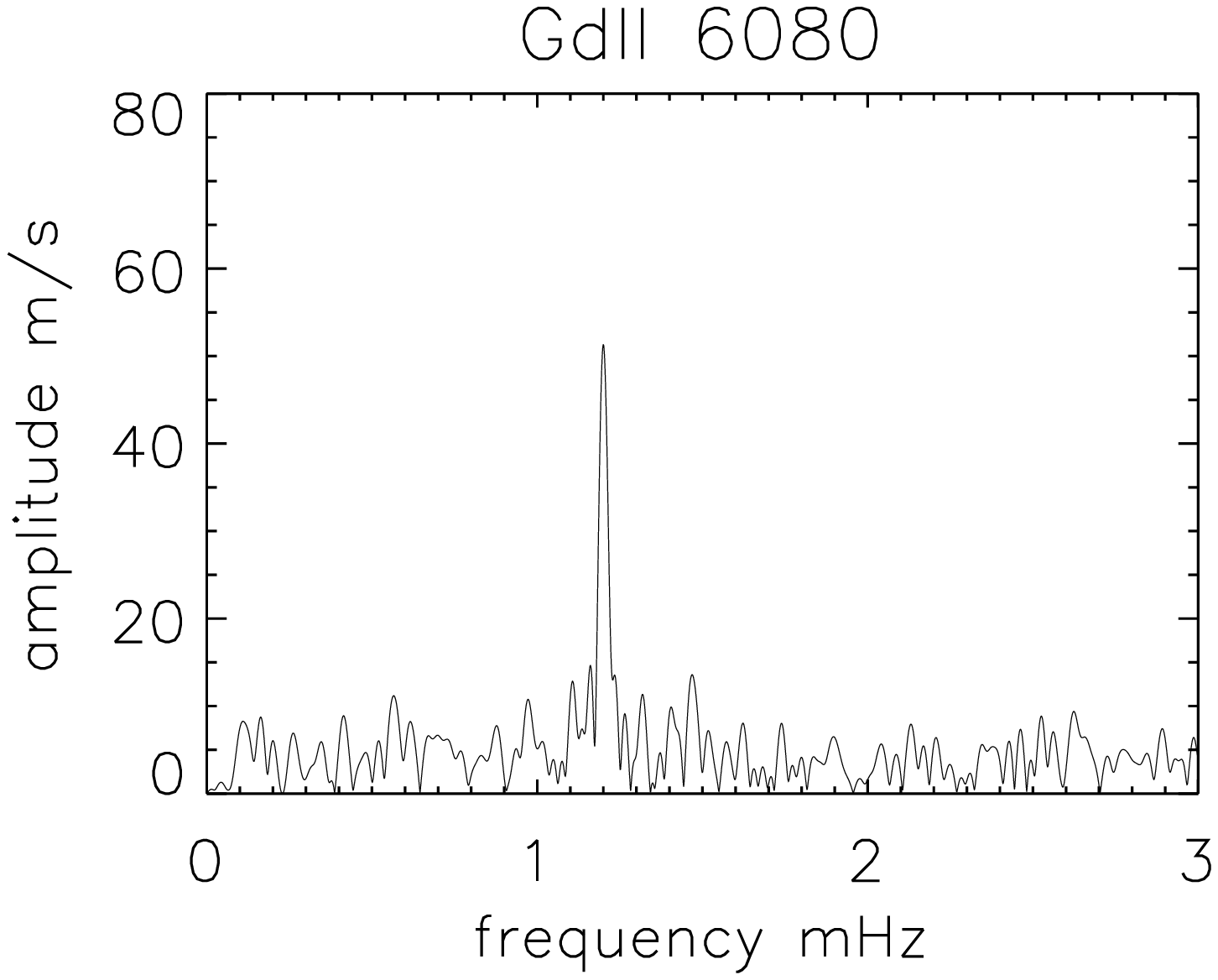}
\epsfxsize 5.cm\epsfbox{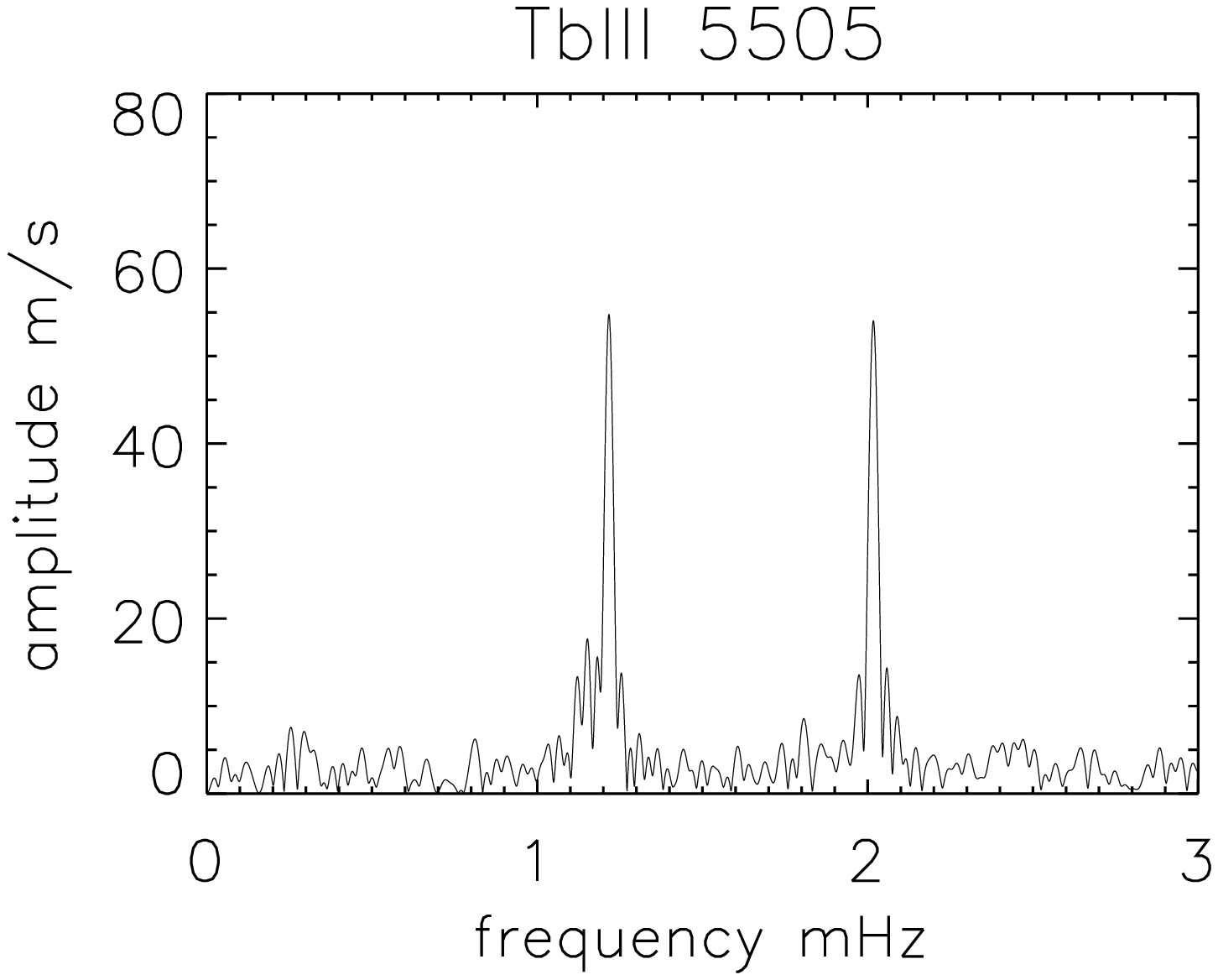}
\epsfxsize 5.cm\epsfbox{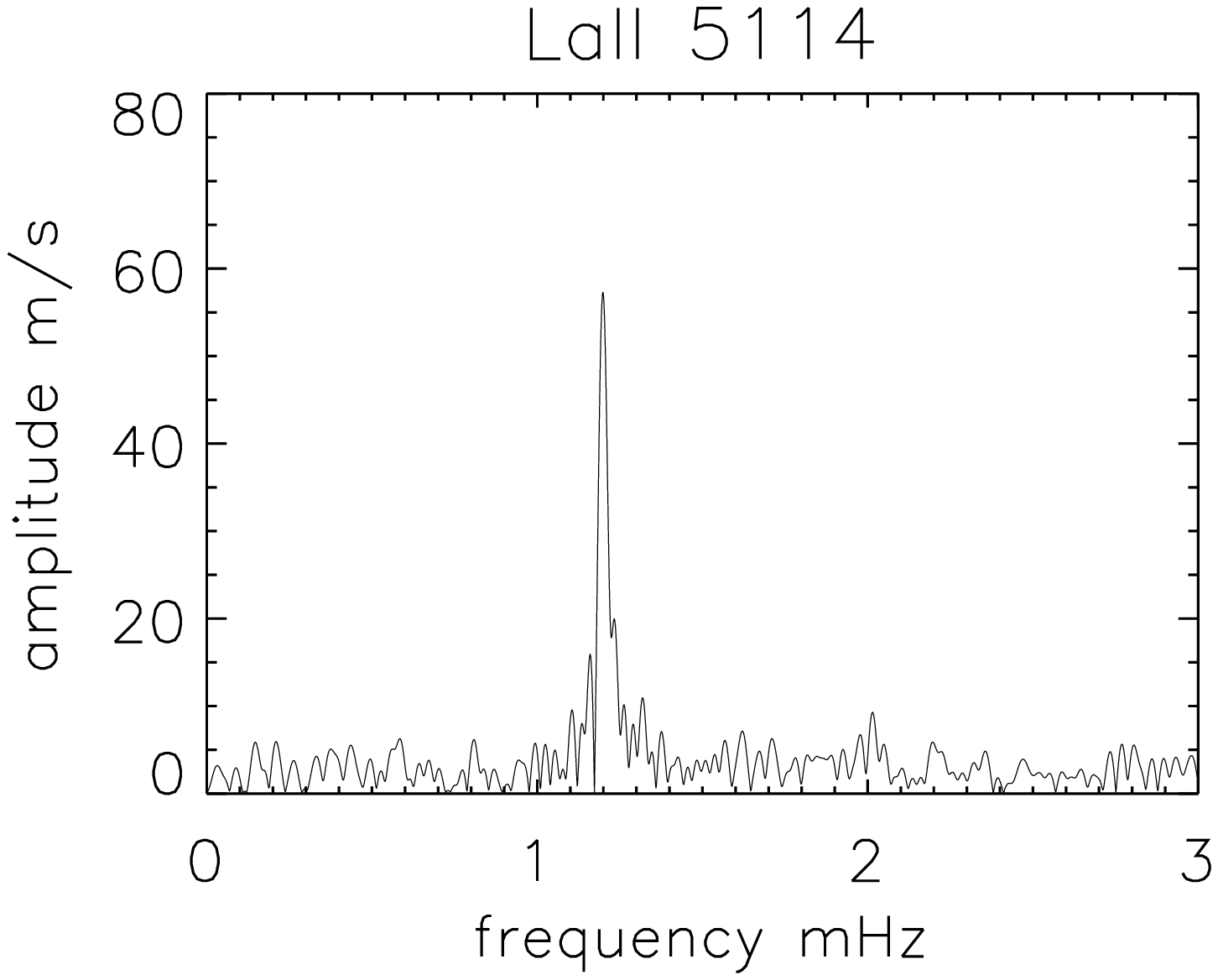}
\epsfxsize 5.cm\epsfbox{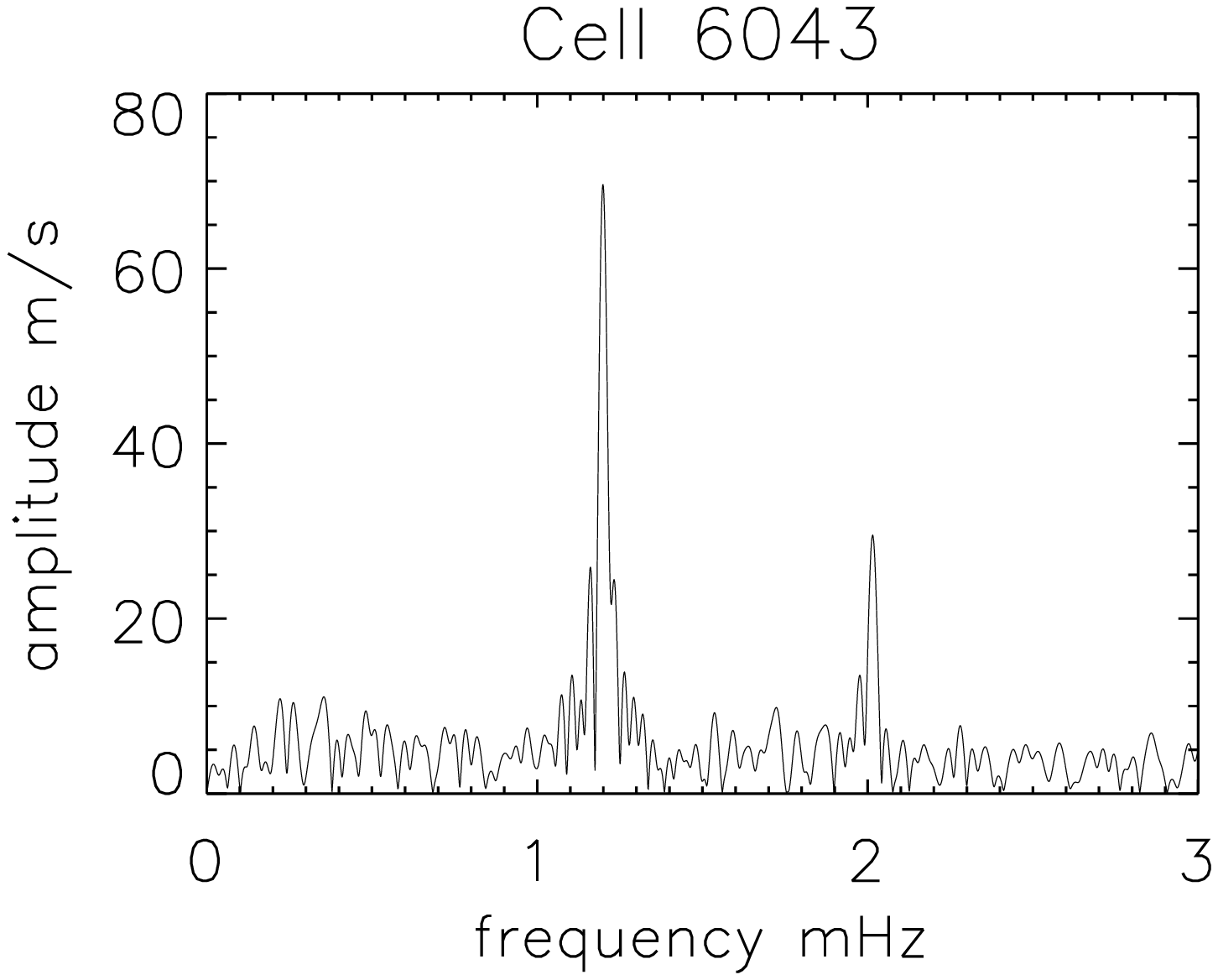}
\epsfxsize 5.cm\epsfbox{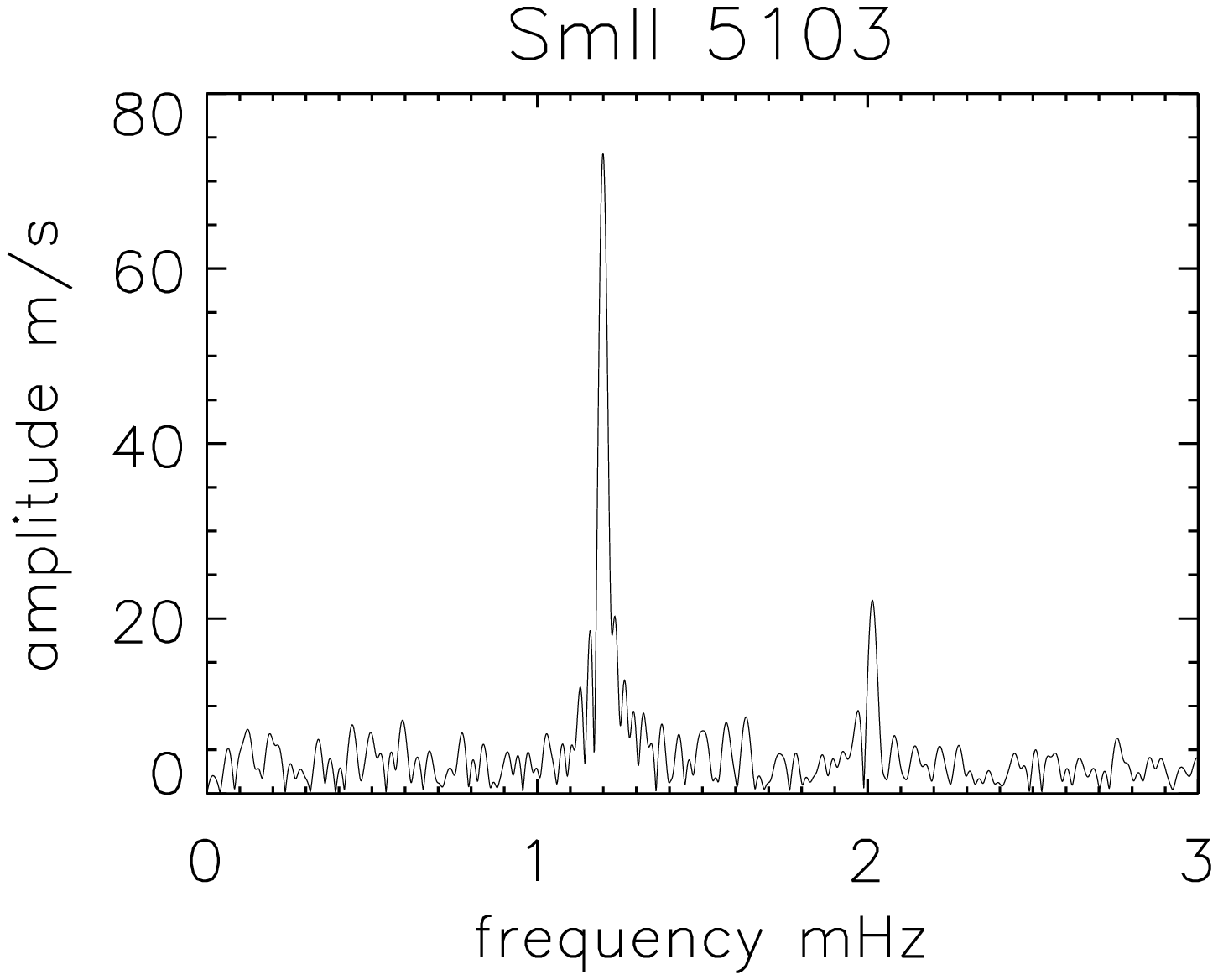}
\caption{\label{fig:lines} Amplitude spectra for sample lines arranged in approximately increasing order of amplitude for $\nu_1$. The amplitude variability for $\nu_1$ and $\nu_2$ as a function of the atmospheric line-forming region is apparent. All amplitude spectra are for the full 10.1-h time spectroscopic series, hence the range of behaviour seen is not temporal, but spatial.}
\end{figure*}

\begin{figure*}
\centering
\epsfxsize 5.cm\epsfbox{5284.ps}
\epsfxsize 5.cm\epsfbox{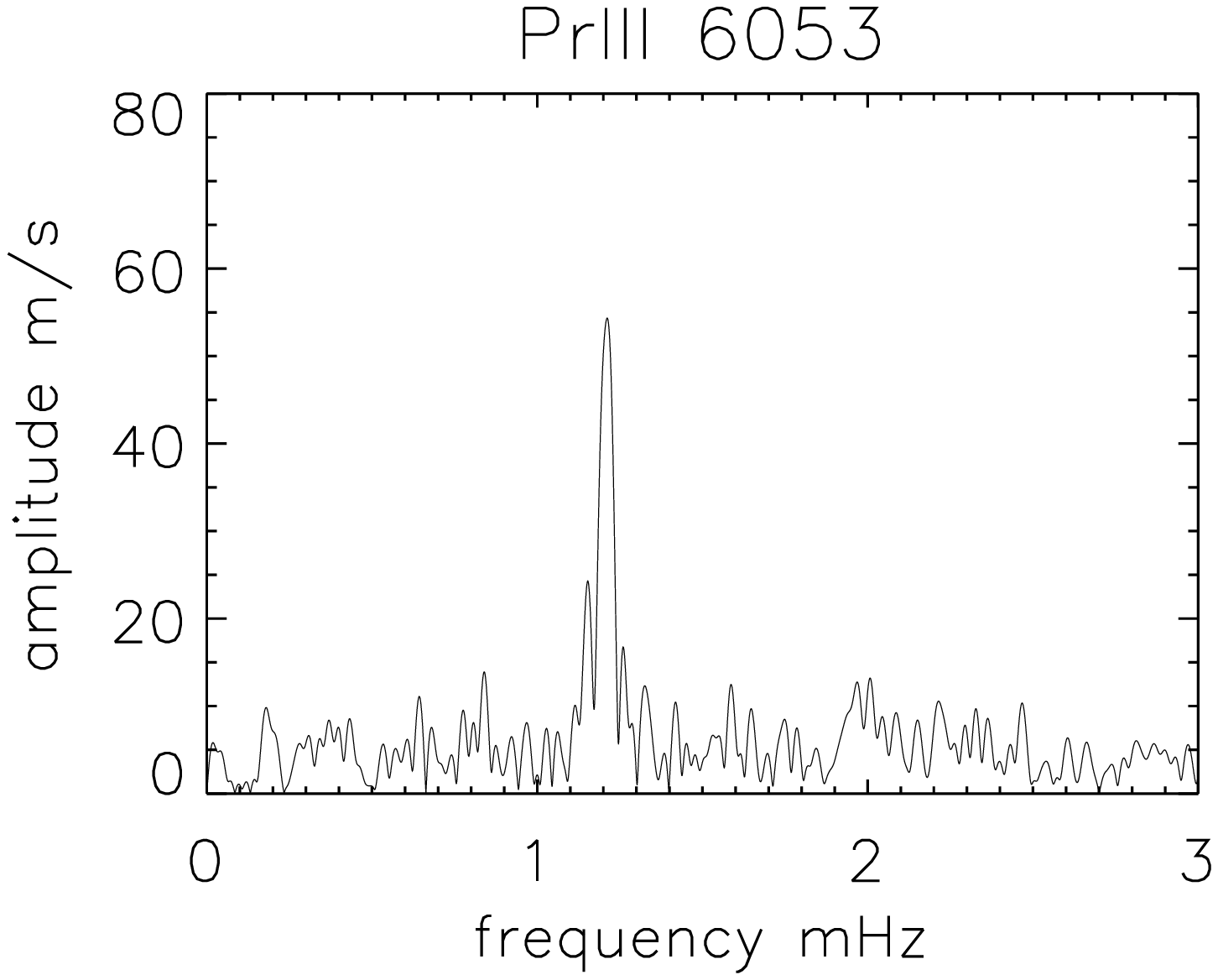}
\epsfxsize 5.cm\epsfbox{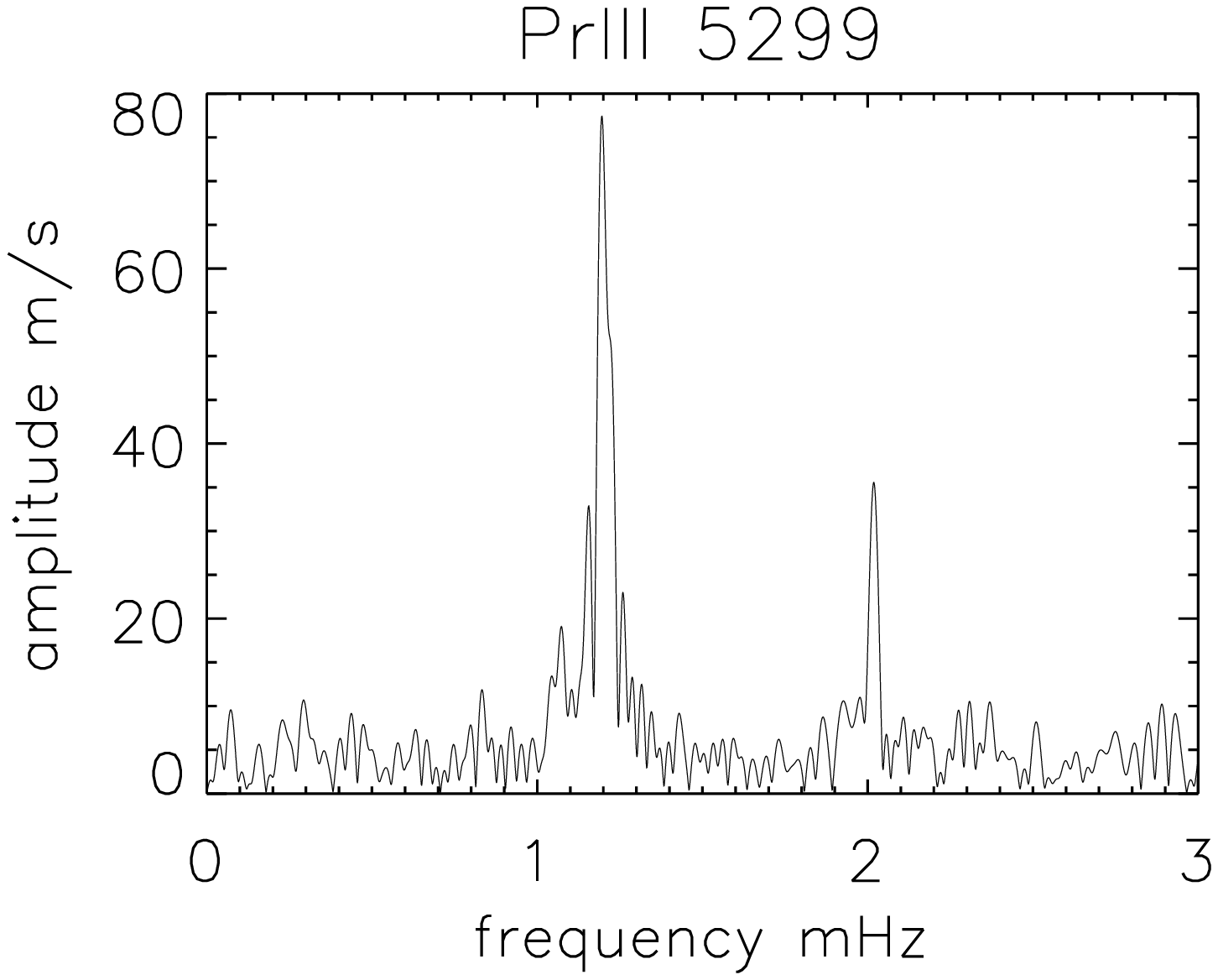}
\epsfxsize 5.cm\epsfbox{5286.ps}
\epsfxsize 5.cm\epsfbox{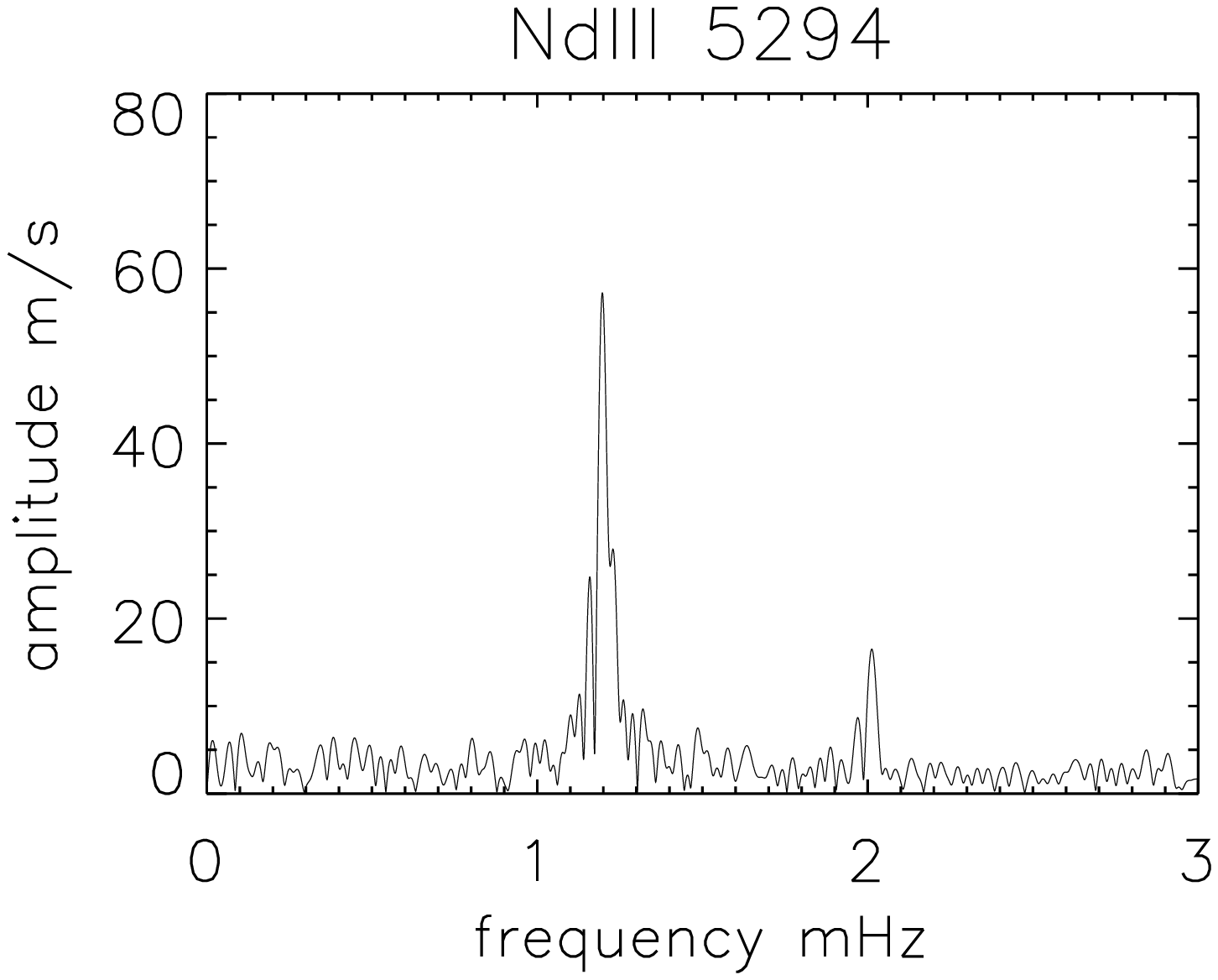}
\epsfxsize 5.cm\epsfbox{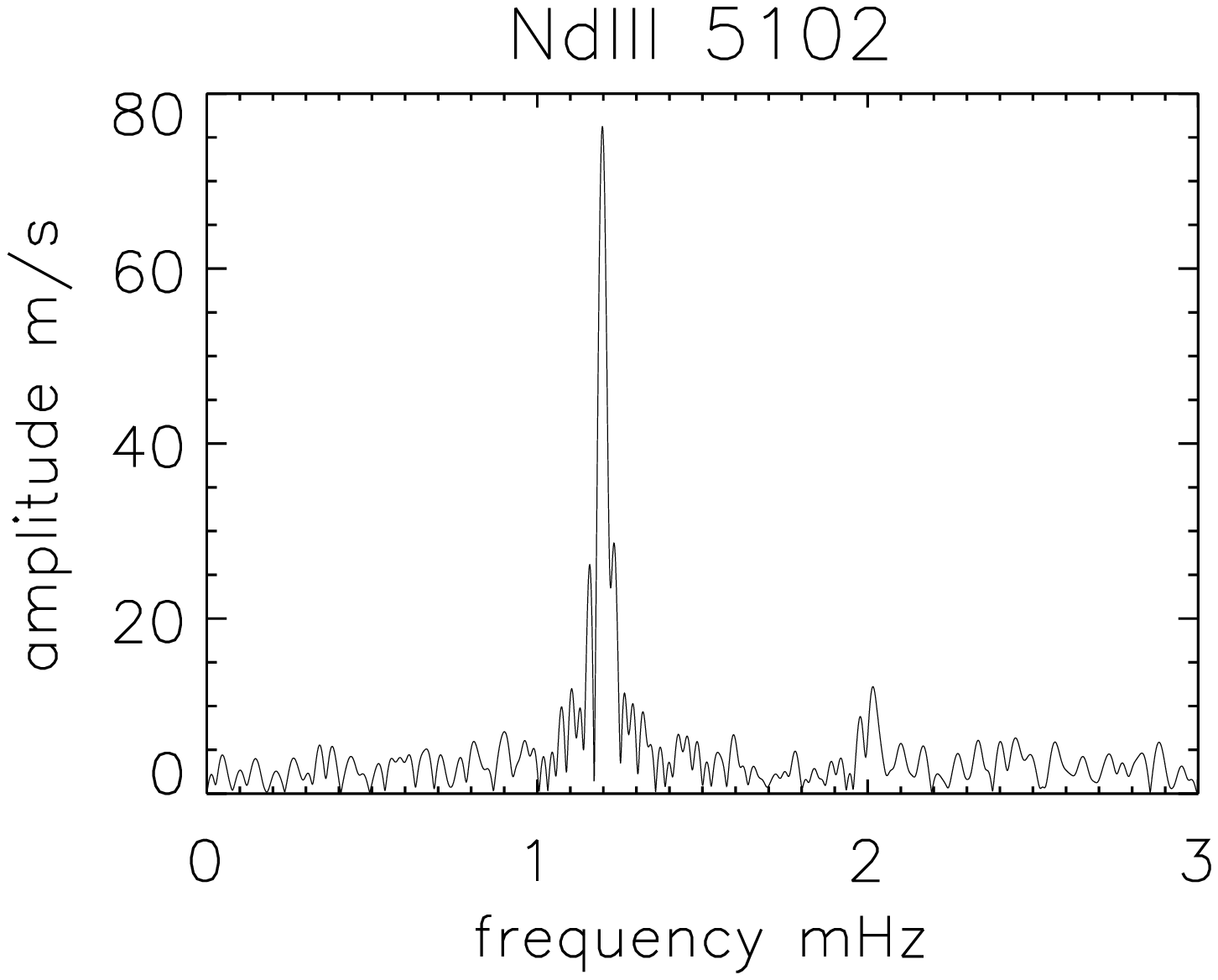}
\caption{\label{fig:lines2} Amplitude spectra for three lines of Pr~\textsc{iii} (top) and three lines of Nd~\textsc{iii} (bottom) showing that differences in the line forming regions even for particular ions result in very different measured pulsation amplitudes. }
\end{figure*}

Fig.~\ref{fig:lines} shows amplitude spectra for a sample of spectral lines arranged approximately in order of increasing amplitude of $\nu_1$. Elemental abundances in roAp stars are both stratified vertically in the atmosphere, and horizontally, as a consequence of abundance spots, or patches. Each spectral line has its own line-forming region in three dimensions, resulting in different pulsation amplitudes for $\nu_1$ and $\nu_2$ for each line. This is even true for lines of the same ion, as seen in Fig.~\ref{fig:lines2}, as a consequence of the different excitation for each line and the extreme sensitivity of pulsation amplitude to position in the atmosphere.

The range of pulsation amplitudes for the different spectral lines is unlikely to be a temporal effect. All the amplitude spectra shown are for the same full 10.1-h spectroscopic data set. While the observed pulsation amplitudes for the two frequencies may be a function of rotation if the degree of the modes is different, the lack of any rotation signal in photometry, along with our measured $v \sin i = 3$~km~s$^{-1}$, suggests that either HD~217522 is seen close to pole-on, or it has a long rotation period. In either case, the aspect of the star is unlikely to have changed significantly in the 10.1~h of our spectroscopic time series. Therefore, the range of amplitudes for both $\nu_1$ and $\nu_2$ seen in Figs~\ref{fig:lines} and \ref{fig:lines2} shows the extreme sensitivity of the pulsation three-dimensional geometry to the line-forming layers. While this ultimately may allow detailed atmospheric abundance models, it cautions against firm conclusions about the relationship of the line forming layers of various ions (e.g., \citealt{kurtz2005}; \citealt{ryabchikova2007}).

Although we have argued that the range of pulsation amplitudes for $\nu_1$ and $\nu_2$ for different lines is unlikely to be a temporal effect, the lines are {\it individually} variable in time. This is critical to our goal to understand the time variability of the photometric amplitudes shown in Section~3 above.

\subsection{The temporal variability of pulsation amplitude and phase for the spectral lines of Ce~\textsc{ii} }

\begin{figure*}
\centering
\epsfxsize 5.cm\epsfbox{6043.ps}
\epsfxsize 5.cm\epsfbox{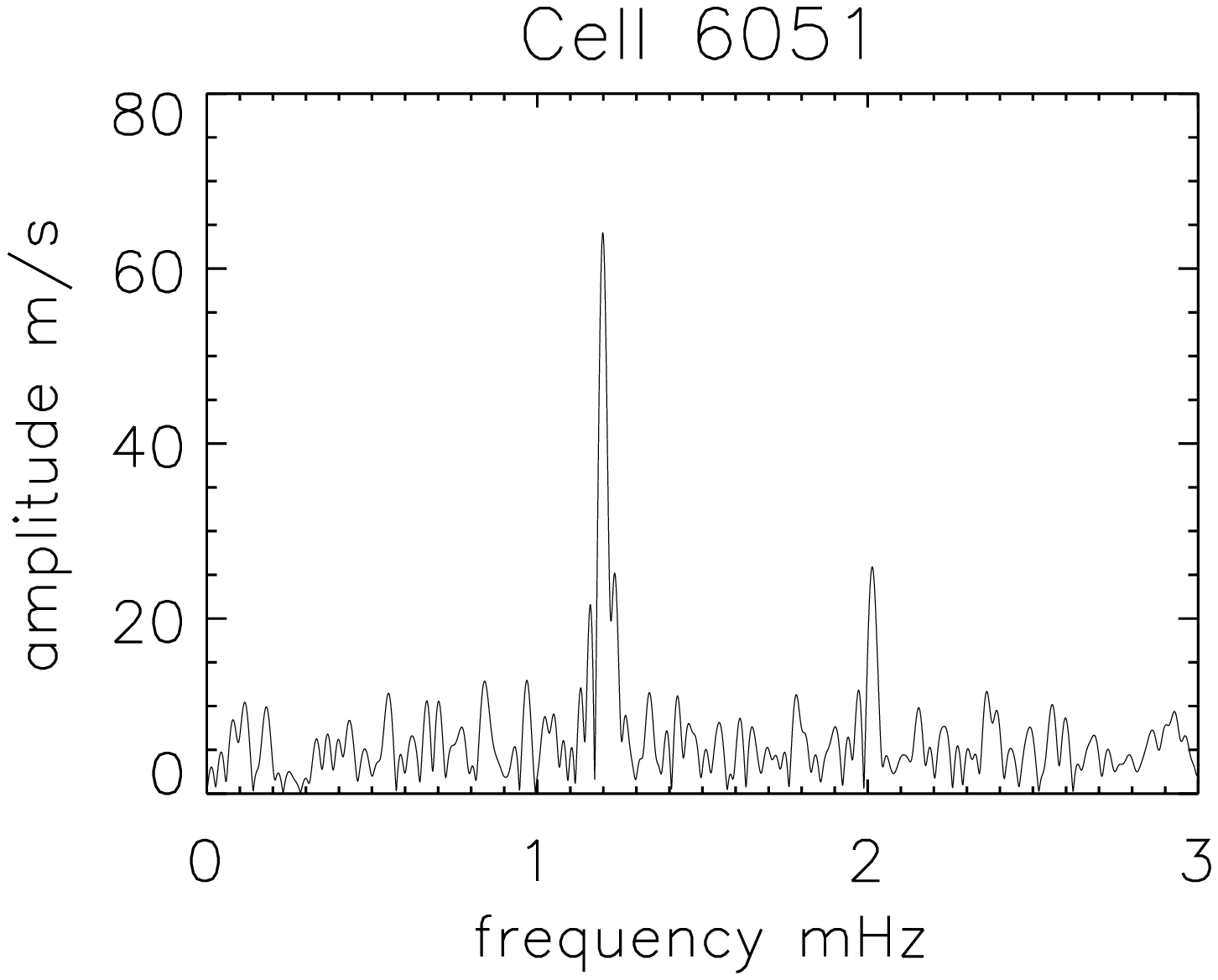}
\epsfxsize 5.cm\epsfbox{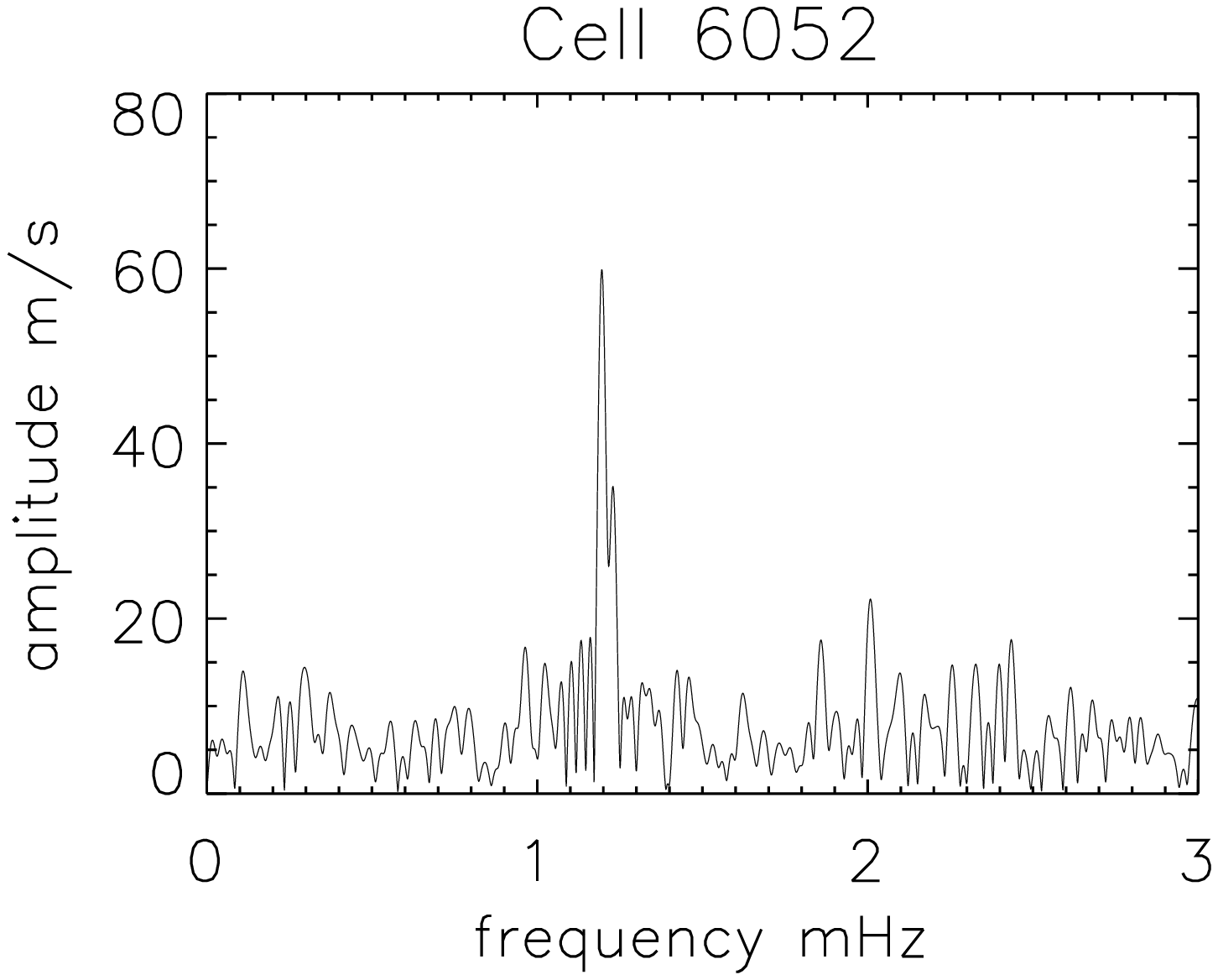}
\epsfxsize 5.cm\epsfbox{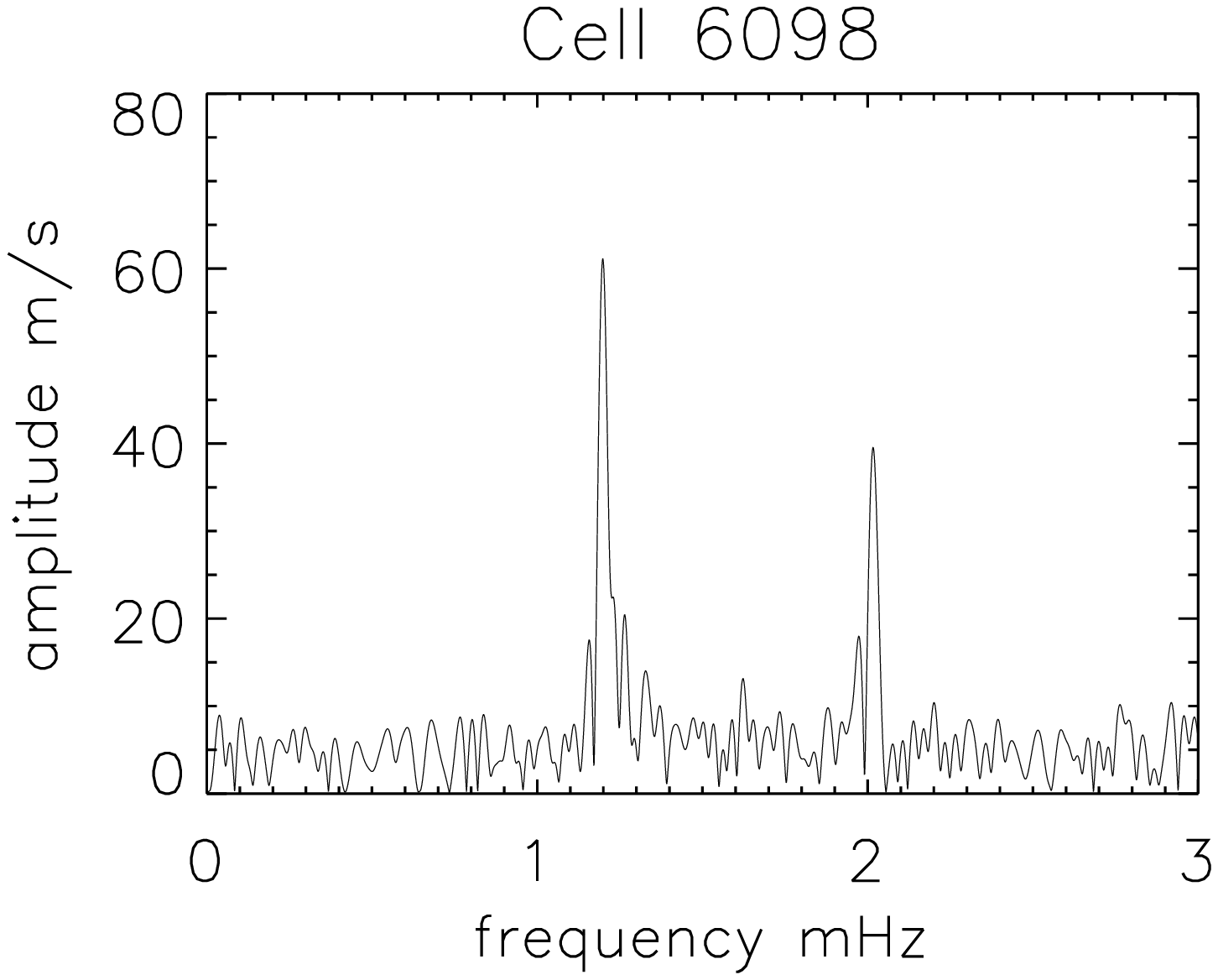}
\epsfxsize 5.cm\epsfbox{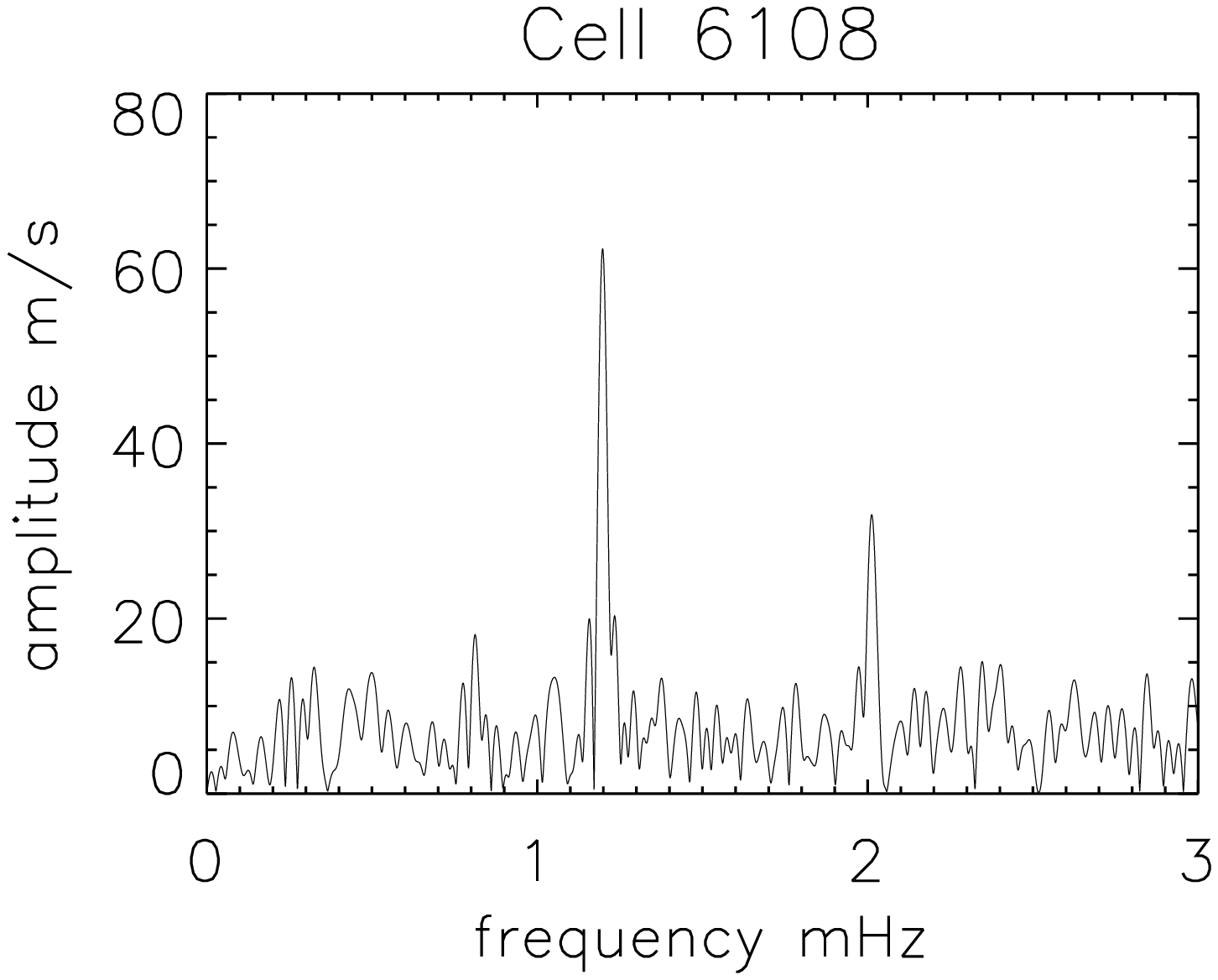}
\epsfxsize 5.cm\epsfbox{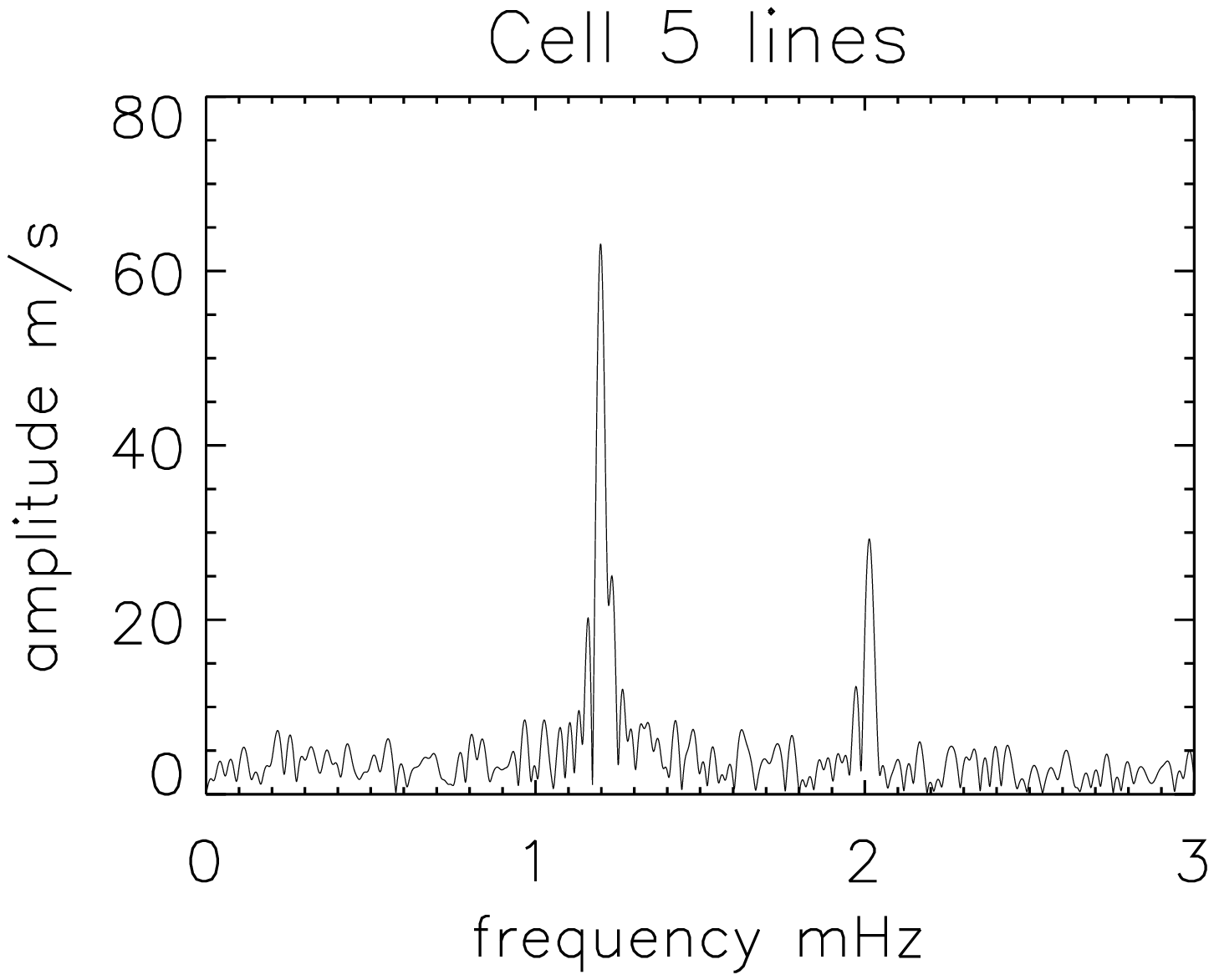}
\caption{\label{fig:ce}
Amplitude spectrum for five lines of Ce~\textsc{ii} showing similar amplitudes for $\nu_1$ and $\nu_2$ for all five lines. The bottom right panel is the amplitude spectrum of all Ce~\textsc{ii} lines combined. }
\end{figure*}

\begin{figure*}
\centering
\includegraphics[width=0.45\linewidth,angle=0]{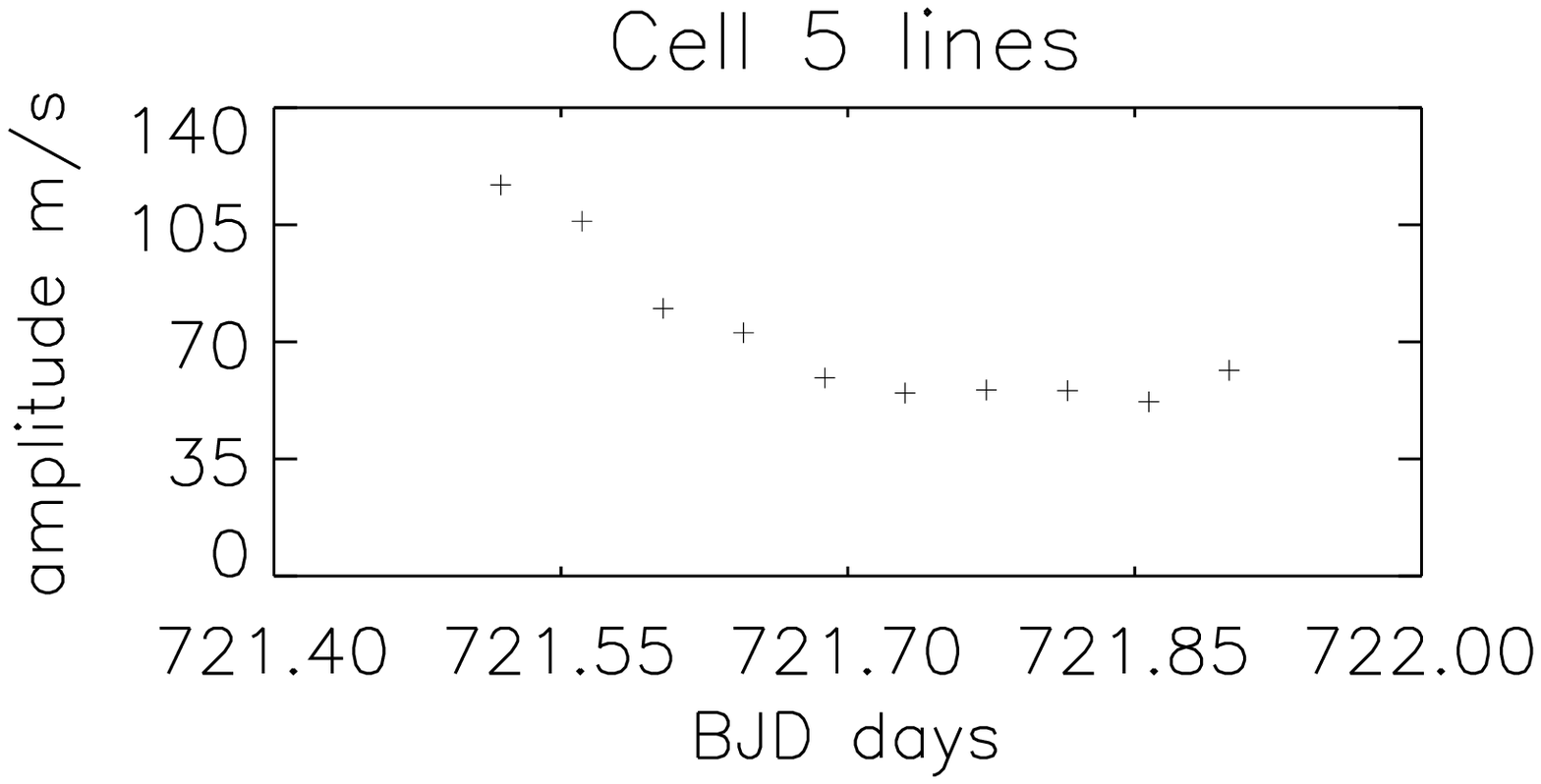}
\vspace{2mm}
\includegraphics[width=0.45\linewidth,angle=0]{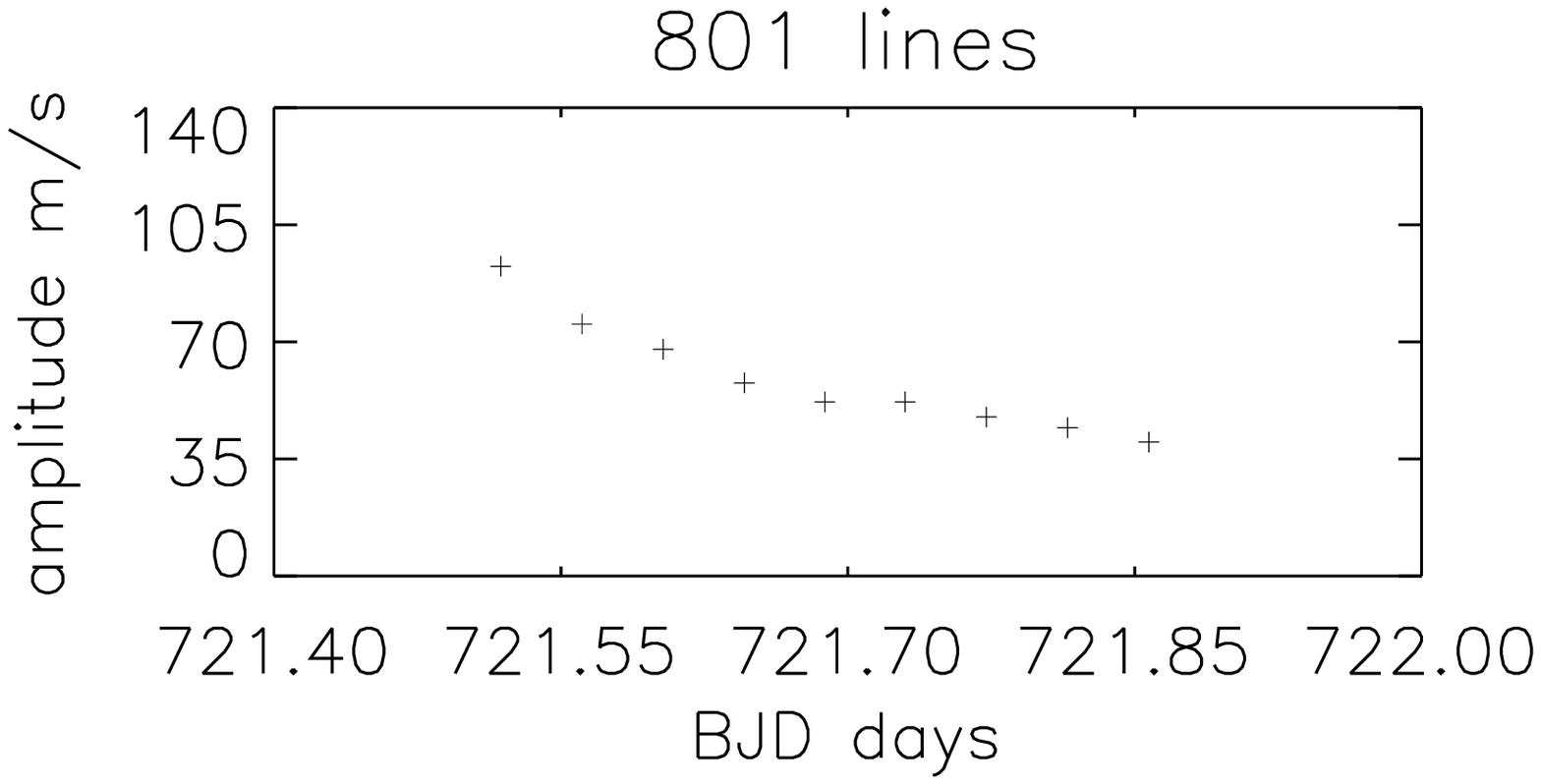}
\includegraphics[width=0.45\linewidth,angle=0]{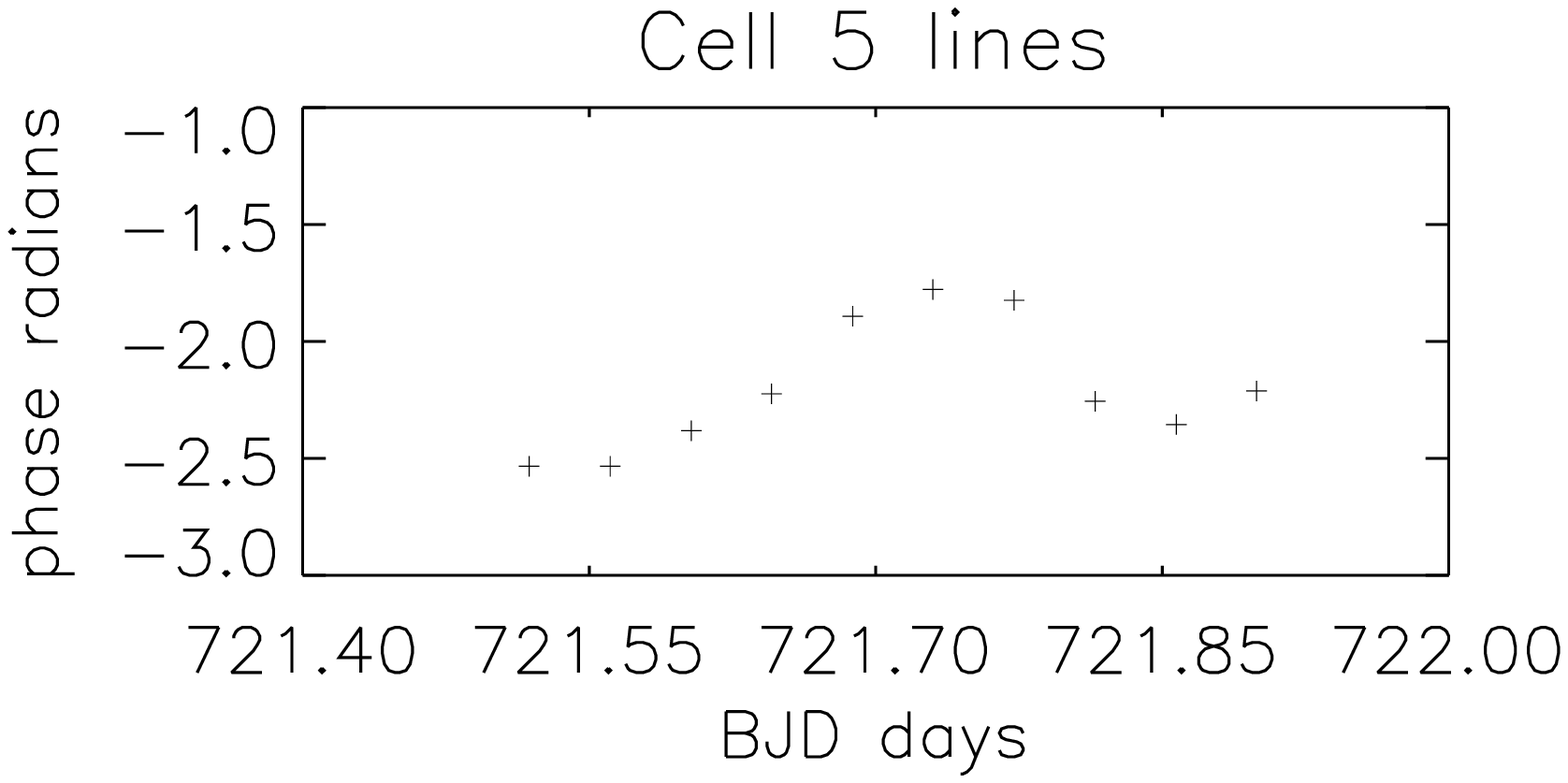}
\vspace{2mm}
\includegraphics[width=0.45\linewidth,angle=0]{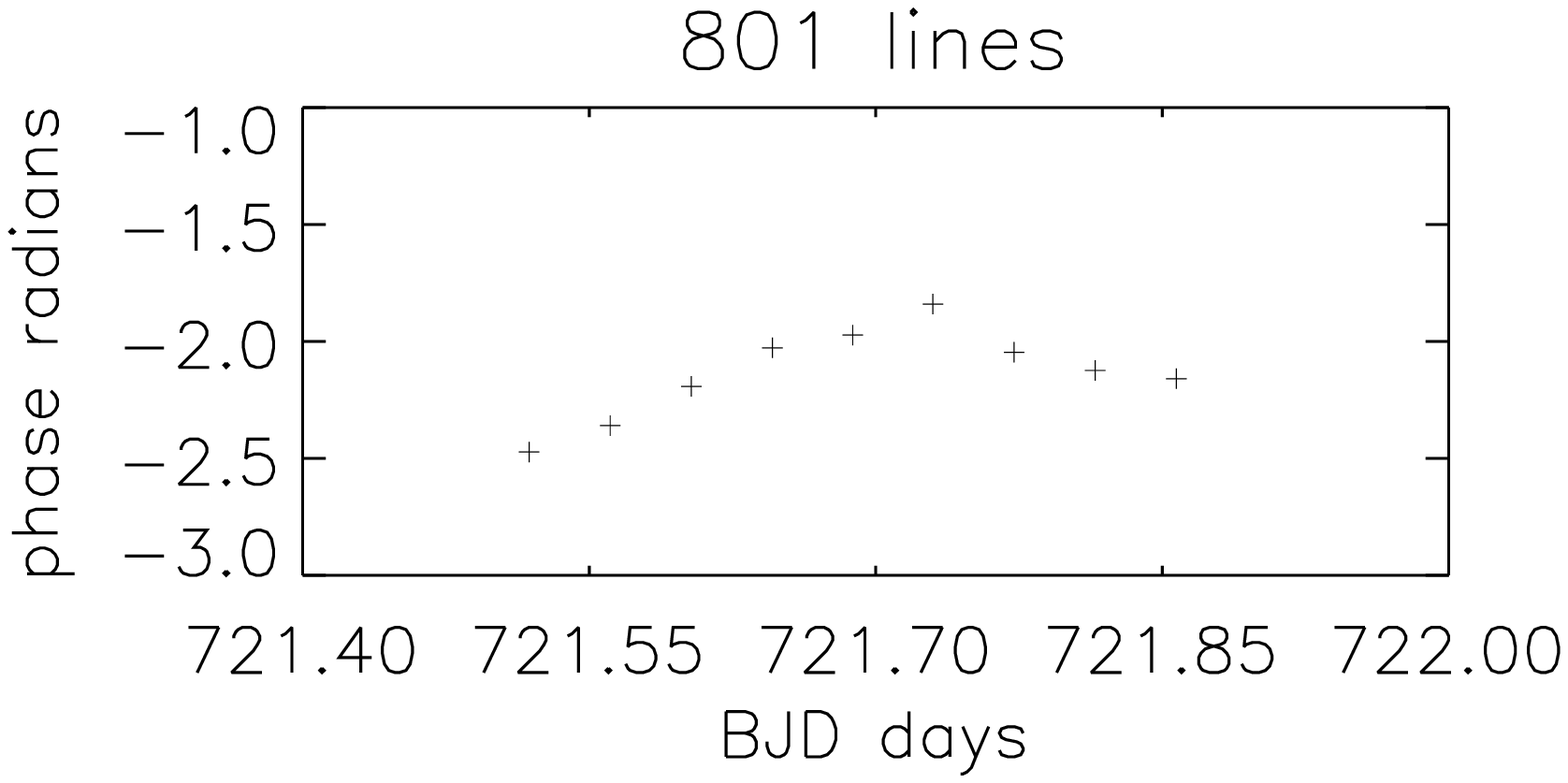}
\caption{The left panels show the radial velocity amplitudes and phases for all five Ce~\textsc{ii} lines as a function of time on the night of 2008 September 11 showing short time scale variability in the amplitude and phase. The right panels show the same for all 801 lines analysed. The time is BJD -- 2454000.0. The formal errors are smaller than the size of the points plotted.}
\label{fig:phamp}
\end{figure*}

Fig.~\ref{fig:ce} shows the amplitude spectra for 5 Ce~\textsc{ii} lines, where it can be seen that the amplitudes for $\nu_1$ and $\nu_2$ are similar for all five lines. To gain better signal-to-noise, we therefore combined the five lines for a frequency analysis and show the amplitude spectrum for that in the bottom right panel. In each case the peak for $\nu_1$ is wider than the spectral window central sinc function as a consequence of amplitude and phase variability over the 10.1-h time span of the data set. We therefore fitted $\nu_1$ to 1-h sections of the data and show the amplitude and phase variability as a function of time in Fig.~\ref{fig:phamp}.  The amplitude of $\nu_1$ drops by a factor of 2, and the phase increases and decreases over a range of 1~rad in the 10.1~h. The phase variations are equivalent to frequency variations. We performed the same analysis on an ensemble of 801 spectral lines from a mixture of ions with the same results, but with a much lower noise in the amplitude and phase determinations as can be seen in the right panels of Fig.~\ref{fig:phamp}, showing that the variations seen are highly significant. 

The question then arises whether the variability of frequency and amplitude of the peaks in the amplitude spectra for $\nu_1$ for both the spectroscopic radial velocities and the photometric variations is a consequence of unresolved stable frequencies, or intrinsic frequency and amplitude variability. Unresolved pulsation modes could be present in the 10.1-h spectroscopic data if the large separation is less than about 30~$\umu$Hz, the frequency resolution of the data set. That is possible for an roAp star; this is the frequency separation for the well-studied case of HR~1217, for example \citep{white2011}. It is also possible for another stable mode to be unresolved in the spectroscopic data if two modes (e.g., ($n, \ell$) and ($n-1, \ell+2$) are separated by the small separation, which is expected to be of the order of a few $\umu$Hz for roAp stars. However, in both of these cases we would be able to resolve the frequencies in the photometric data. Although those suffer from strong daily aliases in the spectral window patterns, stable peaks can be disentangled, even with single-site observations (see \citet{kurtz1982} for examples). We conclude that $\nu_1$ in HD~217522 shows frequency and amplitude variations on a time-scale as short as hours; by inference the lower amplitude $\nu_2$ does also. This conclusion is consistent with the night-to-night variations seen in the amplitude spectra of the photometric data, as well as the more rapid frequency and amplitude variability seen in the spectroscopic radial velocity data. We discuss this in a wider context for roAp stars in Section~\ref{discussion} below.

\section{The missing modes}

\begin{figure}
\centering
\epsfxsize 5.5cm\epsfbox{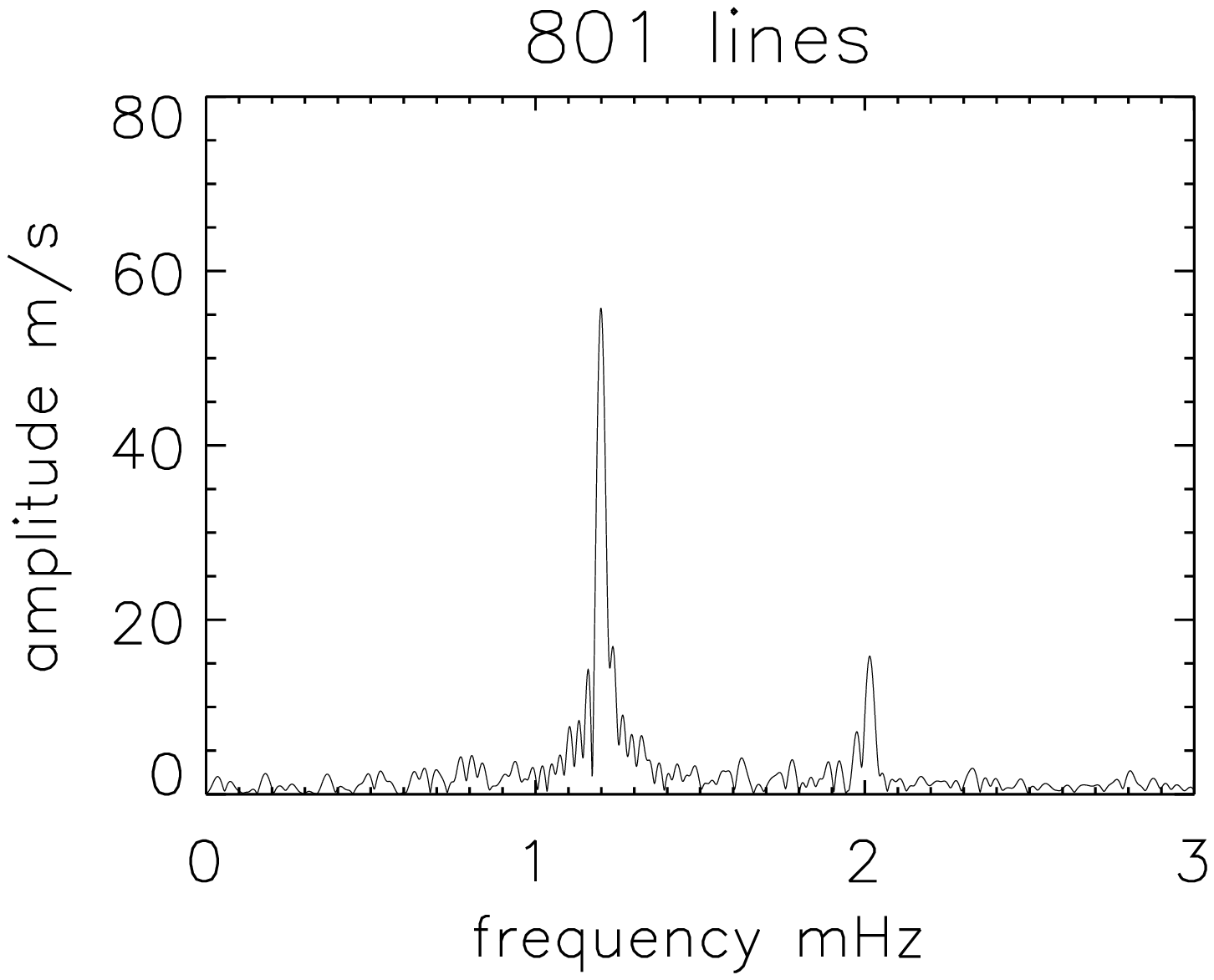}
\epsfxsize 5.5cm\epsfbox{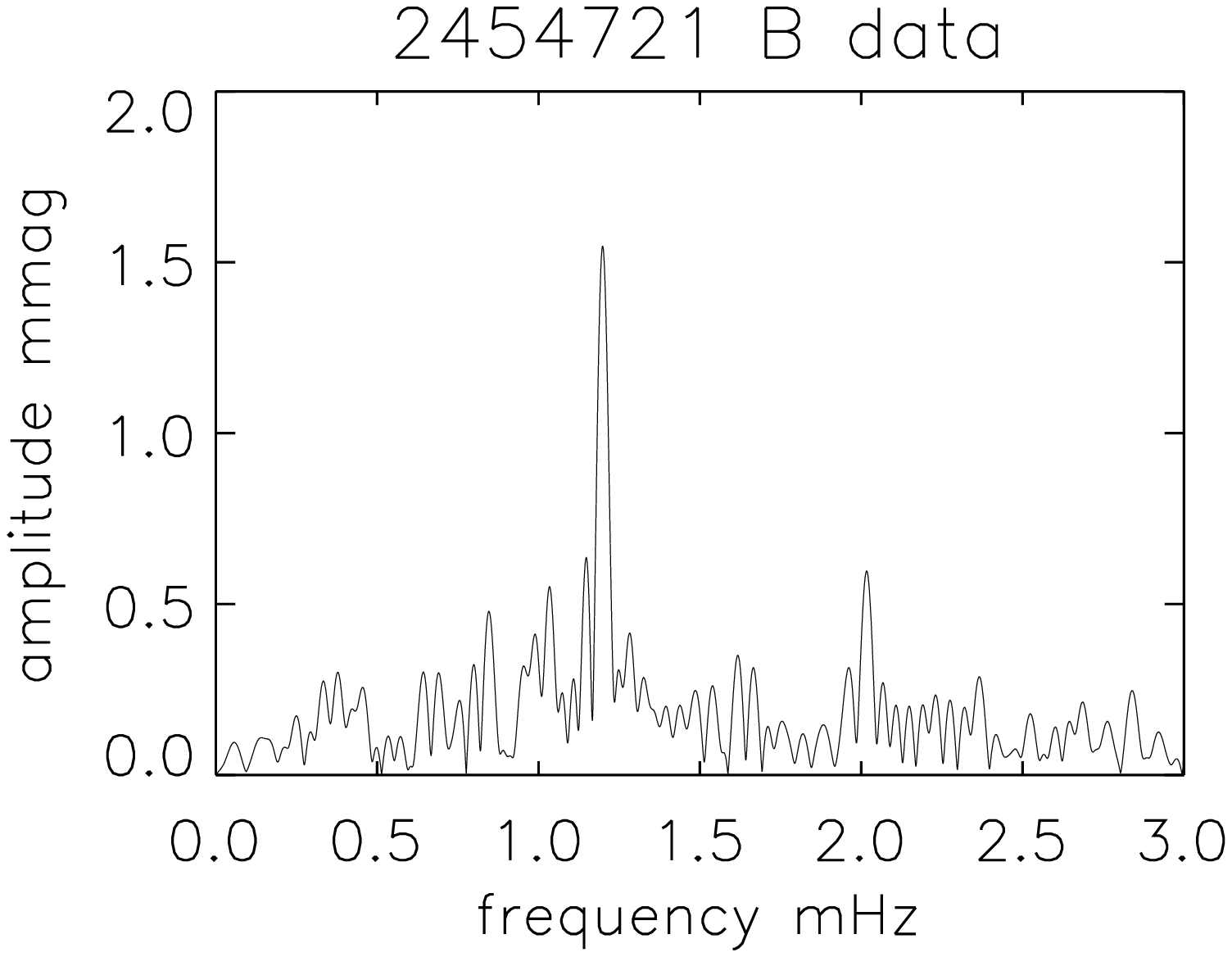}
\caption{\label{fig:801lines}
Top panel: Amplitude spectrum of radial velocity of 801 lines. No missing modes are visible. Bottom panel: the Johnson B amplitude spectrum obtained contemporaneously for comparison.  }
\end{figure}

A goal of our spectroscopic observations was to look for evidence of the missing modes between $\nu_1$ and $\nu_2$. For roAp stars such as HR~1217 \citep{white2011} and HD~60435 \citep{matthews1987} a series of alternating even-and-odd degree modes are seen separated by half the large separation, as expected asymptotically for high overtone p~modes. For HD~217522 we were seeking higher signal-to-noise in radial velocity measurements than can be obtained from ground-based photometric measurements. In Fig.~\ref{fig:ce} the lower right panel shows the amplitude spectrum for 5 Ce~\textsc{ii} lines, where there is no evidence for the missing modes. We take this further by combining the radial velocities for 801 spectral lines of many ions in the top panel of Fig.~\ref{fig:801lines}. There is no evidence for the missing modes at a level of a few m~s$^{-1}$. We conclude that they are not excited, and discuss the implications in Section~\ref{discussion} below.

\section{Discussion}
\label{discussion}

One goal of this investigation was to obtain contemporaneous spectroscopic and photometric observations to test whether the lack of detection of $\nu_2$ in the 1982 photometric data could have been a problem of visibility in the broadband photometry. A comparison of the top and bottom panels of Fig.~\ref{fig:801lines} shows the similarity of the radial velocity and photometric amplitudes for contemporaneous data. Thus we conclude that $\nu_2$ was truly absent in the 1982 photometric data. We now understand that the frequency and amplitude of the pulsation modes associated with $\nu_1$ and $\nu_2$ can modulate on short and long time scales. Long time scale frequency and amplitude variations have been reported in some other roAp stars. \citet{kurtz1997b} found frequency variability in HR~3831 on a time scale of about 1.6~y. In light of frequency in other roAp stars -- including HD~217522 reported in this paper -- their suggestion that this could possibly be cyclic is insecure. \citet{white2011} found variations in pulsation amplitude on times of the order of days in the well-studied roAp star HR~1217. They also found some evidence of phase variations, which is equivalent to frequency variations, hence may be similar to what we have observed for HD~217522. \citet{balona2011} found significant frequency variations with a range of the order 7~$\umu$Hz for the roAp star KIC~10483436 with more than 3~y of {\it Kepler} photometric data with nearly $\umu$mag precision. The best-studied case is now KIC~7582608 for which \citet{holdsworth2014} found frequency variation on a variety of short and long time-scales using a combination of 4~y of {\it Kepler} data and 6~y of SuperWASP data. Significantly, other roAp stars show no frequency variations; the best case for this now is KIC~10195926 which shows no frequency or amplitude variation over the full 4-y {\it Kepler} data set (\citet{balona2011} demonstrates this for the first 3~y of {\it Kepler} data).

It is now apparent that some roAp stars have highly stable pulsation frequencies and amplitudes, even on time scales of years, while other roAp stars show frequency and amplitude variations on time scales as short as hours. Whether this is a result of driving and damping, mode coupling or some instability is not known. Because few of the $\sim$60 known roAp stars have been studied extensively enough to determine whether they have frequency and amplitude variations, progress on this problem will come slowly as data build up. Ultimately, it is important to know where in the roAp instability strip the stable and unstable pulsators lie. HD~217522 is one of the coolest known roAp stars, but the even-cooler HD~101065 has at most much milder changes in frequency and amplitude in various spectroscopic and photometric data sets obtained since its discovery in 1978 (see, e.g. \citealt{mkrtichian2008}). 

\citet{kreidl1991} interpreted the observed frequency change between the 1982 and 1989 data to be due to pulsation mode change. They used models of \citet{hellerkawaler1988} to estimate the luminosity of HD~217522 to be 27 ${\rm L_\odot}$. Using the same interpretation of the frequency separation as \citet{kreidl1991}, \citet{matthews1999} calculated asteroseismic parallax for this star and found it to be inconsistent with that from the Hipparcos catalogue. Matthews et al. also estimated the $T_{\rm eff}$ assuming the Hipparcos parallax and large separation were correct and obtained the value that was different from that obtained from photometric indices by 1500~K! Based on this they concluded that both the large separation and Hipparcos parallax of HD~217522 are wrong. 

Using an analysis similar to \citet{matthews1999} and an improved estimate of $T_{\rm eff}$ measured by fitting the H$\alpha$ line profile in our high signal-to-noise ratio spectra, we find (assuming $M=1.5~M_\odot$) that the large frequency separation consistent with latest parallax measurement for this star by \citep{leeuwen2007} is  $\sim 58~\mu$Hz. The luminosity consistent with the revised Hipparcos parallax is $6.8\pm~0.8~{\rm L_\odot}$. It thus appears that the 15.4~$\umu$Hz change in frequency of $\nu_1$ between 1982 and 1989 was not a change of mode, but was part of the frequency variability of this star on many time scales. It is therefore not surprising that interpreting that frequency difference as half the large separation led to an incorrect value of luminosity. 

\bibliography{HD217522_medupe}

\section*{Acknowledgments}
R.M. thanks the NRF's Thuthuka Grant for funding.

\end{document}